\documentclass[superscriptaddress,twocolumn,twoside,prd,floatfix,letterpaper]{revtex4}

\usepackage{graphicx} 
\usepackage{amsmath}
\usepackage{amssymb}
\usepackage{fancyhdr}
\usepackage{journals}
\usepackage{rotating}
\newcommand{\href}[2]{#2}                   

\def \Vista {{\sc Vista}}
\def \Sleuth {{\sc Sleuth}}

\def \Pythia {{\sc Pythia}}
\def \Herwig {{\sc Herwig}}
\def \MadGraph {{\sc MadGraph}}
\def \MadEvent {{\sc MadEvent}}

\def \CdfSim {{\sc CdfSim}}

\def \DZero {{D\O}}

\def \pmiss {{\ensuremath{{\,/\!\!\!p}}}}
\def \pslash {{\ensuremath{{\,/\!\!\!p}}}}

\def \SumPt {{\ensuremath{\sum{p_T}}}}

\def \scriptP {\ensuremath{{\cal P}}}
\def \tildeScriptP {\ensuremath{\tilde{\cal P}}}
\def \twiddleScriptP {\tildeScriptP}  
\def \detEta {{\ensuremath{\eta_{\text{det}}}}}
\newcommand {\codeBrowserURL}[1]{{http://cdfkits.fnal.gov/SamCode/source/#1?v=Knuteson}}

\def \VistaApproximateDefiniteLuminosity {{927}}
\def \VistaApproximateLuminosity {{\VistaApproximateDefiniteLuminosity}}
\def\clcLuminosity {{902}}
\def \totalNumberOfFudgeFactors {{44}}

\def \pTmin {{17}}
\def \numberOfVistaDiscrepantDistributions {{\ensuremath{384}}}

\newcommand{\highlight}[1]{{#1}}
\newcommand{\highlightB}[1]{{#1}}

\newcommand{\bra}[1]{\langle #1|}
\newcommand{\ket}[1]{|#1\rangle}
\newcommand {\poo}[2]{{\ensuremath{p(#1\!\rightarrow\!#2)}}}
\newcommand {\abs}[1]{\left| #1 \right|}
\newcounter{mylistc}

\newcommand {\godParentComment}[1]{{}}
\newcommand {\cdfSpecific}[1]{{}}
\newcommand {\cdfSpecificHref}[2]{{{#2}}}

\begin{document}

\title{Model-Independent and Quasi-Model-Independent Search for New Physics at CDF}


\affiliation{Institute of Physics, Academia Sinica, Taipei, Taiwan 11529, Republic of China} 
\affiliation{Argonne National Laboratory, Argonne, Illinois 60439} 
\affiliation{Institut de Fisica d'Altes Energies, Universitat Autonoma de Barcelona, E-08193, Bellaterra (Barcelona), Spain} 
\affiliation{Baylor University, Waco, Texas  76798} 
\affiliation{Istituto Nazionale di Fisica Nucleare, University of Bologna, I-40127 Bologna, Italy} 
\affiliation{Brandeis University, Waltham, Massachusetts 02254} 
\affiliation{University of California, Davis, Davis, California  95616} 
\affiliation{University of California, Los Angeles, Los Angeles, California  90024} 
\affiliation{University of California, San Diego, La Jolla, California  92093} 
\affiliation{University of California, Santa Barbara, Santa Barbara, California 93106} 
\affiliation{Instituto de Fisica de Cantabria, CSIC-University of Cantabria, 39005 Santander, Spain} 
\affiliation{Carnegie Mellon University, Pittsburgh, PA  15213} 
\affiliation{Enrico Fermi Institute, University of Chicago, Chicago, Illinois 60637} 
\affiliation{Comenius University, 842 48 Bratislava, Slovakia; Institute of Experimental Physics, 040 01 Kosice, Slovakia} 
\affiliation{Joint Institute for Nuclear Research, RU-141980 Dubna, Russia} 
\affiliation{Duke University, Durham, North Carolina  27708} 
\affiliation{Fermi National Accelerator Laboratory, Batavia, Illinois 60510} 
\affiliation{University of Florida, Gainesville, Florida  32611} 
\affiliation{Laboratori Nazionali di Frascati, Istituto Nazionale di Fisica Nucleare, I-00044 Frascati, Italy} 
\affiliation{University of Geneva, CH-1211 Geneva 4, Switzerland} 
\affiliation{Glasgow University, Glasgow G12 8QQ, United Kingdom} 
\affiliation{Harvard University, Cambridge, Massachusetts 02138} 
\affiliation{Division of High Energy Physics, Department of Physics, University of Helsinki and Helsinki Institute of Physics, FIN-00014, Helsinki, Finland} 
\affiliation{University of Illinois, Urbana, Illinois 61801} 
\affiliation{The Johns Hopkins University, Baltimore, Maryland 21218} 
\affiliation{Institut f\"{u}r Experimentelle Kernphysik, Universit\"{a}t Karlsruhe, 76128 Karlsruhe, Germany} 
\affiliation{Center for High Energy Physics: Kyungpook National University, Taegu 702-701, Korea; Seoul National University, Seoul 151-742, Korea; SungKyunKwan University, Suwon 440-746, Korea; Korea Institute of Science and Technology Information, Daejeon, 305-806, Korea; Chonnam National University, Gwangju, 500-757, Korea} 
\affiliation{Ernest Orlando Lawrence Berkeley National Laboratory, Berkeley, California 94720} 
\affiliation{University of Liverpool, Liverpool L69 7ZE, United Kingdom} 
\affiliation{University College London, London WC1E 6BT, United Kingdom} 
\affiliation{Centro de Investigaciones Energeticas Medioambientales y Tecnologicas, E-28040 Madrid, Spain} 
\affiliation{Massachusetts Institute of Technology, Cambridge, Massachusetts  02139} 
\affiliation{Institute of Particle Physics: McGill University, Montr\'{e}al, Canada H3A~2T8; and University of Toronto, Toronto, Canada M5S~1A7} 
\affiliation{University of Michigan, Ann Arbor, Michigan 48109} 
\affiliation{Michigan State University, East Lansing, Michigan  48824} 
\affiliation{University of New Mexico, Albuquerque, New Mexico 87131} 
\affiliation{Northwestern University, Evanston, Illinois  60208} 
\affiliation{The Ohio State University, Columbus, Ohio  43210} 
\affiliation{Okayama University, Okayama 700-8530, Japan} 
\affiliation{Osaka City University, Osaka 588, Japan} 
\affiliation{University of Oxford, Oxford OX1 3RH, United Kingdom} 
\affiliation{University of Padova, Istituto Nazionale di Fisica Nucleare, Sezione di Padova-Trento, I-35131 Padova, Italy} 
\affiliation{LPNHE, Universite Pierre et Marie Curie/IN2P3-CNRS, UMR7585, Paris, F-75252 France} 
\affiliation{University of Pennsylvania, Philadelphia, Pennsylvania 19104} 
\affiliation{Istituto Nazionale di Fisica Nucleare Pisa, Universities of Pisa, Siena and Scuola Normale Superiore, I-56127 Pisa, Italy} 
\affiliation{University of Pittsburgh, Pittsburgh, Pennsylvania 15260} 
\affiliation{Purdue University, West Lafayette, Indiana 47907} 
\affiliation{University of Rochester, Rochester, New York 14627} 
\affiliation{The Rockefeller University, New York, New York 10021} 
\affiliation{Istituto Nazionale di Fisica Nucleare, Sezione di Roma 1, University of Rome ``La Sapienza,'' I-00185 Roma, Italy} 
\affiliation{Rutgers University, Piscataway, New Jersey 08855} 
\affiliation{Texas A\&M University, College Station, Texas 77843} 
\affiliation{Istituto Nazionale di Fisica Nucleare, University of Trieste/\ Udine, Italy} 
\affiliation{University of Tsukuba, Tsukuba, Ibaraki 305, Japan} 
\affiliation{Tufts University, Medford, Massachusetts 02155} 
\affiliation{Waseda University, Tokyo 169, Japan} 
\affiliation{Wayne State University, Detroit, Michigan  48201} 
\affiliation{University of Wisconsin, Madison, Wisconsin 53706} 
\affiliation{Yale University, New Haven, Connecticut 06520} 
\author{T.~Aaltonen}
\affiliation{Division of High Energy Physics, Department of Physics, University of Helsinki and Helsinki Institute of Physics, FIN-00014, Helsinki, Finland}
\author{A.~Abulencia}
\affiliation{University of Illinois, Urbana, Illinois 61801}
\author{J.~Adelman}
\affiliation{Enrico Fermi Institute, University of Chicago, Chicago, Illinois 60637}
\author{T.~Akimoto}
\affiliation{University of Tsukuba, Tsukuba, Ibaraki 305, Japan}
\author{M.G.~Albrow}
\affiliation{Fermi National Accelerator Laboratory, Batavia, Illinois 60510}
\author{B.~\'{A}lvarez~Gonz\'{a}lez}
\affiliation{Instituto de Fisica de Cantabria, CSIC-University of Cantabria, 39005 Santander, Spain}
\author{S.~Amerio}
\affiliation{University of Padova, Istituto Nazionale di Fisica Nucleare, Sezione di Padova-Trento, I-35131 Padova, Italy}
\author{D.~Amidei}
\affiliation{University of Michigan, Ann Arbor, Michigan 48109}
\author{A.~Anastassov}
\affiliation{Rutgers University, Piscataway, New Jersey 08855}
\author{A.~Annovi}
\affiliation{Laboratori Nazionali di Frascati, Istituto Nazionale di Fisica Nucleare, I-00044 Frascati, Italy}
% ...
\collaboration{CDF Collaboration\footnote{With visitors from $^a$University of Athens, 15784 Athens, Greece, 
$^b$Chinese Academy of Sciences, Beijing 100864, China, 
$^c$University of Bristol, Bristol BS8 1TL, United Kingdom, 
$^d$University Libre de Bruxelles, B-1050 Brussels, Belgium, 
$^e$University of California, Irvine, Irvine, CA  92697, 
$^f$University of California Santa Cruz, Santa Cruz, CA  95064, 
$^g$Cornell University, Ithaca, NY  14853, 
$^h$University of Cyprus, Nicosia CY-1678, Cyprus, 
$^i$University College Dublin, Dublin 4, Ireland, 
$^j$University of Edinburgh, Edinburgh EH9 3JZ, United Kingdom, 
$^k$University of Heidelberg, D-69120 Heidelberg, Germany, 
$^l$Universidad Iberoamericana, Mexico D.F., Mexico, 
$^m$University of Manchester, Manchester M13 9PL, England, 
$^n$Nagasaki Institute of Applied Science, Nagasaki, Japan, 
$^o$University de Oviedo, E-33007 Oviedo, Spain, 
$^p$Queen Mary's College, University of London, London, E1 4NS, England, 
$^q$Texas Tech University, Lubbock, TX  79409, 
$^r$IFIC(CSIC-Universitat de Valencia), 46071 Valencia, Spain.
}}
\noaffiliation



\begin{abstract}
Data collected in Run II of the Fermilab Tevatron are searched for indications of new electroweak scale physics.  Rather than focusing on particular new physics scenarios, CDF data are analyzed for discrepancies with respect to the standard model prediction.  A model-independent approach (\Vista) considers the gross features of the data, and is sensitive to new large cross section physics.  A quasi-model-independent approach (\Sleuth) searches for a significant excess of events with large summed transverse momentum, and is particularly sensitive to new electroweak scale physics that appears predominantly in one final state.  This global search for new physics in over three hundred exclusive final states in \VistaApproximateLuminosity~pb$^{-1}$ of $p\bar{p}$ collisions at $\sqrt{s}=1.96$~TeV reveals no such significant indication of physics beyond the standard model. 
\end{abstract}
\maketitle

\clearpage
\tableofcontents


\vspace{1in}
\section{Introduction}


In the past thirty years of frontier energy collider physics, the only objects discovered were those for which the predictions were definite.  The W and Z bosons (discovered at CERN in 1983~\cite{UA1Wdiscovery:Arnison:1983rp,UA2Wdiscovery:Banner:1983jy,UA1Zdiscovery:Arnison:1983mk,UA2Zdiscovery:Bagnaia:1983zx}) and the top quark (discovered at Fermilab in 1995~\cite{CDFTopDiscovery:Abe:1995hr,D0TopDiscovery:Abachi:1995iq}) were well known objects long before their discoveries, with all quantum numbers but mass uniquely specified.  The present situation is qualitatively different, with plausible predictions for physics lying beyond the standard model running the gamut of possible experimental signatures.  

Searches for physics beyond the standard model typically begin with a particular model.  A region is selected in the data where the model's expected contribution is enhanced, and the extent to which the data (dis)favor the model is determined by comparing the prediction to data.


The state of the theoretical landscape and the vastness of most model spaces suggests the utility of searching in a different space altogether.  The experimental space, defined by the isolated and energetic objects observed in frontier energy collisions, forms a natural space to consider.

This article describes a systematic and model-independent look (\Vista) at gross features of the data, and a quasi-model-independent search (\Sleuth) for new physics at high transverse momentum.  These global algorithms provide a complementary approach to searches optimized for more specific new physics scenarios.

\section{Strategy}
\label{sec:Strategy}

The search for new physics described in this article is designed with the intention of maximizing the chance for discovery, and not excluding model parameter space if no discrepancy is found.  Discrepancies between data and \highlight{a complete} standard model background estimate are identified in a global sample of high transverse momentum (high-$p_T$) collision events.  Three statistics are employed to identify and quantify disagreement: populations of exclusive final states defined by the objects the events contain, shapes of kinematic distributions, and excesses on the tail of summed scalar transverse momentum distributions.

The \Vista~\footnote{The name ``\Vista'' (Spanish or Italian for ``view'' or ``sight'') reflects the goal of obtaining a global, panoramic view of data collected at the energy frontier.} algorithm provides a global study of the standard model prediction and CDF detector response in the bulk of the high-$p_T$ data; an algorithm called \Sleuth\ complements this with a search for possibly small cross section physics in the high-$p_T$ tails. The purpose of these algorithms is to identify discrepancies worthy of further consideration.

A claim of discovery requires convincing arguments that the observed discrepancy between data and standard model prediction
\begin{enumerate}
\item is not a statistical fluctuation,
\item is not due to a mismodeling of the detector response, and
\item is not due to an inadequate implementation of the standard model prediction,
\end{enumerate}
and therefore must be due to new underlying physics. 
Any observed discrepancy is subject to scrutiny, and explanations are sought in terms of the above points.

The \Vista\ and \Sleuth\ algorithms provide a means for making the above three arguments, with a high threshold placed on the statistical significance of a discrepancy in order to minimize the chance of a false discovery  claim. As described later, this threshold is the requirement that the false discovery rate is less than 0.001, after taking into account the total number of final states, distributions, or regions being examined.

This analysis employs a correction model implementing specific hypotheses to account for mismodeling of detector response and imperfect implementation of standard model prediction.  Achieving this on the entire high-$p_T$ dataset requires a framework for quickly implementing and testing modifications to the correction model. The specific details of the correction model are intentionally kept as simple as possible in the interest of transparency in the event of a possible new physics claim. \Vista's toolkit includes a global comparison of data to the standard model prediction, with a check of thousands of kinematic distributions and an easily adjusted correction model allowing a quick fit for values of associated correction factors.  

The traditional notions of signal and control regions are modified.  Without prejudice as to where new physics may appear, all regions of the data are treated as both signal and control.  This analysis is not blind, but rather seeks to identify and understand discrepancies between data and the standard model prediction.  With the goal of discovery, emphasis is placed on examining discrepancies, focusing on outliers rather than global goodness of fit.  Individual discrepancies that are not statistically significant are generally not pursued.

\Vista\ and \Sleuth\ are employed simultaneously, rather than sequentially.  An effect highlighted by \Sleuth\ prompts additional investigation of the discrepancy, usually resulting in a specific hypothesis explaining the discrepancy in terms of a detector effect or adjustment to the standard model prediction that is then fed back and tested for global consistency using \Vista.

Forming hypotheses for the cause of specific discrepancies, implementing those hypotheses to assess their wider consequences, and testing global agreement after the implementation are emphasized as the crucial activities for the investigator throughout the process of data analysis.  This process is constrained by the requirement that all adjustments be physically motivated.  The investigation and resolution of discrepancies highlighted by the algorithms is {\em{the}} defining characteristic of this global analysis~\footnote{It is not possible to systematically simulate the process of constructing, implementing, and testing hypotheses motivated by particular discrepancies, since this process is carried out by individuals.  The statistical interpretation of this analysis is made bearing this process in mind.}.

This search for new physics terminates when one of two conditions are satisfied: either a compelling case for new physics is made, or there remain no statistically significant discrepancies on which a new physics case can be made. In the former case, to quantitatively assess the significance of the potential discovery, a full treatment of systematic uncertainties must be implemented. In the latter case, it is sufficient to demonstrate that all observed effects are not in significant disagreement with an appropriate global Standard Model description.


\section{\Vista}
\label{sec:Vista}

This section describes \Vista: object identification, event selection, estimation of standard model backgrounds, simulation of the CDF detector response, development of a correction model, and results.

\subsection{CDF II detector}
\label{sec:CdfDetector}

CDF II is a general purpose detector~\cite{CdfDetectorTDR,CdfDetector:Acosta:2004yw} designed to detect particles produced in $p\bar{p}$ collisions.  The detector has a cylindrical layout centered on the accelerator beamline.  

CDF uses a cylindrical coordinate system with the $z$-axis along the axis of the colliding beams.  The variable $\theta$ is the polar angle relative to the incoming proton beam, and the variable $\phi$ is the azimuthal angle about the beam axis.  The pseudorapidity of a particle trajectory is defined as $\eta = -\ln(\tan(\theta/2))$.  It is also useful to define detector pseudorapidity $\detEta$, denoting a particle's pseudorapidity in a coordinate system in which the origin lies at the center of the CDF detector rather than at the event vertex.  The transverse momentum $p_T$ is the component of the momentum projected on a plane perpendicular to the beam axis.  

Charged particle tracks are reconstructed in a 3.1~m long open cell drift chamber that performs up to 96 measurements of the track position in the radial region from 0.4~m to 1.4~m. Between the beam pipe and this tracking chamber are multiple layers of silicon microstrip detectors, enabling high precision determination of the impact parameter of a track relative to the primary event vertex.  The tracking detectors are immersed in a uniform 1.4~T solenoidal magnetic field.

Outside the solenoid, calorimeter modules are arranged in a projective tower geometry to provide energy measurements for both charged and neutral particles.  Proportional chambers are embedded in the electromagnetic calorimeters to measure the transverse profile of electromagnetic showers at a depth corresponding to the shower maximum for electrons. The outermost part of the detector consists of a series of drift chambers used to detect and identify muons, minimum ionizing particles that typically pass through the calorimeter. 

A set of forward gas \v{C}erenkov detectors is used to measure the average number of inelastic $p\bar{p}$ collisions per Tevatron bunch crossing, and hence determine the luminosity acquired.  A three level trigger and data acquisition system selects the most interesting collisions for offline analysis.

Here and below the word ``central'' is used to describe objects with $\abs{\detEta}<1.0$; ``plug'' is used to describe objects with $1.0<\abs{\detEta}<2.5$.


\subsection{Object identification}

Energetic and isolated electrons, muons, taus, photons, jets, and $b$-tagged jets with $\abs{\detEta}<2.5$ and $p_T>17$~GeV are identified according to standard criteria.  The same criteria are used for all events.  The isolation criteria employed vary according to object, but roughly require less than 2~GeV of extra energy flow within a cone of $\Delta R = \sqrt{\Delta \eta^2 + \Delta \phi^2} = 0.4$ in $\eta$--$\phi$ space around each object.

Standard CDF criteria~\cite{WandZCrossSectionPaper:Abulencia:2005ix} are used to identify electrons ($e^\pm$) in the central and plug regions of the CDF detector.  Electrons are characterized by a narrow shower in the central or plug electromagnetic calorimeter and a matching isolated track in the central gas tracking chamber or a matching plug track in the silicon detector.  

Standard CDF muons ($\mu^\pm$) are identified using three separate subdetectors in the regions $\abs{\detEta}<0.6$, $0.6<\abs{\detEta}<1.0$, and $1.0<\abs{\detEta}<1.5$~\cite{WandZCrossSectionPaper:Abulencia:2005ix}.  Muons are characterized by a track in the central tracking chamber matched to a track segment in the central muon detectors, with energy consistent with minimum ionizing deposition in the electromagnetic and hadronic calorimeters along the muon trajectory.

Narrow central jets with a single charged track are identified as tau leptons ($\tau^\pm$) that have decayed hadronically~\cite{Ztautau}.  Taus are distinguished from electrons by requiring a substantial fraction of their energy to be deposited in the hadron calorimeter; taus are distinguished from muons by requiring no track segment in the muon detector coinciding with the extrapolated track of the tau.  Track and calorimeter isolation requirements are imposed.

Standard CDF criteria requiring the presence of a narrow electromagnetic cluster with no associated tracks are used to identify photons ($\gamma$) in the central and plug regions of the CDF detector~\cite{CdfPhotonId:Acosta:2004sn}.

Jets ($j$) are reconstructed using the JetClu~\cite{JetClu:Abe:1991ui} clustering algorithm with a cone of size $\Delta R = 0.4$, unless the event contains one or more jets with $p_T>200$~GeV and no leptons or photons, in which case cones of $\Delta R = 0.7$ are used.~\cdfSpecific{\footnote{Jet energies are corrected to level 7, using {\tt jetCorr04b}.}}Jet energies are appropriately corrected to the parton level~\cite{jetEnergyScale:Bhatti:2005ai}. Since uncertainties in the standard model prediction grow with increasing jet multiplicity, up to the four largest $p_T$ jets are used to characterize the event; any reconstructed jets with $p_T$-ordered ranking of five or greater are neglected, except in final states with small summed scalar transverse momentum containing only jets.

A secondary vertex $b$-tagging algorithm is used to identify jets likely resulting from the fragmentation of a bottom quark ($b$) produced in the hard scattering~\cite{CdfBtagging:Neu:2006rs}.

Momentum visible in the detector but not clustered into an electron, muon, tau, photon, jet, or $b$-tagged jet is referred to as unclustered momentum ({\tt uncl}).

Missing momentum ($\pmiss$) is calculated as the negative vector sum of the 4-vectors of all identified objects and unclustered momentum.  An event is said to contain a $\pmiss$ object if the transverse momentum of this object exceeds \pTmin~GeV, and if additional quality criteria discriminating against fake missing momentum due to jet mismeasurement are satisfied~\footnote{An additional quality criterion is applied to the significance of the missing transverse momentum $\vec{\pmiss}_T$ in an event, requiring that the energies of hadronic objects cannot be adjusted within resolution to reduce the missing transverse momentum to less than 10~GeV.  The transverse components of all hadronic energy clusters $\vec{p}_{Ti}$ in the event are projected onto the unit missing transverse momentum vector $\hat{\pmiss}_T=\vec{\pmiss}_T/\abs{\vec{\pmiss}_T}$, and a ``conservative'' missing transverse momentum ${\pmiss_T}'=\pmiss_T - 2.5 \sqrt{ \sum_i{ \abs{ \vec{p}_{Ti} \cdot \hat{\pmiss}_T} } }$ is defined, where the sum is over hadronic energy clusters in the event, and the hadronic energy resolution of the CDF detector has been approximated as $100\% \sqrt{{p_T}_i}$, expressed in GeV.  An event is said to contain missing transverse momentum if $\pmiss_T>\pTmin$~GeV and ${\pmiss_T}'>10$~GeV.}.


\subsection{Event selection}
\label{sec:Vista:OfflineTrigger}

Events containing an energetic and isolated electron, muon, tau, photon, or jet are selected.  A set of three level online triggers requires:
\begin{itemize}
\item a central electron candidate with $p_T>18$~GeV passing level 3, with an associated track having $p_T>8$~GeV and an electromagnetic energy cluster with $p_T>16$~GeV at levels 1 and 2; or
\item a central muon candidate with $p_T>18$~GeV passing level 3, with an associated track having $p_T>15$~GeV and muon chamber track segments at levels 1 and 2; or
\item a central or plug photon candidate with $p_T>25$~GeV passing level 3, with hadronic to electromagnetic energy less than 1:8 and with energy surrounding the photon to the photon's energy less than 1:7 at levels 1 and 2; or
\item a central or plug jet with $p_T>20$~GeV passing level 3, with 15~GeV of transverse momentum required at levels 1 and 2, with corresponding prescales of 50 and 25, respectively; or
\item a central or plug jet with $p_T>100$~GeV passing level 3, with energy clusters of 20~GeV and 90~GeV required at levels 1 and 2; or
\item a central electron candidate with $p_T>4$~GeV and a central muon candidate with $p_T>4$~GeV passing level 3, with a muon segment, electromagnetic cluster, and two tracks with $p_T>4$~GeV required at levels 1 and 2; or
\item a central electron or muon candidate with $p_T>4$~GeV and a plug electron candidate with $p_T>8$~GeV, requiring a central muon segment and track or central electromagnetic energy cluster and track at levels 1 and 2, together with an isolated plug electromagnetic energy cluster; or
\item two central or plug electromagnetic clusters with $p_T>18$~GeV passing level 3, with hadronic to electromagnetic energy less than 1:8 at levels 1 and 2; or
\item two central tau candidates with $p_T>10$~GeV passing level 3, each with an associated track having $p_T>10$~GeV and a calorimeter cluster with $p_T>5$~GeV at levels 1 and 2.
\end{itemize}

Events satisfying one or more of these online triggers are recorded for further study.  
Offline event selection for this analysis uses a variety of further filters.
Single object requirements keep events containing:
\begin{itemize}
\item a central electron with $p_T>25$~GeV, or
\item a plug electron with $p_T>40$~GeV, or
\item a central muon with $p_T>25$~GeV, or
\item a central photon with $p_T>60$~GeV, or
\item a central jet or $b$-tagged jet with $p_T>200$~GeV, or
\item a central jet or $b$-tagged jet with $p_T>40$~GeV (prescaled by a factor of roughly $10^4$),
\end{itemize}
possibly with other objects present.  
Multiple object criteria select events containing:
\begin{itemize}
\item two electromagnetic objects (electron or photon) with $\abs{\eta}<2.5$ and $p_T>25$~GeV, or
\item two taus with $\abs{\eta}<1.0$ and $p_T>17$~GeV, or
\item a central electron or muon with $p_T>17$~GeV and a central or plug electron, central muon, or central tau with $p_T>17$~GeV, or
\item a central photon with $p_T>40$~GeV and a central electron or muon with $p_T>17$~GeV, or
\item a central or plug photon with $p_T>40$~GeV and a central tau with $p_T>40$~GeV, or
\item a central photon with $p_T>40$~GeV and a central $b$-jet with $p_T>25$~GeV, or
\item a central jet or $b$-tagged jet with $p_T>40$~GeV and a central tau with $p_T>17$~GeV (prescaled by a factor of roughly $10^3$), or
\item a central or plug photon with $p_T>40$~GeV and two central taus with $p_T>17$~GeV, or
\item a central or plug photon with $p_T>40$~GeV and two central $b$-tagged jets with $p_T>25$~GeV, or
\item a central or plug photon with $p_T>40$~GeV, a central tau with $p_T>25$~GeV, and a central $b$-tagged jet with $p_T>25$~GeV,
\end{itemize}
possibly with other objects present.  
Explicit online triggers feeding this offline selection are required.  The $p_T$ thresholds for these criteria are chosen to be sufficiently above the online trigger turn-on curves  that trigger efficiencies can be treated as roughly independent of object $p_T$.

Good run criteria are imposed, requiring the operation of all major subdetectors.  To reduce contributions from cosmic rays and events from beam halo, standard CDF cosmic ray and beam halo filters are applied~\cite{CdfCosmicFilter}.

These selections result in a sample of roughly two million high-$p_T$ data events in an integrated luminosity of \VistaApproximateLuminosity~pb$^{-1}$.


\subsection{Event generation}
\label{sec:Vista:EventGeneration}

Standard model backgrounds are estimated by generating a large sample of Monte Carlo events, using the \Pythia~\cite{Pythia:Sjostrand:2000wi}, \MadEvent~\cite{MadEvent:Maltoni:2002qb2}, and \Herwig~\cite{Herwig:Corcella:2002jc} generators.  \MadEvent\ performs an exact leading order matrix element calculation, and provides 4-vector information corresponding to the outgoing legs of the underlying Feynman diagrams, together with color flow information.  \Pythia\ 6.218 is used to handle showering and fragmentation.  The CTEQ5L~\cite{CTEQ5L:Lai:1999wy} parton distribution functions are used.  

\paragraph*{$\text{QCD jets}$.}
QCD dijet and multijet production are estimated using \Pythia.  Samples are generated with Tune A~\cite{RickFieldPythiaTunes} with lower cuts on $\hat{p}_T$, the transverse momentum of the scattered partons in the center of momentum frame of the incoming partons, of 0, 10, 18, 40, 60, 90, 120, 150, 200, 300, and 400~GeV.  These samples are combined to provide a complete estimation of QCD jet production, using the sample with greatest statistics in each range of $\hat{p}_T$.  
\paragraph*{$\gamma\text{+jets}$.}
The estimation of QCD single prompt photon production comes from \Pythia.  Five samples are generated with Tune A corresponding to lower cuts on $\hat{p}_T$ of 8, 12, 22, 45, and 80~GeV.  These samples are combined to provide a complete estimation of single prompt photon production in association with one or more jets, placing cuts on $\hat{p}_T$ to avoid double counting.  
\paragraph*{$\gamma\gamma\text{+jets}$.}
QCD diphoton production is estimated using \Pythia.  
\paragraph*{$V\text{+jets}$.}
The estimation of $V$+jets processes (with $V$ denoting $W$ or $Z$), where the $W$ or $Z$ decays to first or second generation leptons, comes from \MadEvent, with \Pythia\ employed for showering.  Tune AW~\cite{RickFieldPythiaTunes} is used within \Pythia\ for these samples.  The CKKW matching prescription~\cite{CKKW:Krauss:2002up} with a matching scale of 15~GeV is used to combine these samples and avoid double counting.  Additional statistics are generated on the high-$p_T$ tails using the MLM matching prescription~\cite{MrennaMatching:Mrenna:2003if}.  The factorization scale is set to the vector boson mass; the renormalization scale for each vertex is set to the $p_T$ of the jet.  $W$+4 jets are  generated inclusively in the number of jets; $Z$+3 jets are generated inclusively in the number of jets.  
\paragraph*{$VV\text{+jets}$.}
The estimation of $WW$, $WZ$, and $ZZ$ production with zero or more jets comes from \Pythia.  
\paragraph*{$V\gamma\text{+jets}$.}
The estimation of $W\gamma$ and $Z\gamma$ production comes from \MadEvent, with showering provided by \Pythia.  These samples are inclusive in the number of jets.
\paragraph*{$W(\rightarrow\tau\nu)\text{+jets}$.}
Estimation of $W\rightarrow\tau\nu$ with zero or more jets comes from \Pythia.
\paragraph*{$Z(\rightarrow\tau\tau)\text{+jets}$.}
Estimation of $Z\rightarrow\tau\tau$ with zero or more jets comes from \Pythia.  
\paragraph*{$t\bar{t}$.}
Top quark pair production is estimated using \Herwig\ assuming a top quark mass of 175~GeV and NNLO cross section of $6.77\pm 0.42$~pb~\cite{Kidonakis:2003qe}.
 
Remaining processes, including for example $Z(\rightarrow\nu\bar{\nu})\gamma$ and $Z(\rightarrow\ell^+\ell^-)b\bar{b}$, are generated by systematically looping over possible final state partons, using \MadGraph~\cite{MadGraph:Stelzer:1994ta} to determine all relevant diagrams, and using \MadEvent\ to perform a Monte Carlo integration over the final state phase space and to generate events.    The MLM matching prescription is employed to combine samples with different numbers of final state jets.

A higher statistics estimate of the high-$p_T$ tails is obtained by computing the thresholds in $\sum{p_T}$ corresponding to the top 10\% and 1\% of each process, where $\sum{p_T}$ denotes the scalar summed transverse momentum of all identified objects in an event.  Roughly ten times as many events are generated for the top 10\%, and roughly one hundred times as many events are generated for the top 1\%.  

\paragraph*{Cosmic rays.}
Backgrounds from cosmic ray or beam halo muons that interact with the hadronic or electromagnetic calorimeters, producing objects that look like a photon or jet, are estimated using a sample of data events containing fewer than three reconstructed tracks. This procedure is described in more detail in Appendix~\ref{sec:CorrectionModelDetails:CosmicRays}. 

\paragraph*{Minimum bias.}
Minimum bias events are overlaid according to run-dependent instantaneous luminosity in some of the Monte Carlo samples, including those used for inclusive $W$ and $Z$ production.  In all samples not containing overlaid minimum bias events, including those used to estimate QCD dijet production, additional unclustered momentum is added to events to mimic the effect of the majority of multiple interactions, in which a soft dijet event accompanies the rare hard scattering of interest.  A random number is drawn from a Gaussian centered at 0 with width 1.5 GeV for each of the $x$ and $y$ components of the added unclustered momentum.  Backgrounds due to two rare hard scatterings occurring in the same bunch crossing are estimated by forming overlaps of events, as described in Appendix~\ref{sec:Overlaps}.

Each generated standard model event is assigned a weight, calculated as the cross section for the process (in units of picobarns) divided by the number of events generated for that process, representing the number of such events expected in a data sample corresponding to an integrated luminosity of 1~pb$^{-1}$.  When multiplied by the integrated luminosity of the data sample used in this analysis, the weight gives the predicted number of such events in this analysis.


\subsection{Detector simulation}
\label{sec:Vista:DetectorSimulation}

The response of the CDF detector is simulated using a {\sc{geant}}-based detector simulation (\CdfSim)~\cite{Gerchtein:2003ba}, with {\sc{gflash}}~\cite{GFLASH:Grindhammer:1989zg} used to simulate shower development in the calorimeter.

In $p\bar{p}$ collisions there is an ordering of frequency with which objects of different types are produced:  many more jets ($j$) are produced than $b$-jets ($b$) or photons ($\gamma$), and many more of these are produced than charged leptons ($e$, $\mu$, $\tau$).  The CDF detectors and reconstruction algorithms have been designed such that the probability of misreconstructing a frequently produced object as an infrequently produced object is small.  The fraction of central jets that \CdfSim\ misreconstructs as photons, electrons, and muons is $\sim 10^{-3}$, $\sim 10^{-4}$, and $\sim 10^{-5}$, respectively.  Due to these small numbers, the use of \CdfSim\ to model these fake processes would require generating samples with prohibitively large statistics.  Instead, the modeling of a frequently produced object faking a less frequently produced object (specifically: $j$ faking $b$, $\gamma$, $e$, $\mu$, or $\tau$; or $b$ or $\gamma$ faking $e$, $\mu$, or $\tau$) is obtained by the application of a misidentification probability, a particular type of correction factor in the \Vista\ correction model, described in the next section.

In Monte Carlo samples passed through \CdfSim, reconstructed leptons and photons are required to match to a corresponding generator level object.  This procedure removes reconstructed leptons or photons that arise from a misreconstructed quark or gluon jet.


\subsection{Correction model}
\label{sec:Vista:CorrectionModel}

\begin{table*}
\begin{tabular}{lllllr}
{\bf Code } & {\bf Category } & {\bf Explanation } & {\bf Value } & {\bf Error } & {\bf Error(\%)} \\ \hline 
0001 & luminosity & CDF integrated luminosity & 927 & 20 & 2.2 \\ 
0002 & $k$-factor & cosmic $\gamma$ & 0.69 & 0.05 & 7.3 \\ 
0003 & $k$-factor & cosmic $j$ & 0.446 & 0.014 & 3.1 \\ 
0004 & $k$-factor & 1$\gamma$1$j$ & 0.95 & 0.04 & 4.2 \\ 
0005 & $k$-factor & 1$\gamma$2$j$ & 1.2 & 0.05 & 4.1 \\ 
0006 & $k$-factor & 1$\gamma$3$j$ & 1.48 & 0.07 & 4.7 \\ 
0007 & $k$-factor & 1$\gamma$4$j$+ & 1.97 & 0.16 & 8.1 \\ 
0008 & $k$-factor & 2$\gamma$0$j$ & 1.81 & 0.08 & 4.4 \\ 
0009 & $k$-factor & 2$\gamma$1$j$ & 3.42 & 0.24 & 7.0 \\ 
0010 & $k$-factor & 2$\gamma$2$j$+ & 1.3 & 0.16 & 12.3 \\ 
0011 & $k$-factor & $W$0$j$  & 1.453 & 0.027 & 1.9 \\ 
0012 & $k$-factor & $W$1$j$ & 1.06 & 0.03 & 2.8 \\ 
0013 & $k$-factor & $W$2$j$ & 1.02 & 0.03 & 2.9 \\ 
0014 & $k$-factor & $W$3$j$+ & 0.76 & 0.05 & 6.6 \\ 
0015 & $k$-factor & $Z$0$j$  & 1.419 & 0.024 & 1.7 \\ 
0016 & $k$-factor & $Z$1$j$ & 1.18 & 0.04 & 3.4 \\ 
0017 & $k$-factor & $Z$2$j$+ & 1.03 & 0.05 & 4.8 \\ 
0018 & $k$-factor & 2$j$, $\hat{p}_T<150$ & 0.96 & 0.022 & 2.3 \\ 
0019 & $k$-factor & 2$j$, $150<\hat{p}_T$ & 1.256 & 0.028 & 2.2 \\ 
0020 & $k$-factor & 3$j$, $\hat{p}_T<150$ & 0.921 & 0.021 & 2.3 \\ 
0021 & $k$-factor & 3$j$, $150<\hat{p}_T$ & 1.36 & 0.03 & 2.4 \\ 
0022 & $k$-factor & 4$j$, $\hat{p}_T<150$ & 0.989 & 0.025 & 2.5 \\ 
0023 & $k$-factor & 4$j$, $150<\hat{p}_T$ & 1.7 & 0.04 & 2.3 \\ 
0024 & $k$-factor & 5$j$+ & 1.25 & 0.05 & 4.0 \\ 
0025 & ID eff & \poo{e}{e} central & 0.986 & 0.006 & 0.6 \\ 
0026 & ID eff & \poo{e}{e} plug & 0.933 & 0.009 & 1.0 \\ 
0027 & ID eff & \poo{\mu}{\mu}, $\abs{\eta}<0.6$ & 0.845 & 0.008 & 0.9 \\ 
0028 & ID eff & \poo{\mu}{\mu}, $0.6<\abs{\eta}$ & 0.915 & 0.011 & 1.2 \\ 
0029 & ID eff & \poo{\gamma}{\gamma} central & 0.974 & 0.018 & 1.8 \\ 
0030 & ID eff & \poo{\gamma}{\gamma} plug & 0.913 & 0.018 & 2.0 \\ 
0031 & ID eff & \poo{b}{b} central & 1 & 0.04 & 4.0 \\ 
0032 & fake rate & \poo{e}{\gamma} plug & 0.045 & 0.012 & 27.0 \\ 
0033 & fake rate & \poo{q}{e} central & 9.71$\times 10^{-5}$ & 1.9$\times 10^{-6}$ & 2.0 \\ 
0034 & fake rate & \poo{q}{e} plug & 0.000876 & 1.8$\times 10^{-5}$ & 2.1 \\ 
0035 & fake rate & \poo{q}{\mu} & 1.157$\times 10^{-5}$ & 2.7$\times 10^{-7}$ & 2.3 \\ 
0036 & fake rate & \poo{j}{b} & 0.01684 & 0.00027 & 1.6 \\ 
0037 & fake rate & \poo{q}{\tau}, $p_T<60$ & 0.00341 & 0.00012 & 3.5 \\ 
0038 & fake rate & \poo{q}{\tau}, $60<p_T$ & 0.00038 & 4$\times 10^{-5}$ & 10.5 \\ 
0039 & fake rate & \poo{q}{\gamma} central & 0.000265 & 1.5$\times 10^{-5}$ & 5.7 \\ 
0040 & fake rate & \poo{q}{\gamma} plug & 0.00159 & 0.00013 & 8.2 \\ 
0041 & trigger & \poo{e}{\text{trig}} central, $p_T>25$ & 0.976 & 0.007 & 0.7 \\ 
0042 & trigger & \poo{e}{\text{trig}} plug, $p_T>25$ & 0.835 & 0.015 & 1.8 \\ 
0043 & trigger & \poo{\mu}{\text{trig}} $\abs{\eta}<0.6$, $p_T>25$ & 0.917 & 0.007 & 0.8 \\ 
0044 & trigger & \poo{\mu}{\text{trig}} $0.6<\abs{\eta}<1.0$, $p_T>25$ & 0.96 & 0.01 & 1.0 \\ 
\end{tabular}

\caption{The \totalNumberOfFudgeFactors\ correction factors introduced in the \Vista\ correction model.  The leftmost column ({\tt Code}) shows correction factor codes.  The second column ({\tt Category}) shows correction factor categories.  The third column ({\tt Explanation}) provides a short description.  The correction factor best fit value ({\tt Value}) is given in the fourth column.  The correction factor error ({\tt Error}) resulting from the fit is shown in the fifth column.  The fractional error ({\tt Error(\%)}) is listed in the sixth column.  All values are dimensionless with the exception of code {\tt 0001} (luminosity), which has units of pb$^{-1}$.  The values and uncertainties of these correction factors are valid within the context of this correction model.}
\label{tbl:CorrectionFactorDescriptionValuesSigmas}
\end{table*}

Unfortunately some numbers that cannot be determined from first principles enter the comparison between data and the standard model prediction.  These numbers are referred to as ``correction factors'' in the \Vista\ correction model.  This correction model is applied to generated Monte Carlo events to obtain the standard model prediction across all final states.

Correction factors must be obtained from the data themselves.  These factors may be thought of as Bayesian nuisance parameters.  The actual values of the correction factors are not directly of interest.  Of interest is the comparison of data to standard model prediction, with correction factors adjusted to whatever they need to be, consistent with external constraints, to bring the standard model into closest agreement with the data.

The traditional prescription for determining these correction factors is to measure them in a control region in which no signal is expected.  This procedure encounters difficulty when the entire high-$p_T$ data sample is considered to be a signal region. The approach adopted instead is to ask whether a consistent set of correction factors can be chosen so that the standard model prediction is in agreement with the CDF high-$p_T$ data.  

The correction model is obtained by an iterative procedure informed by observed inadequacies in modeling.  The process of correction model improvement, motivated by observed discrepancies, may allow a real signal to be artificially suppressed.  If adjusting correction factor values within allowed bounds removes a signal, then the case for the signal disappears, since it can be explained in terms of known physics.  This is true in any analysis.  The stronger the constraints on the correction model, the more difficult it is to artificially suppress a real signal.  By requiring a consistent interpretation of hundreds of final states, \Vista\ is less likely to mistakenly explain away new physics than if it had more limited scope.

The \totalNumberOfFudgeFactors\ correction factors currently included in the \Vista\ correction model are shown in Table~\ref{tbl:CorrectionFactorDescriptionValuesSigmas}.  These factors can be classified into two categories: theoretical and experimental.  A more detailed description of each individual correction factor is provided in Appendix~\ref{sec:VistaCorrectionModel:CorrectionFactorValues}.

Theoretical correction factors reflect the practical difficulty of calculating accurately within the framework of the standard model.  These factors take the form of $k$-factors, so-called ``knowledge factors,'' representing the ratio of the unavailable all order cross section to the calculable leading order cross section.  Twenty-three $k$-factors are used for standard model processes including QCD multijet production, W+jets, Z+jets, and (di)photon+jets production.

Experimental correction factors include the integrated luminosity of the data, efficiencies associated with triggering on electrons and muons, efficiencies associated with the correct identification of physics objects, and fake rates associated with the mistaken identification of physics objects.  Obtaining an adequate description of object misidentification has required an understanding of the underlying physical mechanisms by which objects are misreconstructed, as described in Appendix~\ref{sec:MisidentificationMatrix}.

In the interest of simplicity, correction factors representing $k$-factors, efficiencies, and fake rates are generally taken to be constants, independent of kinematic quantities such as object $p_T$, with only five exceptions. The $p_T$ dependence of three fake rates is too large to be treated as approximately constant: the jet faking electron rate $\poo{j}{e}$ in the plug region of the CDF detector; the jet faking $b$-tagged jet rate $\poo{j}{b}$, which increases steadily with increasing $p_T$; and the jet faking tau rate $\poo{j}{\tau}$, which decreases steadily with increasing $p_T$.  Two other fake rates possess geometrical features in $\eta$--$\phi$ due to the construction of the CDF detector: the jet faking electron rate $\poo{j}{e}$ in the central region, because of the fiducial tower geometry of the electromagnetic calorimeter; and the jet faking muon rate $\poo{j}{\mu}$, due to the non-trivial fiducial geometry of the muon chambers.  After determining appropriate functional forms, a single overall multiplicative correction factor is used.

Correction factor values are obtained from a global fit to the data.  The procedure is outlined here, with further details relegated to Appendix~\ref{sec:CorrectionFactorFitDetails}.

Events are first partitioned into final states according to the number and types of objects present.  Each final state is then subdivided into bins according to each object's detector pseudorapidity ($\detEta$) and transverse momentum ($p_T$), as described in Appendix~\ref{sec:CorrectionFactorFitDetails:chi_k}.

Generated Monte Carlo events, adjusted by the correction model, provide the standard model prediction for each bin.  The standard model prediction in each bin is therefore a function of the correction factor values.  A figure of merit is defined to quantify global agreement between the data and the standard model prediction, and correction factor values are chosen to maximize this agreement, consistent with external experimental constraints.

Letting $\vec{s}$ represent a vector of correction factors, for the $k^\text{th}$ bin 
\begin{equation}
\chi^2_k(\vec{s})=\frac{(\text{Data}[k]-\text{SM}[k])^2}{\sqrt{\text{SM}[k]}^2 + \delta\text{SM}[k]^2},
\label{eq:chi_k}
\end{equation}
where $\text{Data}[k]$ is the number of data events observed in the $k^\text{th}$ bin, $\text{SM}[k]$ is the number of events predicted by the standard model in the $k^\text{th}$ bin, $\delta\text{SM}[k]$ is the Monte Carlo statistical uncertainty on the standard model prediction in the $k^\text{th}$ bin~\footnote{Given a set of Monte Carlo events with individual weights $w_j$, so that the total standard model prediction from these Monte Carlo events is $\text{SM}=\sum_j{w_j}$ events, the ``effective weight'' $w_{\text{eff}}$ of these events can be taken to be the weighted average of the weights: $w_{\text{eff}}=\frac{\sum_j{w_j w_j}}{\sum_j{w_j}}$.  The ``effective number of Monte Carlo events'' is $N_{\text{eff}}=\text{SM}/w_{\text{eff}}$, and the error on the standard model prediction is $\delta{\text{SM}}=\text{SM}/\sqrt{N_{\text{eff}}}$.}, and $\sqrt{\text{SM}[k]}$ is the statistical uncertainty on the expected data in the $k^\text{th}$ bin.  The standard model prediction $\text{SM}[k]$ in the $k^{\text{th}}$ bin is a function of $\vec{s}$. 

Relevant information external to the \Vista\ high-$p_T$ data sample provides additional constraints in this global fit.  The CDF luminosity counters measure the integrated luminosity of the sample described in this article to be \clcLuminosity~pb$^{-1} \pm 6\%$ by measuring the fraction of bunch crossings in which zero inelastic collisions occur~\cite{CdfLuminosityCountersCLC:Acosta:2002hx}.  The integrated luminosity of the sample measured by the luminosity counters enters in the form of a Gaussian constraint on the luminosity correction factor.  Higher order theoretical calculations exist for some standard model processes, providing constraints on corresponding $k$-factors, and some CDF experimental correction factors are also constrained from external information.  In total, 26 of the \totalNumberOfFudgeFactors\ correction factors are constrained.  The specific constraints employed are provided in Appendix~\ref{sec:CorrectionFactorFitDetails:chi_constraints}.

The overall function to be minimized takes the form
\begin{equation}
\chi^2(\vec{s}) = \left(\sum_{k\in\text{bins}}{\chi^2_k}(\vec{s})\right) + \chi^2_{\text{constraints}}(\vec{s}),
\label{eqn:chiSqd}
\end{equation}
where the sum in the first term is over bins in the CDF high-$p_T$ data sample with $\chi^2_k(\vec{s})$ defined in Eq.~\ref{eq:chi_k}, and the second term is the contribution from explicit constraints.

Minimization of $\chi^2(\vec{s})$ in Eq.~\ref{eqn:chiSqd} as a function of the vector of correction factors $\vec{s}$ results in a set of correction factor values $\vec{s}_0$ providing the best global agreement between the data and the standard model prediction.  The best fit correction factor values are shown in Table~\ref{tbl:CorrectionFactorDescriptionValuesSigmas}, together with absolute and fractional uncertainties.  The determined uncertainties are not used explicitly in the subsequent analysis, but rather provide information used implicitly to assist in appropriate adjustment to the correction model in light of observed discrepancies.  The uncertainties are verified by subdividing the data into thirds, performing separate fits on each third, and noting that the correction factor values obtained with each subset are consistent within quoted uncertainties.  Further details on the correlation matrix and other technical aspects of this global fit can be found in Appendix~\ref{sec:CorrectionFactorCovarianceMatrix}.

Although the correction factors are determined from a global fit, in practice the determination of many correction factors' values are dominated by one recognizable subsample.  The rate $\poo{j}{e}$ for a jet to fake an electron is determined largely by the number of events in the $ej$ final state, since the largest contribution to this final state is from dijet events with one jet misreconstructed as an electron.  Similarly, the rates $\poo{j}{b}$ and $\poo{j}{\tau}$ for a jet to fake a $b$-tagged jet and tau lepton are determined largely by the number of events in the $bj$ and $\tau j$ final states, respectively.  The determination of the fake rate $\poo{j}{\gamma}$, photon efficiency $\poo{\gamma}{\gamma}$, and $k$-factors for prompt photon production and prompt diphoton production are dominated by the $\gamma j$, $\gamma jj$, and $\gamma \gamma$ final states.  Additional knowledge incorporated in the determination of fake rates is described in Appendix~\ref{sec:MisidentificationMatrix}.

The global fit $\chi^2$ per number of bins is $288.1 / 133 + 27.9$, where the last term is the contribution to the $\chi^2$ from the imposed constraints.  A $\chi^2$ per degree of freedom larger than unity is expected, since the limited set of correction factors in this correction model is not expected to provide a complete description of all features of the data.  Emphasis is placed on individual outlying discrepancies that may motivate a new physics claim, rather than overall goodness of fit. 

Corrections to object identification efficiencies are typically less than 10\%; fake rates are consistent with an understanding of the underlying physical mechanisms responsible; $k$-factors range from slightly less than unity to greater than two for some processes with multiple jets.  All values obtained are physically reasonable.  Further analysis is provided in Appendix~\ref{sec:VistaCorrectionModel:CorrectionFactorValues}.

\begin{table*}
{\mbox{\hspace{-2.5cm} 
\tiny
\begin{minipage}{9in}
\begin{tabular}{l@{ }r@{ }r@{ $\pm$ }l@{ }l}
{\bf Final State} & {\bf Data} & \multicolumn{2}{c}{\bf Background} & \multicolumn{1}{c}{\bf $\sigma$} \\ \hline 
3j$\tau^\pm$ & $71$ & $113.7$ & $3.6$ & $-2.3$ \\ 
5j & $1661$ & $1902.9$ & $50.8$ & $-1.7$ \\ 
2j$\tau^\pm$ & $233$ & $296.5$ & $5.6$ & $-1.6$ \\ 
2j$2\tau^\pm$ & $6$ & $27$ & $4.6$ & $-1.4$ \\ 
b$e^\pm$j & $2207$ & $2015.4$ & $28.7$ & $+1.4$ \\ 
3j, high $\SumPt$ & $35436$ & $37294.6$ & $524.3$ & $-1.1$ \\ 
$e^\pm$3j$p\!\!\!/$ & $1954$ & $1751.6$ & $42$ & $+1.1$ \\ 
b$e^\pm$2j & $798$ & $695.3$ & $13.3$ & $+1.1$ \\ 
3j$p\!\!\!/$, low $\SumPt$ & $811$ & $967.5$ & $38.4$ & $-0.8$ \\ 
$e^\pm$$\mu^\pm$ & $26$ & $11.6$ & $1.5$ & $+0.8$ \\ 
$e^\pm$$\gamma$ & $636$ & $551.2$ & $11.2$ & $+0.7$ \\ 
$e^\pm$3j & $28656$ & $27281.5$ & $405.2$ & $+0.6$ \\ 
b5j & $131$ & $95$ & $4.7$ & $+0.5$ \\ 
j$2\tau^\pm$ & $50$ & $85.6$ & $8.2$ & $-0.4$ \\ 
j$\tau^\pm$$\tau^\mp$ & $74$ & $125$ & $13.6$ & $-0.4$ \\ 
b$p\!\!\!/$, low $\SumPt$ & $10$ & $29.5$ & $4.6$ & $-0.4$ \\ 
$e^\pm$j$\gamma$ & $286$ & $369.4$ & $21.1$ & $-0.3$ \\ 
$e^\pm$j$p\!\!\!/$$\tau^\mp$ & $29$ & $14.2$ & $1.8$ & $+0.2$ \\ 
2j, high $\SumPt$ & $96502$ & $92437.3$ & $1354.5$ & $+0.1$ \\ 
b$e^\pm$3j & $356$ & $298.6$ & $7.7$ & $+0.1$ \\ \hline
8j & $11$ & $6.1$ & $2.5$ & \\ 
7j & $57$ & $35.6$ & $4.9$ & \\ 
6j & $335$ & $298.4$ & $14.7$ & \\ 
4j, low $\SumPt$ & $39665$ & $40898.8$ & $649.2$ & \\ 
4j, high $\SumPt$ & $8241$ & $8403.7$ & $144.7$ & \\ 
4j$2\gamma$ & $38$ & $57.5$ & $11$ & \\ 
4j$\tau^\pm$ & $20$ & $36.9$ & $2.4$ & \\ 
4j$p\!\!\!/$, low $\SumPt$ & $516$ & $525.2$ & $34.5$ & \\ 
4j$\gamma$$p\!\!\!/$ & $28$ & $53.8$ & $11$ & \\ 
4j$\gamma$ & $3693$ & $3827.2$ & $112.1$ & \\ 
4j$\mu^\pm$ & $576$ & $568.2$ & $26.1$ & \\ 
4j$\mu^\pm$$p\!\!\!/$ & $232$ & $224.7$ & $8.5$ & \\ 
4j$\mu^\pm$$\mu^\mp$ & $17$ & $20.1$ & $2.5$ & \\ 
$3\gamma$ & $13$ & $24.2$ & $3$ & \\ 
3j, low $\SumPt$ & $75894$ & $75939.2$ & $1043.9$ & \\ 
3j$2\gamma$ & $145$ & $178.1$ & $7.4$ & \\ 
3j$p\!\!\!/$, high $\SumPt$ & $20$ & $30.9$ & $14.4$ & \\ 
3j$\gamma$$\tau^\pm$ & $13$ & $11$ & $2$ & \\ 
3j$\gamma$$p\!\!\!/$ & $83$ & $102.9$ & $11.1$ & \\ 
3j$\gamma$ & $11424$ & $11506.4$ & $190.6$ & \\ 
3j$\mu^\pm$$p\!\!\!/$ & $1114$ & $1118.7$ & $27.1$ & \\ 
3j$\mu^\pm$$\mu^\mp$ & $61$ & $84.5$ & $9.2$ & \\ 
3j$\mu^\pm$ & $2132$ & $2168.7$ & $64.2$ & \\ 
3bj, low $\SumPt$ & $14$ & $9.3$ & $1.9$ & \\ 
$2\tau^\pm$ & $316$ & $290.8$ & $24.2$ & \\ 
$2\gamma$$p\!\!\!/$ & $161$ & $176$ & $9.1$ & \\ 
$2\gamma$ & $8482$ & $8349.1$ & $84.1$ & \\ 
2j, low $\SumPt$ & $93408$ & $92789.5$ & $1138.2$ & \\ 
2j$2\gamma$ & $645$ & $612.6$ & $18.8$ & \\ 
2j$\tau^\pm$$\tau^\mp$ & $15$ & $25$ & $3.5$ & \\ 
2j$p\!\!\!/$, low $\SumPt$ & $74$ & $106$ & $7.8$ & \\ 
2j$p\!\!\!/$, high $\SumPt$ & $43$ & $37.7$ & $100.2$ & \\ 
2j$\gamma$ & $33684$ & $33259.9$ & $397.6$ & \\ 
2j$\gamma$$\tau^\pm$ & $48$ & $41.4$ & $3.4$ & \\ 
2j$\gamma$$p\!\!\!/$ & $403$ & $425.2$ & $29.7$ & \\ 
2j$\mu^\pm$$p\!\!\!/$ & $7287$ & $7320.5$ & $118.9$ & \\ 
2j$\mu^\pm$$\gamma$$p\!\!\!/$ & $13$ & $12.6$ & $2.7$ & \\ 
2j$\mu^\pm$$\gamma$ & $41$ & $35.7$ & $6.1$ & \\ 
2j$\mu^\pm$$\mu^\mp$ & $374$ & $394.2$ & $24.8$ & \\ 
\end{tabular}
\begin{tabular}{l@{ }r@{ }r@{ $\pm$ }l@{ }}
{\bf Final State} & {\bf Data} & \multicolumn{2}{c}{\bf Background} \\ \hline 
2j$\mu^\pm$ & $9513$ & $9362.3$ & $166.8$ \\ 
$2e^\pm$j & $13$ & $9.8$ & $2.2$ \\ 
$2e^\pm$$e^\mp$ & $12$ & $4.8$ & $1.2$ \\ 
$2e^\pm$ & $23$ & $36.1$ & $3.8$ \\ 
2b, low $\SumPt$ & $327$ & $335.8$ & $7$ \\ 
2b, high $\SumPt$ & $187$ & $173.1$ & $7.1$ \\ 
2b3j, high $\SumPt$ & $28$ & $33.5$ & $5.5$ \\ 
2b2j, low $\SumPt$ & $355$ & $326.3$ & $8.4$ \\ 
2b2j, high $\SumPt$ & $56$ & $80.2$ & $5$ \\ 
2b2j$\gamma$ & $16$ & $15.4$ & $3.6$ \\ 
2b$\gamma$ & $37$ & $31.7$ & $4.8$ \\ 
2bj, low $\SumPt$ & $415$ & $393.8$ & $9.1$ \\ 
2bj, high $\SumPt$ & $161$ & $195.8$ & $8.3$ \\ 
2bj$p\!\!\!/$, low $\SumPt$ & $28$ & $23.2$ & $2.6$ \\ 
2bj$\gamma$ & $25$ & $24.7$ & $4.3$ \\ 
2b$e^\pm$2j$p\!\!\!/$ & $15$ & $12.3$ & $1.6$ \\ 
2b$e^\pm$2j & $30$ & $30.5$ & $2.5$ \\ 
2b$e^\pm$j & $28$ & $29.1$ & $2.8$ \\ 
2b$e^\pm$ & $48$ & $45.2$ & $3.7$ \\ 
$\tau^\pm$$\tau^\mp$ & $498$ & $428.5$ & $22.7$ \\ 
$\gamma$$\tau^\pm$ & $177$ & $204.4$ & $5.4$ \\ 
$\gamma$$p\!\!\!/$ & $1952$ & $1945.8$ & $77.1$ \\ 
$\mu^\pm$$\tau^\pm$ & $18$ & $19.8$ & $2.3$ \\ 
$\mu^\pm$$\tau^\mp$ & $151$ & $179.1$ & $4.7$ \\ 
$\mu^\pm$$p\!\!\!/$ & $321351$ & $320500$ & $3475.5$ \\ 
$\mu^\pm$$p\!\!\!/$$\tau^\mp$ & $22$ & $25.8$ & $2.7$ \\ 
$\mu^\pm$$\gamma$ & $269$ & $285.5$ & $5.9$ \\ 
$\mu^\pm$$\gamma$$p\!\!\!/$ & $269$ & $282.2$ & $6.6$ \\ 
$\mu^\pm$$\mu^\mp$$p\!\!\!/$ & $49$ & $61.4$ & $3.5$ \\ 
$\mu^\pm$$\mu^\mp$$\gamma$ & $32$ & $29.9$ & $2.6$ \\ 
$\mu^\pm$$\mu^\mp$ & $10648$ & $10845.6$ & $96$ \\ 
j$2\gamma$ & $2196$ & $2200.3$ & $35.2$ \\ 
j$2\gamma$$p\!\!\!/$ & $38$ & $27.3$ & $3.2$ \\ 
j$\tau^\pm$ & $563$ & $585.7$ & $10.2$ \\ 
j$p\!\!\!/$, low $\SumPt$ & $4183$ & $4209.1$ & $56.1$ \\ 
j$\gamma$ & $49052$ & $48743$ & $546.3$ \\ 
j$\gamma$$\tau^\pm$ & $106$ & $104$ & $4.1$ \\ 
j$\gamma$$p\!\!\!/$ & $913$ & $965.2$ & $41.5$ \\ 
j$\mu^\pm$ & $33462$ & $34026.7$ & $510.1$ \\ 
j$\mu^\pm$$\tau^\mp$ & $29$ & $37.5$ & $4.5$ \\ 
j$\mu^\pm$$p\!\!\!/$$\tau^\mp$ & $10$ & $9.6$ & $2.1$ \\ 
j$\mu^\pm$$p\!\!\!/$ & $45728$ & $46316.4$ & $568.2$ \\ 
j$\mu^\pm$$\gamma$$p\!\!\!/$ & $78$ & $69.8$ & $9.9$ \\ 
j$\mu^\pm$$\gamma$ & $70$ & $98.4$ & $12.1$ \\ 
j$\mu^\pm$$\mu^\mp$ & $1977$ & $2093.3$ & $74.7$ \\ 
$e^\pm$4j & $7144$ & $6661.9$ & $147.2$ \\ 
$e^\pm$4j$p\!\!\!/$ & $403$ & $363$ & $9.9$ \\ 
$e^\pm$3j$\tau^\mp$ & $11$ & $7.6$ & $1.6$ \\ 
$e^\pm$3j$\gamma$ & $27$ & $21.7$ & $3.4$ \\ 
$e^\pm$$2\gamma$ & $47$ & $74.5$ & $5$ \\ 
$e^\pm$2j & $126665$ & $122457$ & $1672.6$ \\ 
$e^\pm$2j$\tau^\mp$ & $53$ & $37.3$ & $3.9$ \\ 
$e^\pm$2j$\tau^\pm$ & $20$ & $24.7$ & $2.3$ \\ 
$e^\pm$2j$p\!\!\!/$ & $12451$ & $12130.1$ & $159.4$ \\ 
$e^\pm$2j$\gamma$ & $101$ & $88.9$ & $6.1$ \\ 
$e^\pm$$\tau^\mp$ & $609$ & $555.9$ & $10.2$ \\ 
$e^\pm$$\tau^\pm$ & $225$ & $211.2$ & $4.7$ \\ 
$e^\pm$$p\!\!\!/$ & $476424$ & $479572$ & $5361.2$ \\ 
$e^\pm$$p\!\!\!/$$\tau^\mp$ & $48$ & $35$ & $2.7$ \\ 
\end{tabular}
\begin{tabular}{l@{ }r@{ }r@{ $\pm$ }l@{ }}
{\bf Final State} & {\bf Data} & \multicolumn{2}{c}{\bf Background} \\ \hline 
$e^\pm$$p\!\!\!/$$\tau^\pm$ & $20$ & $18.7$ & $1.9$ \\ 
$e^\pm$$\gamma$$p\!\!\!/$ & $141$ & $144.2$ & $6$ \\ 
$e^\pm$$\mu^\mp$$p\!\!\!/$ & $54$ & $42.6$ & $2.7$ \\ 
$e^\pm$$\mu^\pm$$p\!\!\!/$ & $13$ & $10.9$ & $1.3$ \\ 
$e^\pm$$\mu^\mp$ & $153$ & $127.6$ & $4.2$ \\ 
$e^\pm$j & $386880$ & $392614$ & $5031.8$ \\ 
$e^\pm$j$2\gamma$ & $14$ & $15.9$ & $2.9$ \\ 
$e^\pm$j$\tau^\pm$ & $79$ & $79.3$ & $2.9$ \\ 
$e^\pm$j$\tau^\mp$ & $162$ & $148.8$ & $7.6$ \\ 
$e^\pm$j$p\!\!\!/$ & $58648$ & $57391.7$ & $661.6$ \\ 
$e^\pm$j$\gamma$$p\!\!\!/$ & $52$ & $76.2$ & $9$ \\ 
$e^\pm$j$\mu^\mp$$p\!\!\!/$ & $22$ & $13.1$ & $1.7$ \\ 
$e^\pm$j$\mu^\mp$ & $28$ & $26.8$ & $2.3$ \\ 
$e^\pm$$e^\mp$4j & $103$ & $113.5$ & $5.9$ \\ 
$e^\pm$$e^\mp$3j & $456$ & $473$ & $14.6$ \\ 
$e^\pm$$e^\mp$2j$p\!\!\!/$ & $30$ & $39$ & $4.6$ \\ 
$e^\pm$$e^\mp$2j & $2149$ & $2152$ & $40.1$ \\ 
$e^\pm$$e^\mp$$\tau^\pm$ & $14$ & $11.1$ & $2$ \\ 
$e^\pm$$e^\mp$$p\!\!\!/$ & $491$ & $487.9$ & $12$ \\ 
$e^\pm$$e^\mp$$\gamma$ & $127$ & $132.3$ & $4.2$ \\ 
$e^\pm$$e^\mp$j & $10726$ & $10669.3$ & $123.5$ \\ 
$e^\pm$$e^\mp$j$p\!\!\!/$ & $157$ & $144$ & $11.2$ \\ 
$e^\pm$$e^\mp$j$\gamma$ & $26$ & $45.6$ & $4.7$ \\ 
$e^\pm$$e^\mp$ & $58344$ & $58575.6$ & $603.9$ \\ 
b6j & $24$ & $15.5$ & $2.3$ \\ 
b4j, low $\SumPt$ & $13$ & $9.2$ & $1.8$ \\ 
b4j, high $\SumPt$ & $464$ & $499.2$ & $12.4$ \\ 
b3j, low $\SumPt$ & $5354$ & $5285$ & $72.4$ \\ 
b3j, high $\SumPt$ & $1639$ & $1558.9$ & $24.1$ \\ 
b3j$p\!\!\!/$, low $\SumPt$ & $111$ & $116.8$ & $11.2$ \\ 
b3j$\gamma$ & $182$ & $194.1$ & $8.8$ \\ 
b3j$\mu^\pm$$p\!\!\!/$ & $37$ & $34.1$ & $2$ \\ 
b3j$\mu^\pm$ & $47$ & $52.2$ & $3$ \\ 
b$2\gamma$ & $15$ & $14.6$ & $2.1$ \\ 
b2j, low $\SumPt$ & $8812$ & $8576.2$ & $97.9$ \\ 
b2j, high $\SumPt$ & $4691$ & $4646.2$ & $57.7$ \\ 
b2j$p\!\!\!/$, low $\SumPt$ & $198$ & $209.2$ & $8.3$ \\ 
b2j$\gamma$ & $429$ & $425.1$ & $13.1$ \\ 
b2j$\mu^\pm$$p\!\!\!/$ & $46$ & $40.1$ & $2.7$ \\ 
b2j$\mu^\pm$ & $56$ & $60.6$ & $3.4$ \\ 
b$\tau^\pm$ & $19$ & $19.9$ & $2.2$ \\ 
b$\gamma$ & $976$ & $1034.8$ & $15.6$ \\ 
b$\gamma$$p\!\!\!/$ & $18$ & $16.7$ & $3.1$ \\ 
b$\mu^\pm$ & $303$ & $263.5$ & $7.9$ \\ 
b$\mu^\pm$$p\!\!\!/$ & $204$ & $218.1$ & $6.4$ \\ 
bj, low $\SumPt$ & $9060$ & $9275.7$ & $87.8$ \\ 
bj, high $\SumPt$ & $7236$ & $7030.8$ & $74$ \\ 
bj$2\gamma$ & $13$ & $17.6$ & $3.3$ \\ 
bj$\tau^\pm$ & $13$ & $12.9$ & $1.8$ \\ 
bj$p\!\!\!/$, low $\SumPt$ & $53$ & $60.4$ & $19.9$ \\ 
bj$\gamma$ & $937$ & $989.4$ & $20.6$ \\ 
bj$\gamma$$p\!\!\!/$ & $34$ & $30.5$ & $4$ \\ 
bj$\mu^\pm$$p\!\!\!/$ & $104$ & $112.6$ & $4.4$ \\ 
bj$\mu^\pm$ & $173$ & $141.4$ & $4.8$ \\ 
b$e^\pm$3j$p\!\!\!/$ & $68$ & $52.2$ & $2.2$ \\ 
b$e^\pm$2j$p\!\!\!/$ & $87$ & $65$ & $3.3$ \\ 
b$e^\pm$$p\!\!\!/$ & $330$ & $347.2$ & $6.9$ \\ 
b$e^\pm$j$p\!\!\!/$ & $211$ & $176.6$ & $5$ \\ 
b$e^\pm$$e^\mp$j & $22$ & $34.6$ & $2.6$ \\ 
\end{tabular}
\end{minipage}
}}
\caption{A subset of the \Vista\ comparison between Tevatron Run II data and standard model prediction, showing the twenty most discrepant final states and all final states populated with ten or more data events.  Events are partitioned into exclusive final states based on standard CDF object identification criteria.  Final states are labeled in this table according to the number and types of objects present, and whether (high $\SumPt$) or not (low $\SumPt$) the summed scalar transverse momentum of all objects in the events exceeds 400~GeV, for final states not containing leptons or photons.  Final states are ordered according to decreasing discrepancy between the total number of events expected, taking into account the error from Monte Carlo statistics and the total number observed in the data.  Final states exhibiting mild discrepancies are shown together with the significance of the discrepancy in units of standard deviations ($\sigma$) after accounting for a trials factor corresponding to the number of final states considered.  Final states that do not exhibit even mild discrepancies are listed below the horizontal line in inverted alphabetical order.  Only Monte Carlo statistical uncertainties on the background prediction are included.
}
\label{tbl:VistaCdf}  
\end{table*}

\begin{figure}
\begin{tabular}{cc}
\includegraphics[width=3.5in]{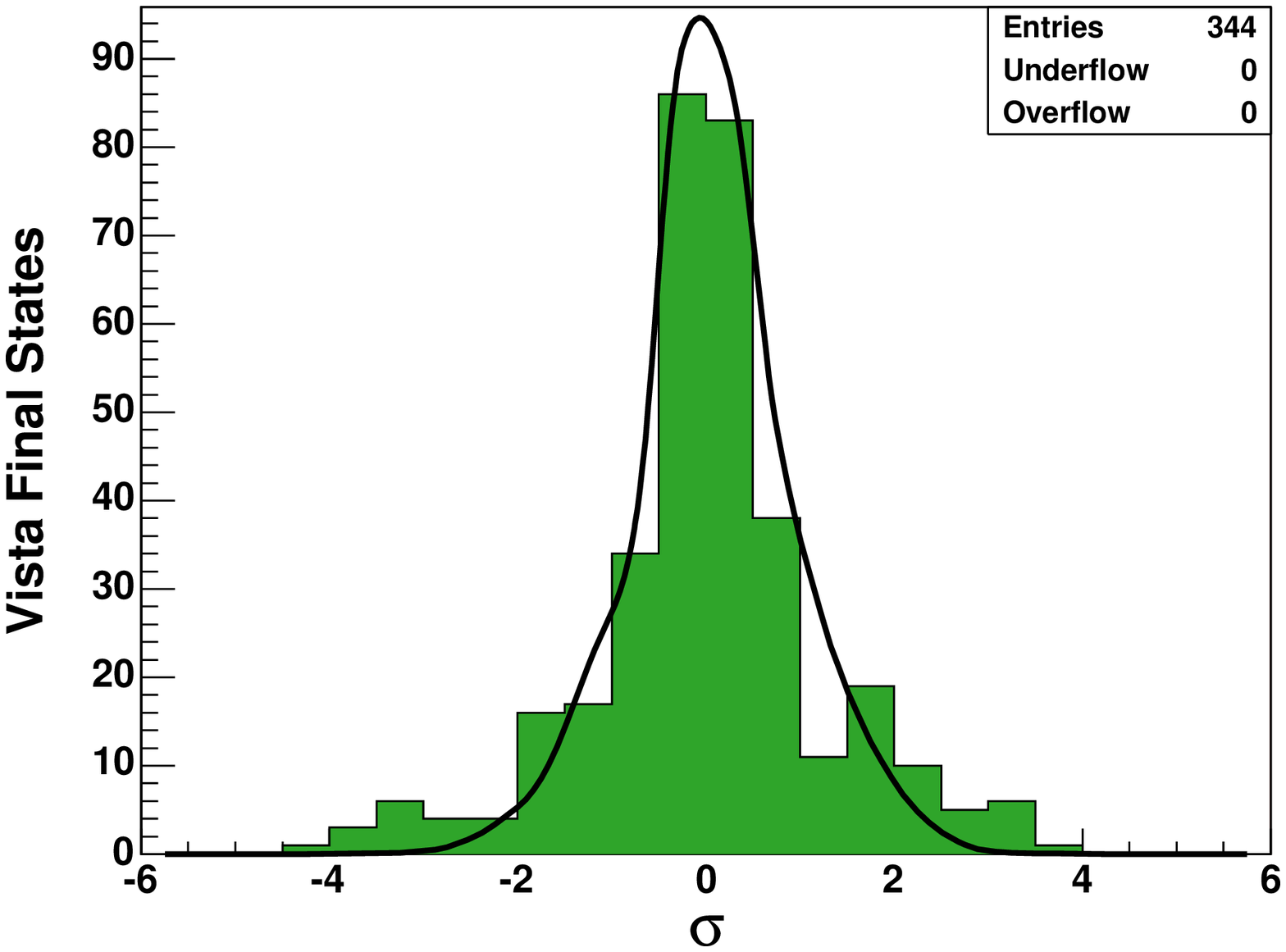} \\
\includegraphics[width=3.5in]{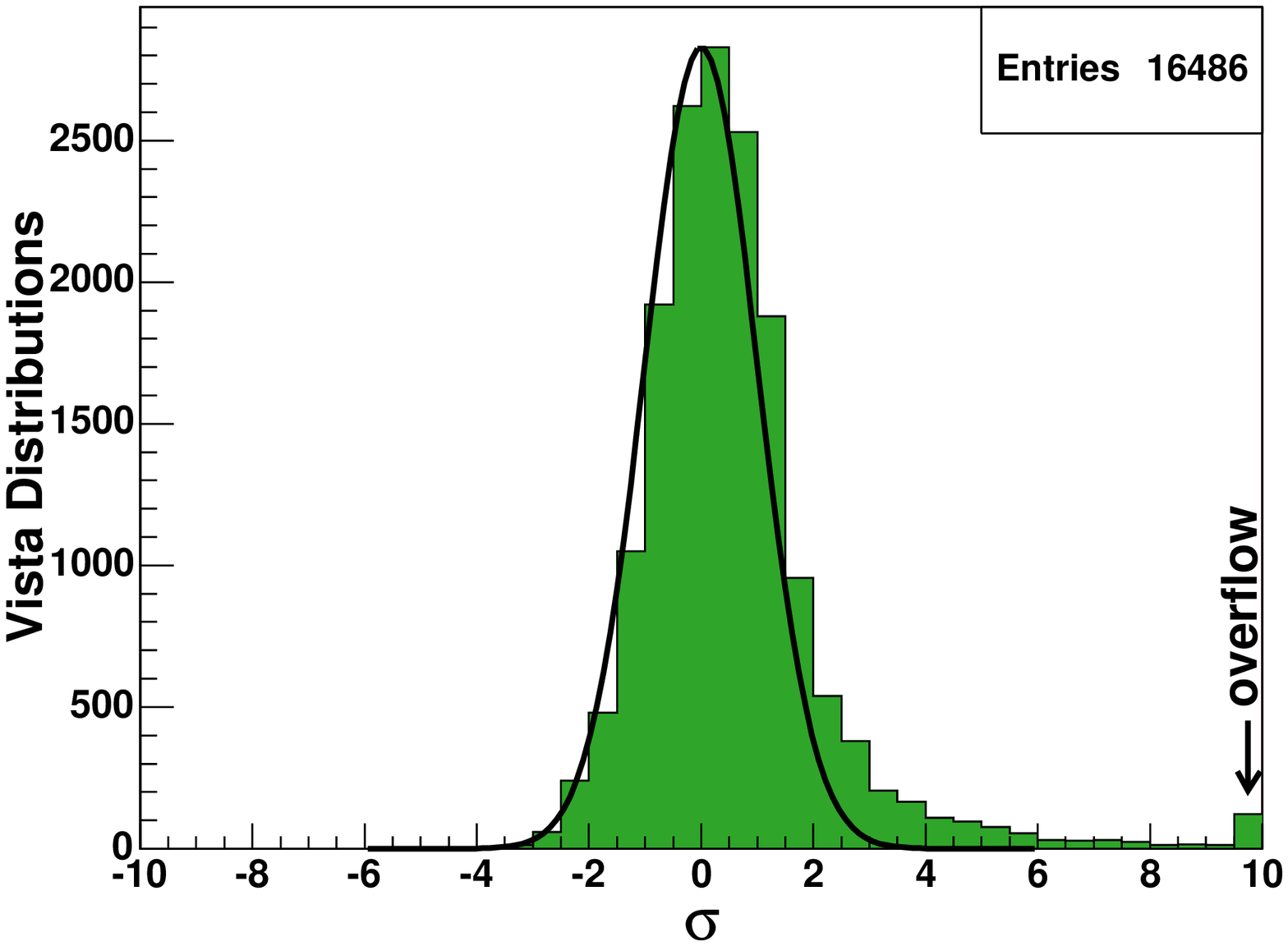} 
\end{tabular}
\caption{Distribution of observed discrepancy between data and the standard model prediction, measured in units of standard deviation ($\sigma$), shown as the solid (green) histogram, \highlight{before accounting for the trials factor.}  The upper pane shows the distribution of discrepancies between the total number of events observed and predicted in the 344 populated final states considered.  Negative values on the horizontal axis correspond to a deficit of data compared to standard model prediction; positive values indicate an excess of data compared to standard model prediction.  The lower pane shows the distribution of discrepancies between the observed and predicted shapes in 16,486 kinematic distributions.  Distributions in which the shapes of data and standard model prediction are in relative disagreement correspond to large positive $\sigma$.  The solid (black) curves indicate expected distributions, if the data were truly drawn from the standard model background.  Interest is focused on the entries in the tails of the upper distribution and the high tail of the lower distribution.  The final state entering the upper histogram at $-4.03\sigma$ is the \Vista\ $3j\,\tau$ final state, which heads Table~\ref{tbl:VistaCdf}.  Most of the distributions entering the lower histogram with $>4\sigma$ derive from the $3j$ $\Delta R(j_2,j_3)$ discrepancy, discussed in the text.}
\label{fig:VistaSummaryCdf}
\end{figure}

\begin{figure}
\includegraphics[width=2.6in,angle=270]{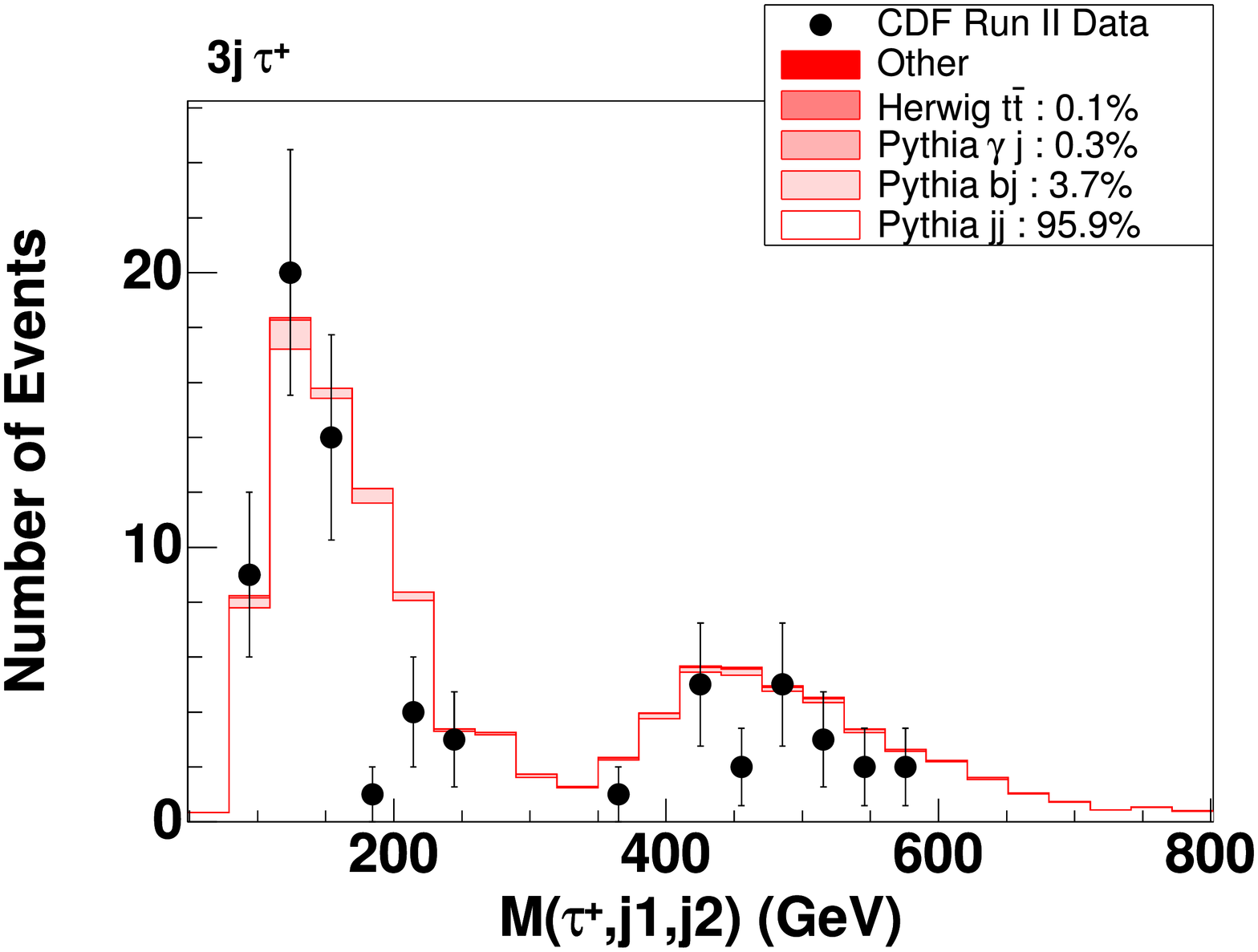}
\caption{The invariant mass of the tau lepton and two leading jets in the final state consisting of three jets and one positively or negatively charged tau.  (The \Vista\ final state naming convention gives the tau lepton a positive charge.)  Data are shown as filled (black) circles, with the standard model prediction shown as the shaded (red) histogram.  This is the most discrepant kinematic distribution in the final state exhibiting the largest population discrepancy.}
\label{fig:3j1tau_mostDiscrepant}
\end{figure}

With the details of the correction model in place, the complete standard model prediction can be obtained.  For each Monte Carlo event after detector simulation, the event weight is multiplied by the value of the luminosity correction factor and the $k$-factor for the relevant standard model process.  The single Monte Carlo event can be misreconstructed in a number of ways, producing a set of Monte Carlo events derived from the original, with weights multiplied by the probability of each misreconstruction.  The weight of each resulting event is multiplied by the probability the event satisfies trigger criteria.  The resulting standard model prediction, corrected as just described, is referred to as ``the standard model prediction'' throughout the rest of this paper, with ``corrected'' implied in all cases.

\subsection{Results}

 Data and standard model events are partitioned into exclusive final states.  This partitioning is orthogonal, with each event ending up in one and only one final state.  Data are compared to standard model prediction in each final state, considering the total number of events observed and predicted, and the shapes of relevant kinematic distributions.

In a data driven search, it is crucial to explicitly account for the {\em{trials factor}}, quantifying the number of places an interesting signal could appear.  Fluctuations at the level of three or more standard deviations are expected to appear simply because a large number of regions are considered.  A reasonably rigorous accounting of this trials factor is possible as long as the measures of interest and the regions to which these measures are applied are specified {\em{a priori}}, as is done here.  \highlight{In this analysis a discrepancy at the level of $3\sigma$ or greater after accounting for the trials factor (typically corresponding to a discrepancy at the level of $5\sigma$ or greater before accounting for the trials factor) is considered ``significant.''}

Discrepancy in the total number of events in a final state ($\text{fs}$) is measured by the Poisson probability $p_{\text{fs}}$ that the number of predicted events would fluctuate up to or above (or down to or below) the number of events observed.  To account for the trials factor due to the 344 \Vista\ final states examined, the quantity $p=1-(1-p_{\text{fs}})^{344}$ is calculated for each final state.  The result is the probability $p$ of observing a discrepancy corresponding to a probability less than $p_{\text{fs}}$ in the total sample studied.  This probability $p$ can then be converted into units of standard deviations by solving for $\sigma$ such that $\int_{\sigma}^{\infty}\, \frac{1}{\sqrt{2\pi}}e^{-\frac{x^2}{2}} dx = p$~\footnote{Final states for which $p>0.5$ after accounting for the trials factor are not even mildly interesting, and the corresponding $\sigma$ after accounting for the trials factor is not quoted.  For the mildly interesting final states with $p<0.5$ after accounting for the trials factor, $\sigma$ is quoted as positive if the number of observed data events exceeds the standard model prediction, and negative if the number of observed data events is less than the standard model prediction.}.  \highlight{A final state exhibiting a population discrepancy greater than 3$\sigma$ after the trials factor is thus accounted for is considered significant.}

Many kinematic distributions are considered in each final state, including the transverse momentum, pseudorapidity, detector pseudorapidity, and azimuthal angle of all objects, masses of individual jets and $b$-jets, invariant masses of all object combinations, transverse masses of object combinations including $\pmiss$, angular separation $\Delta\phi$ and $\Delta R$ of all object pairs, and several other more specialized variables.  A Kolmogorov-Smirnov (KS) test is used to quantify the difference in shape of each kinematic distribution between data and standard model prediction.  As with populations, a trials factor is assessed to account for the 16,486 distributions examined, and the resulting probability is converted into units of standard deviations.  A distribution with KS statistic greater than 0.02 and probability corresponding to greater than 3$\sigma$ after assessing the trials factor is considered significant.

Table~\ref{tbl:VistaCdf} shows a subset of the \Vista\ comparison of data to standard model prediction.  Shown are all final states containing ten or more data events, with the most discrepant final states in population heading the list.  After accounting for the trials factor, no final state has a statistically significant \highlight{($>3\sigma$)} population discrepancy.  The most discrepant final state ($3j\,\tau^\pm$) contains 71 data events and $113.7\pm3.6$ events expected from the standard model.  The Poisson probability for $113.7\pm3.6$ expected events to result in 71 or fewer events observed in this final state is $2.8\times10^{-5}$, corresponding to an entry at $-4.03\sigma$ in Fig.~\ref{fig:VistaSummaryCdf}.  The probability for one or more of the 344 populated final states considered to display disagreement in population corresponding to a probability less than $2.8\times10^{-5}$ is 1\%.  The $3j\,\tau^\pm$ population discrepancy is thus not statistically significant.  The most discrepant kinematic distribution in this final state is the invariant mass of the tau lepton and the two highest transverse momentum jets, shown in Fig.~\ref{fig:3j1tau_mostDiscrepant}.

The six final states with largest population discrepancy are $3j\,\tau$, $5j$, $2j\,\tau$, $2j\,2\tau$, $b\,e\,j$, and the low-$p_T$ $3j$ final state, with $b\,e\,j$ being the only one of these six to exhibit an excess of data.  The $3j\,\tau$, $2j\,\tau$, and $2j\,2\tau$ final states appear to reflect an incomplete understanding of the rate of jets faking taus ($\poo{j}{\tau}$) as a function of the number of jets in the event, at the level of $\sim 30\%$ difference between the total number of observed and predicted events in the most populated of these final states.  The value of $\poo{j}{\tau}$ is primarily determined by the $j\,\tau$ final state.  Interestingly, although the underlying physical mechanism for $\poo{j}{e}$ is very similar to that for $\poo{j}{\tau}$, as discussed in Appendix~\ref{sec:MisidentificationMatrix}, a significant dependence on the presence of additional jets is not observed for $\poo{j}{e}$.

The $5j$ discrepancy results from a tension with the $e\,4j$ final state, whose dominant contribution comes from $5j$ production convoluted with $\poo{j}{e}$.  The low-$p_T$ $3j$ discrepancy results from a tension with the $e\,2j$ final state, whose dominant contribution comes from $3j$ production convoluted with $\poo{j}{e}$.  The $b\,e\,j$ final state is predominantly $3j$ production convoluted with $\poo{j}{b}$ and $\poo{j}{e}$; this discrepancy also arises from a tension with the low-$p_T$ $3j$ and $e\,2j$ final states.  The $b\,e\,j$ final state is the \Vista\ final state in which the largest excess of data over standard model prediction is seen.  The fraction of hypothetical similar CDF experiments that would produce a \Vista\ normalization excess as significant as the excess observed in this final state is $8\%$.  The $5j$, $b\,e\,j$, and low-$p_T$ $3j$ discrepancies correspond to a difference of $\sim 10\%$ between the total number of observed and predicted events in these final states.

Figure~\ref{fig:VistaSummaryCdf} summarizes in a histogram the measured discrepancies between data and the standard model prediction for CDF high-$p_T$ final state populations and kinematic distributions.  Values in this figure represent individual discrepancies, and do not account for the trials factor associated with examining many possibilities.

\begin{figure}
\includegraphics[width=2.6in,angle=270]{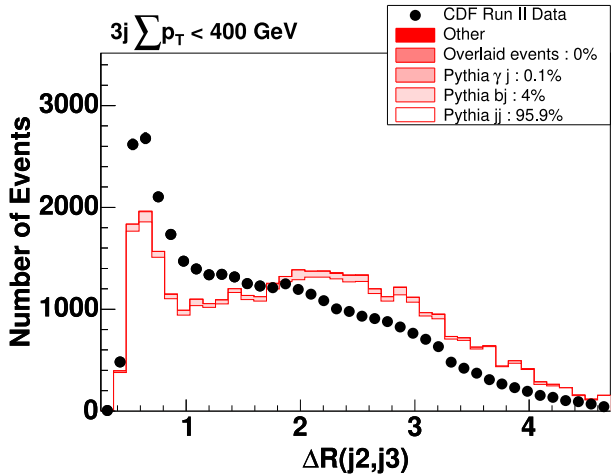}
\caption{A shape discrepancy highlighted by \Vista\ in the final state consisting of exactly three reconstructed jets with $\abs{\eta}<2.5$ and $p_T>17$~GeV, and with one of the jets satisfying $\abs{\eta}<1$ and $p_T>40$~GeV.  This distribution illustrates the effect underlying most of the \Vista\ shape discrepancies.  Filled (black) circles show CDF data, with the shaded (red) histogram showing the prediction of \Pythia.  The discrepancy is clearly statistically significant, with statistical error bars smaller than the size of the data points.  The vertical axis shows the number of events per bin, with the horizontal axis showing the angular separation ($\Delta R=\sqrt{\Delta\eta^2+\delta\phi^2}$) between the second and third jets, where the jets are ordered according to decreasing transverse momentum.  In the region $\Delta R(j_2,j_3)\gtrsim2$, populated primarily by initial state radiation, the standard model prediction can to some extent be adjusted.  The region $\Delta R(j_2,j_3)\lesssim2$ is dominated by final state radiation, the description of which is constrained by data from LEP\,1.}
\label{fig:3j_deltaR_j2j3}
\end{figure}

\begin{figure}
\includegraphics[width=2.5in,angle=270]{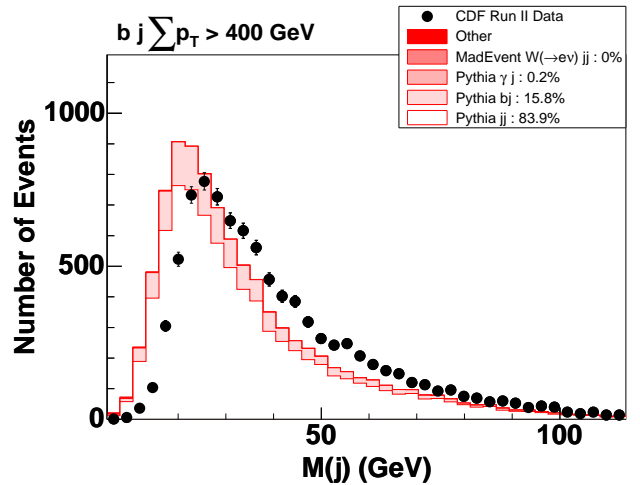}
\caption{The jet mass distribution in the $bj$ final state with $\SumPt>400$~GeV. The $3j$ $\Delta R(j_2,j_3)$ discrepancy illustrated in Fig.~\ref{fig:3j_deltaR_j2j3} manifests itself also by producing jets more massive in data than predicted by \Pythia's showering algorithm.  \highlight{The mass of a jet is determined by treating energy deposited in each calorimeter tower as a massless 4-vector, summing the 4-vectors of all towers within the jet, and computing the mass of the resulting (massive) 4-vector.}}
\label{fig:plots_1b1j_sumPt400+_mass_j}
\end{figure}

\begin{figure}
\includegraphics[width=2.6in,angle=270]{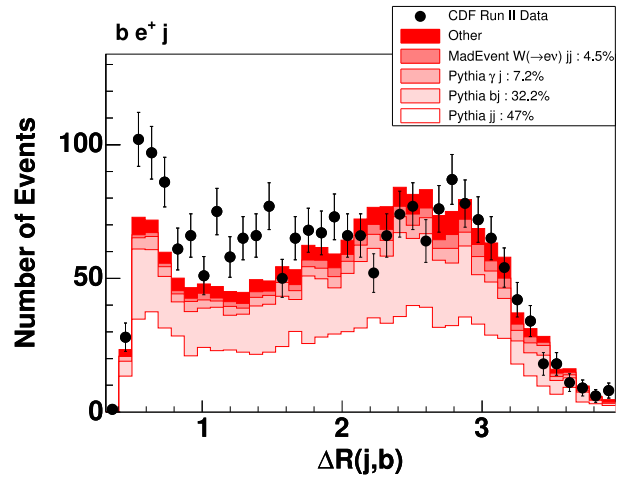}
\caption{The distribution of $\Delta R$ between the jet and $b$-tagged jet in the final state $b\,e\,j$.  The primary standard model contribution to this final state is QCD three jet production with one jet misreconstructed as an electron.  The similarity to the $3j$ $\Delta R(j_2,j_3)$ discrepancy illustrated in Fig.~\ref{fig:3j_deltaR_j2j3} in the region $\Delta R(j,b)<2$ is clear.  Less clear is the underlying explanation for the difference with respect to Fig.~\ref{fig:3j_deltaR_j2j3} in the region $\Delta R(j,b)>2$.}
\label{fig:1b1e+1j_deltaR_jb}
\end{figure}

Of the 16,484 kinematic distributions considered, \numberOfVistaDiscrepantDistributions\ distributions are found to correspond to a discrepancy greater than 3$\sigma$ after accounting for the trials factor, entering with a KS probability of roughly $5\sigma$ or greater in Fig.~\ref{fig:VistaSummaryCdf}.  Of these \numberOfVistaDiscrepantDistributions\ discrepant distributions, 312 are attributed to modeling parton radiation, deriving from the $3j$ $\Delta R(j_2,j_3)$ discrepancy shown in Fig.~\ref{fig:3j_deltaR_j2j3}, with 186 of these 312 shape discrepancies pointing out that individual jet masses are larger in data than in the prediction, as shown in Fig.~\ref{fig:plots_1b1j_sumPt400+_mass_j}.  A careful reading of the literature reveals that the same effect was observed (but not emphasized) by both CDF~\cite{Geer:CdfJetMass:Abe:1996nn,Geer:CdfJetMass:Abe:1997yb} and \DZero~\cite{D0JetMass:Abachi:1995zw} in Tevatron Run I.  The $3j$ $\Delta R(j_2,j_3)$ and jet mass discrepancies appear to be two different views of a single underlying discrepancy, noting that two sufficiently nearby distinct jets correspond to a pattern of calorimetric energy deposits similar to a single massive jet.  The underlying $3j$ $\Delta R(j_2,j_3)$ discrepancy is manifest in many other final states.  The final state $b\,e\,j$, arising primarily from QCD production of three jets with one misreconstructed as an electron, shows a similar discrepancy in $\Delta R(j,b)$ in Fig.~\ref{fig:1b1e+1j_deltaR_jb}.

While these discrepancies are clearly statistically significant, basing a new physics claim around them is difficult.  In the kinematic regime of the discrepancy, different algorithms to match exact leading order calculations with a parton shower lead to different predictions~\cite{MadEventAlpgenComparison:Alwall:2007fs}.  Newer predictions have not been systematically compared to LEP\,1 data, which provide constraints on parton showering reflected in \Pythia's tuning.  Further investigation into obtaining an adequate QCD-based description of this discrepancy continues.

An additional 59 discrepant distributions reflect an inadequate modeling of the overall transverse boost of the system.  The overall transverse boost of the primary physics objects in the event is attributed to two sources:  the intrinsic Fermi motion of the colliding partons within the proton, and soft or collinear radiation of the colliding partons as they approach collision.  Together these effects are here referred to as ``intrinsic $k_T$,'' representing an overall momentum kick to the hard scattering.  Further discussion appears in Appendix~\ref{sec:CorrectionModelDetails:IntrinsicKt}.

The remaining 13 discrepant distributions are seen to be due to the coarseness of the \Vista\ correction model.  Most of these discrepancies, which are at the level of 10\% or less when expressed as $({\text{data}}-{\text{theory}})/{\text{theory}}$, arise from modeling most fake rates as independent of transverse momentum. 

In summary, this global analysis of the bulk features of the high-$p_T$ data has not yielded a discrepancy motivating a new physics claim.  There are no statistically significant population discrepancies in the 344 populated final states considered, and although there are several statistically significant discrepancies among the 16,486 kinematic distributions investigated, the nature of these discrepancies makes it difficult to use them to support a new physics claim.

This global analysis of course cannot conclude with certainty that there is no new physics hiding in the CDF data.  The \Vista\ population and shape statistics may be insensitive to a small excess of events appearing at large $\SumPt$ in a highly populated final state.  For such signals another algorithm is required.


\section{\Sleuth}
\label{sec:Sleuth}

Taking a broad view of all proposed models that might extend the standard model, a profound commonality is noted:  nearly all predict an excess of events at high $p_T$, concentrated in a particular final state.  The second stage of this research program involves the systematic search for such physics using an algorithm called \Sleuth~\cite{KnutesonThesis}.  \Sleuth\ is quasi model independent, where ``quasi'' refers to the assumption that the first sign of new physics will appear as an excess of events in some final state at large summed scalar transverse momentum ($\SumPt$).

The \Sleuth\ algorithm used by CDF in Tevatron Run II is essentially that developed by \DZero\ in Tevatron Run I~\cite{SleuthPRD1:Abbott:2000fb, SleuthPRD2:Abbott:2000gx, SleuthPRL:Abbott:2001ke}, and subsequently improved by H1 in HERA Run I~\cite{H1GeneralSearch:Aktas:2004pz}, with small modifications.  

\Sleuth's definition of interest relies on the following assumptions.
\begin{enumerate}
\item The data can be categorized into exclusive final states in such a way that any signature of new physics is apt to appear predominantly in one of these final states.  
\item New physics will appear with objects at high summed transverse momentum ($\SumPt$) relative to standard model and instrumental background.  
\item New physics will appear as an excess of data over standard model and instrumental background.  
\end{enumerate}

\subsection{Algorithm}
\label{sec:SleuthAlgorithm}
The \Sleuth\ algorithm consists of three steps, following the above three assumptions.

\subsubsection{Final states}
\label{sec:finalStateDefinitions}

In the first step of the algorithm, all events are placed into exclusive final states as in \Vista, with the following modifications.

\begin{itemize}

\item
Jets are identified as pairs, rather than individually, to reduce the total number of final states and to keep signal events with one additional radiated gluon within the same final state.  Final state names include ``$n$ $jj$'' if $n$ jet pairs are identified, with possibly one unpaired jet assumed to have originated from a radiated gluon.

\item
The present understanding of quark flavor suggests that $b$ quarks should be produced in pairs. Bottom quarks are identified as pairs, rather than individually, to increase the robustness of identification and to reduce the total number of final states.  Final state names include ``$n$ $bb$'' if $n$ $b$ pairs are identified.

\item
  Final states related through global charge conjugation are considered to be equivalent.  Thus $e^+e^-\gamma$ is a different final state than $e^+e^+\gamma$, but $e^+e^+\gamma$ and $e^-e^-\gamma$ together make up a single \Sleuth\ final state.  

\item
  Final states related through global interchange of the first and second generation are considered to be equivalent.  Thus $e^+\pmiss\gamma$ and $\mu^+\pmiss\gamma$ together make up a single \Sleuth\ final state.  The decision to consider third generation objects ($b$ quarks and $\tau$ leptons) differently from first and second generation objects reflects theoretical prejudice that the third generation may be special, and the experimental ability (in the case of $b$ quarks) and experimental challenge (in the case of $\tau$ leptons) in the identification of third generation objects. 

\end{itemize}

The symbol $\ell$ is used to denote electron or muon.  The symbol $W$ is used in naming final states containing one electron or muon, significant missing momentum, and perhaps other non-leptonic objects.  Thus the final states $e^+\pmiss\gamma$, $e^-\pmiss\gamma$, $\mu^+\pmiss\gamma$, and $\mu^-\pmiss\gamma$ are combined into the \Sleuth\ final state $W\gamma$.  A table showing the relationship between \Vista\ and \Sleuth\ final states is provided in Appendix~\ref{sec:Sleuth:Partitioning}.

\subsubsection{Variable}

The second step of the algorithm considers a single variable in each exclusive final state:  the summed scalar transverse momentum of all objects in the event ($\sum{p_T}$).  Assuming momentum conservation in the plane transverse to the axis of the colliding beams,
\begin{equation}
\sum_i{\vec{p}_i} + \overrightarrow{\text{uncl}} + \vec{\pmiss} = \vec{0},
\end{equation}
where the sum over $i$ represents a sum over all identified objects in the event, the $i^\text{th}$ object has momentum $\vec{p}_i$, $\overrightarrow{\text{uncl}}$ denotes the vector sum of all momentum visible in the detector but not clustered into an identified object, $\vec{\pmiss}$ denotes the missing momentum, and the equation is a two-component vector equality for the components of the momentum along the two spatial directions transverse to the axis of the colliding beams.  The \Sleuth\ variable \SumPt\ is then defined by
\begin{equation}
\SumPt \equiv \sum_i{\abs{\vec{p}_i}} + \abs{\overrightarrow{\text{uncl}}} + \abs{\vec{\pmiss}},
\end{equation}
where only the momentum components transverse to the axis of the colliding beams are considered when computing magnitudes.

\subsubsection{Regions}
\label{sec:Sleuth:Regions}

The algorithm's third step involves searching for regions in which more events are seen in the data than expected from standard model and instrumental background.  This search is performed in the variable space defined in the second step of the algorithm, for each of the exclusive final states defined in the first step.  

The steps of the search can be sketched as follows.
\begin{itemize}
\item  In each final state, the regions considered are the one dimensional intervals in $\sum{p_T}$ extending from each data point up to infinity.  A region is required to contain at least three data events, as described in Appendix~\ref{sec:Sleuth:MinimumNumberOfEvents}.

\item 
In a particular final state, the data point with the $d^{\text{th}}$ largest value of $\SumPt$ defines an interval in the variable $\SumPt$ extending from this data point up to infinity.  This semi-infinite interval contains $d$ data events.  The standard model prediction in this interval, estimated from the \Vista\ comparison described above, integrates to $b$ predicted events.  In this final state, the interest of the $d^{\text{th}}$ region is defined as the Poisson probability $p_d = \sum_{i=d}^{\infty}\frac{b^i}{i!}e^{-b}$ that the standard model background $b$ would fluctuate up to or above the observed number of data events $d$ in this region.  The most interesting region in this final state is the one with smallest Poisson probability.
\item For this final state, pseudo experiments are generated, with pseudo data pulled from the standard model background.  For each pseudo experiment, the interest of the most interesting region is calculated.  An ensemble of pseudo experiments determines the fraction $\scriptP$ of pseudo experiments in this final state in which the most interesting region is more interesting than the most interesting region in this final state observed in the data.  If there is no new physics in this final state, $\scriptP$ is expected to be a random number pulled from a uniform distribution in the unit interval.  If there is new physics in this final state, $\scriptP$ is expected to be small.
\item  Looping over all final states, $\scriptP$ is computed for each final state.  The minimum of these values is denoted $\scriptP_{\text{min}}$.  The most interesting region in the final state with smallest $\scriptP$ is denoted ${\cal R}$.
\item The interest of the most interesting region ${\cal R}$ in the most interesting final state is defined by $\twiddleScriptP = 1-\prod_a(1-\hat{p}_a)$, where the product is over all \Sleuth\ final states $a$, and $\hat{p}_a$ is the lesser of $\scriptP_{\text{min}}$ and the probability for the total number of events predicted by the standard model in the final state $a$ to fluctuate up to or above three data events. The quantity $\twiddleScriptP$ represents the fraction of hypothetical similar CDF experiments that would produce a final state with $\scriptP < \scriptP_{\text{min}}$. The range of $\twiddleScriptP$ is the unit interval.  If the data are distributed according to standard model prediction, $\twiddleScriptP$ is expected to be a random number pulled from a uniform distribution in the unit interval.  If new physics is present, $\twiddleScriptP$ is expected to be small.
\end{itemize}

\subsubsection{Output}
\label{sec:Sleuth:Output}

The output of the algorithm is the most interesting region ${\cal R}$ observed in the data, and a number $\twiddleScriptP$ quantifying the interest of ${\cal R}$.  A reasonable threshold for discovery is $\twiddleScriptP \lesssim 0.001$, which corresponds loosely to a local $5\sigma$ effect after the trials factor is accounted for.

Although no integration over systematic errors is performed in computing $\twiddleScriptP$, systematic uncertainties do affect the final \Sleuth\ result.  If \Sleuth\ highlights a discrepancy in a particular final state, explanations in terms of a correction to the background estimate are considered.  \highlight{This process necessarily requires physics judgement.}  A reasonable explanation of a \Sleuth\ discrepancy in terms of an inadequacy in the modeling of the detector response or standard model prediction that is consistent with external information is fed back into the \Vista\ correction model and tested for global consistency.  In this way, plausible explanations for discrepancies observed by \Sleuth\ are incorporated into the \Vista\ correction model.  This iteration continues until either all reasonable explanations for a significant \Sleuth\ discrepancy are exhausted, resulting in a possible new physics claim, or no significant \Sleuth\ discrepancy remains.

\begin{figure}
\mbox{\includegraphics[width=2.6in,angle=270]{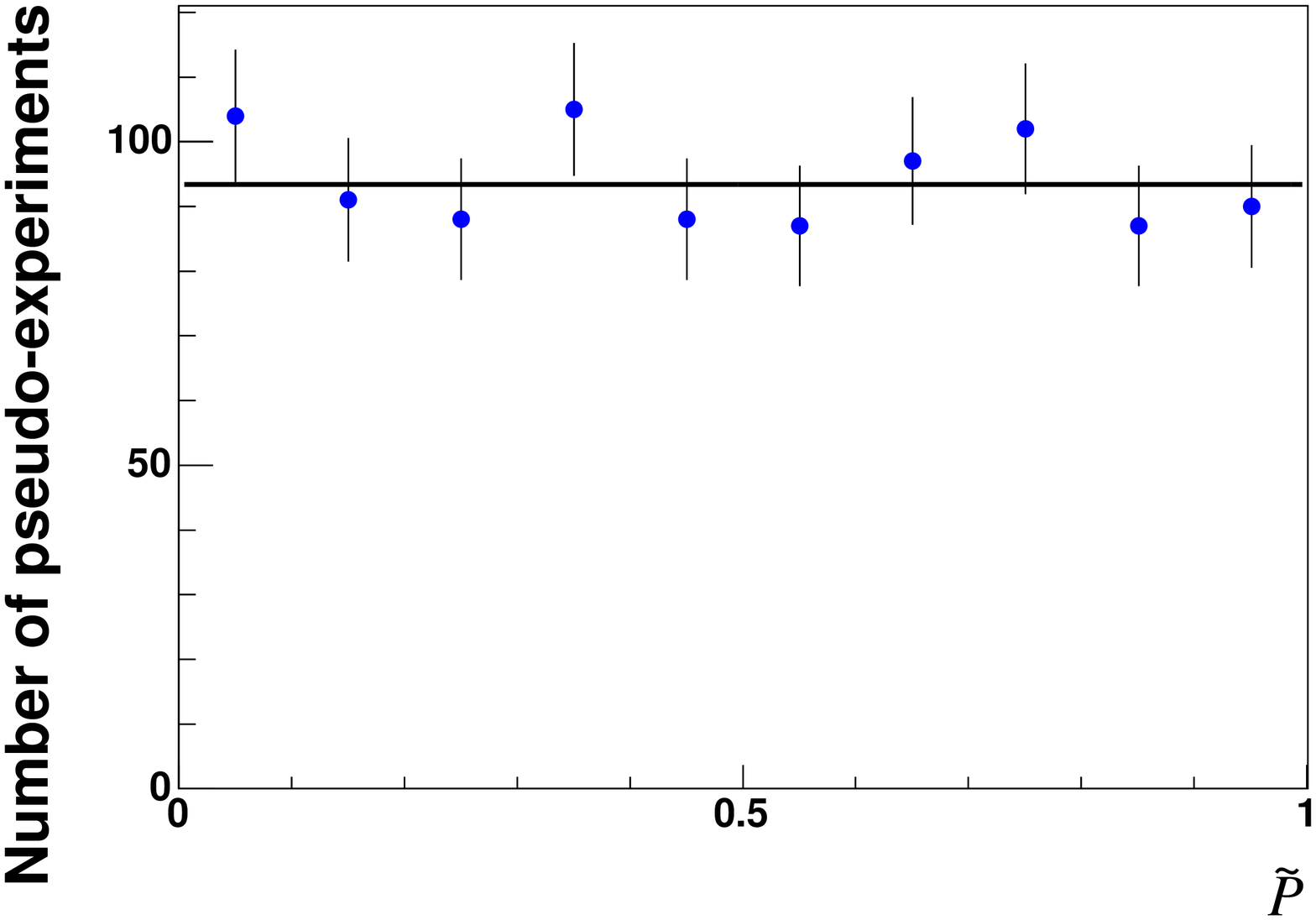} }
\mbox{\includegraphics[width=2.2in,angle=270]{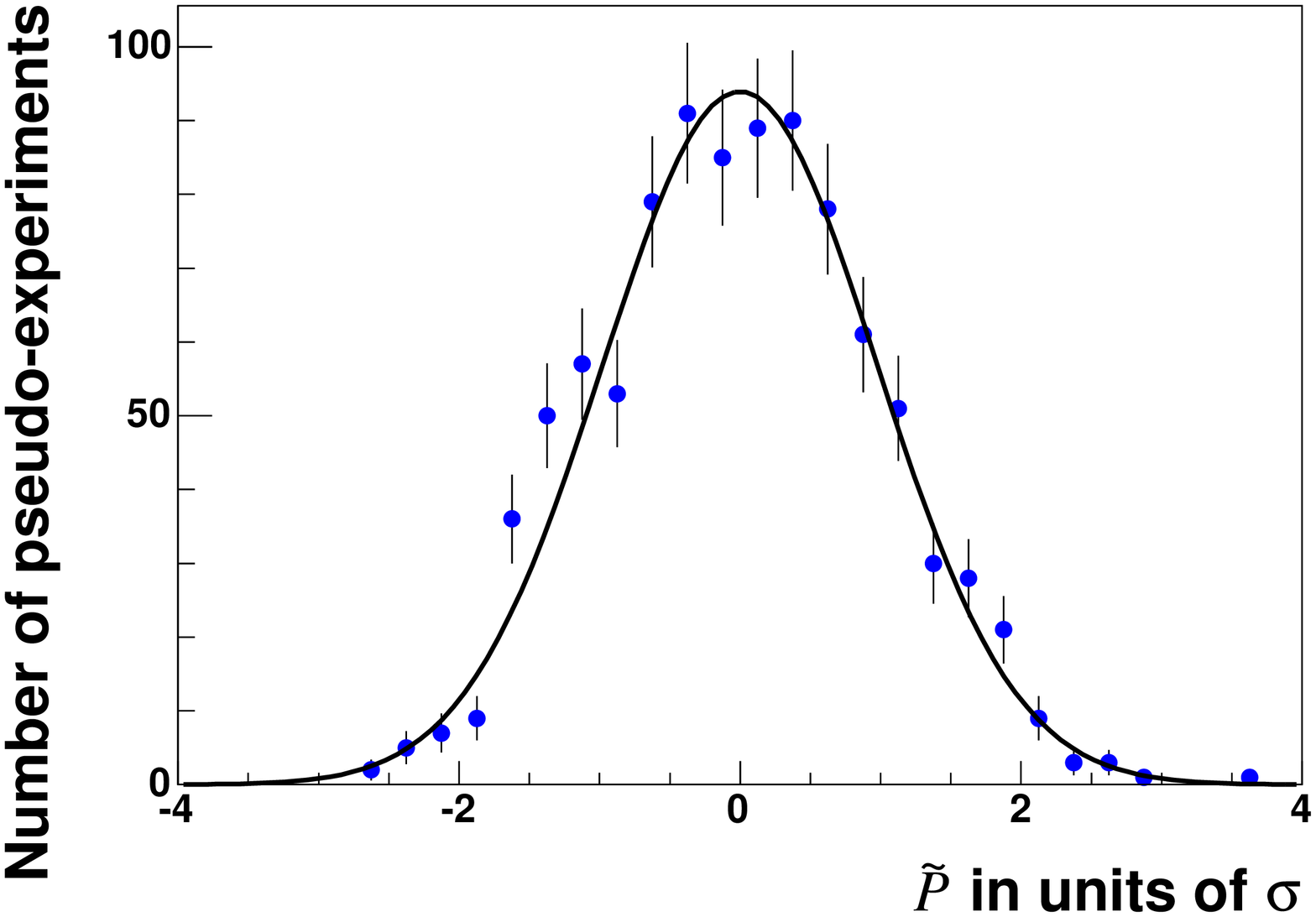} }
\caption{Distribution of $10^3$ \twiddleScriptP\ values from $10^3$ CDF pseudo experiments, in which pseudo data are pulled from the standard model prediction.  The distribution of \twiddleScriptP\ is shown in the unit interval (upper), with one entry for each of the CDF pseudo experiments.  The distribution of \twiddleScriptP\ translated into units of standard deviations is also shown (lower).  The distribution of \twiddleScriptP\ from pseudo experiments is consistent with flat (upper), and consistent with a Gaussian when translated into units of standard deviations (lower), as expected.}
\label{fig:tildeScriptPsPlotsPseudo}
\end{figure}


\begin{figure*}
\begin{tabular}{cc}
\hspace*{-0.15in}\includegraphics[width=2.75in,angle=270]{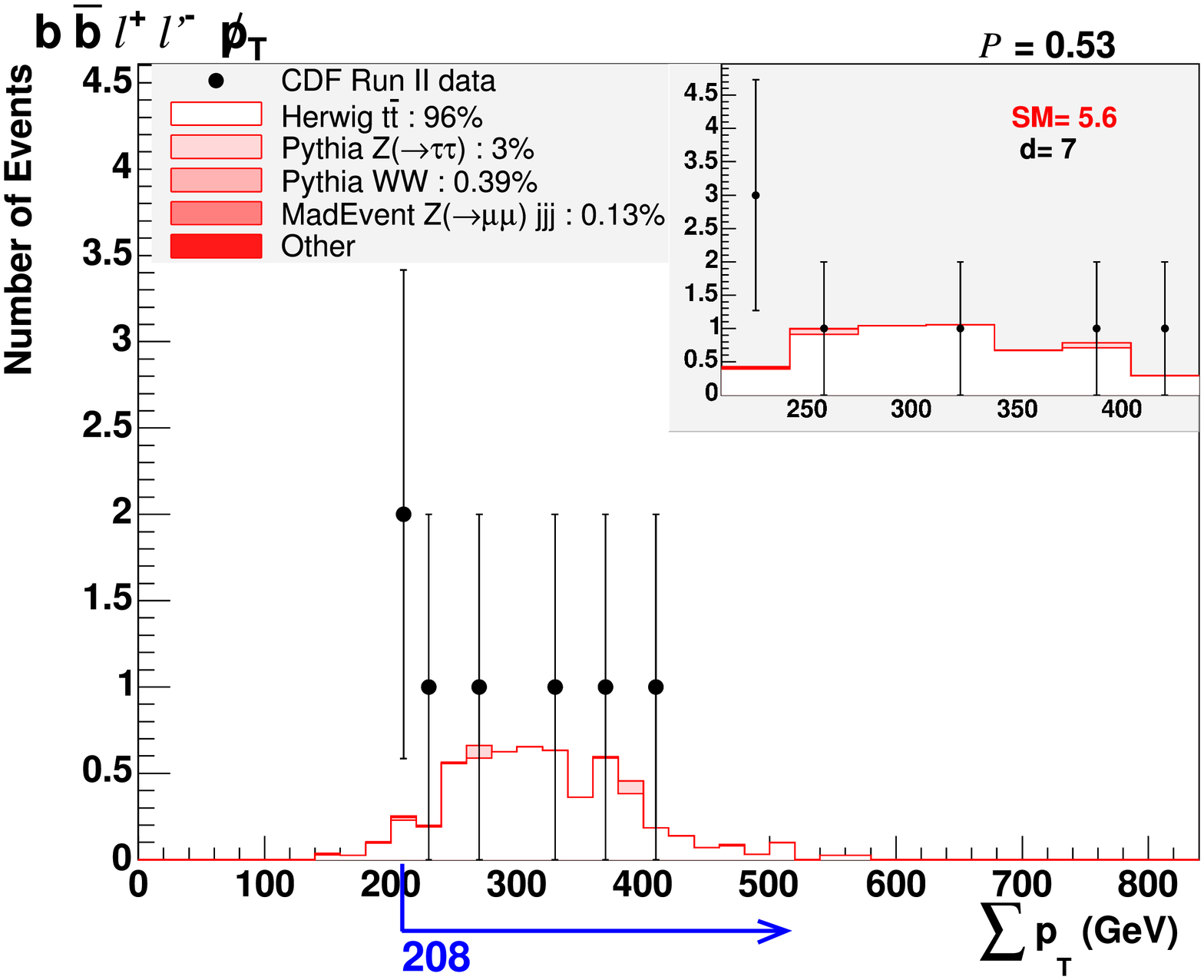} & 
\hspace*{-0.15in}\includegraphics[width=2.75in,angle=270]{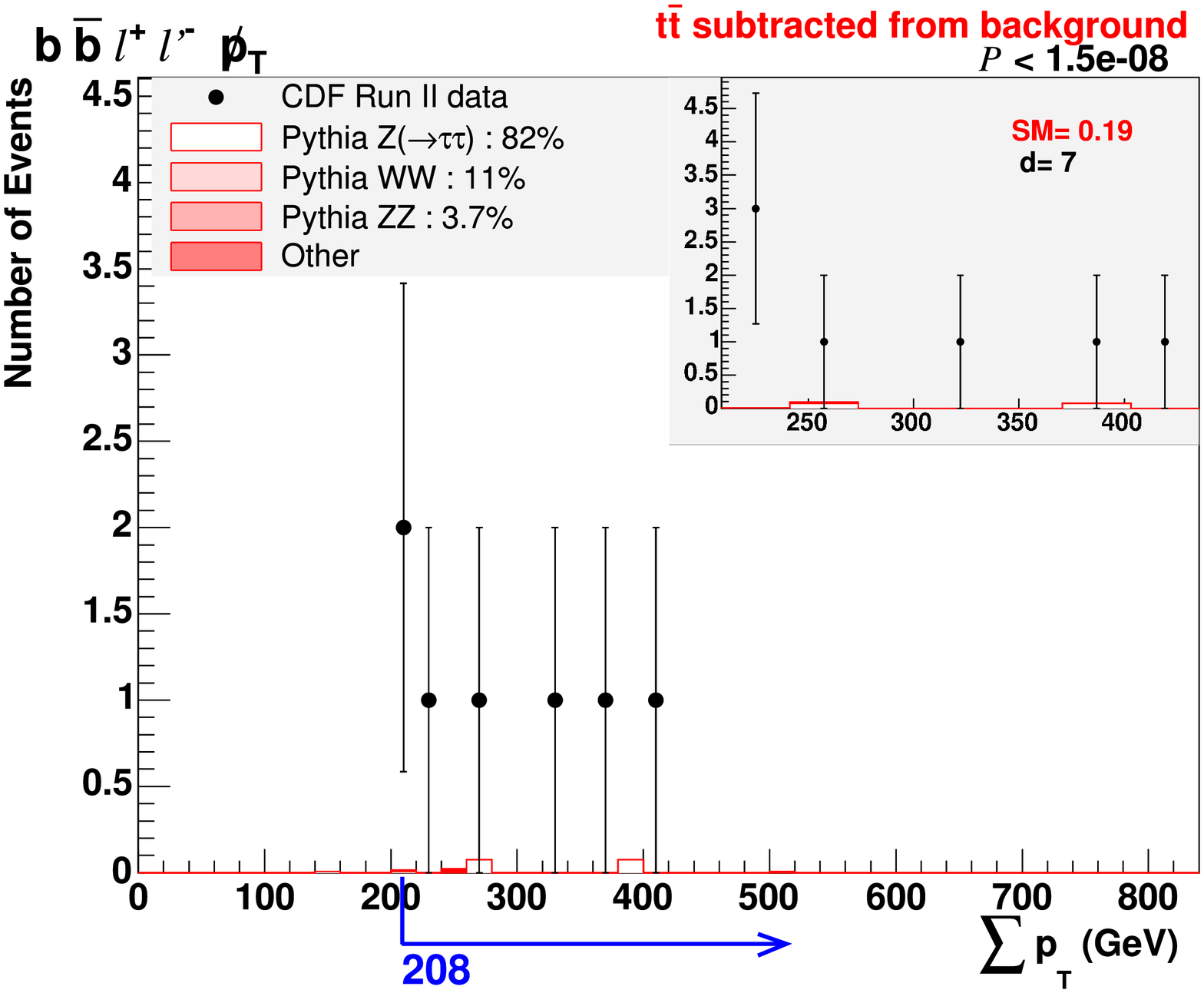} \\
\hspace*{-0.15in}\includegraphics[width=2.75in,angle=270]{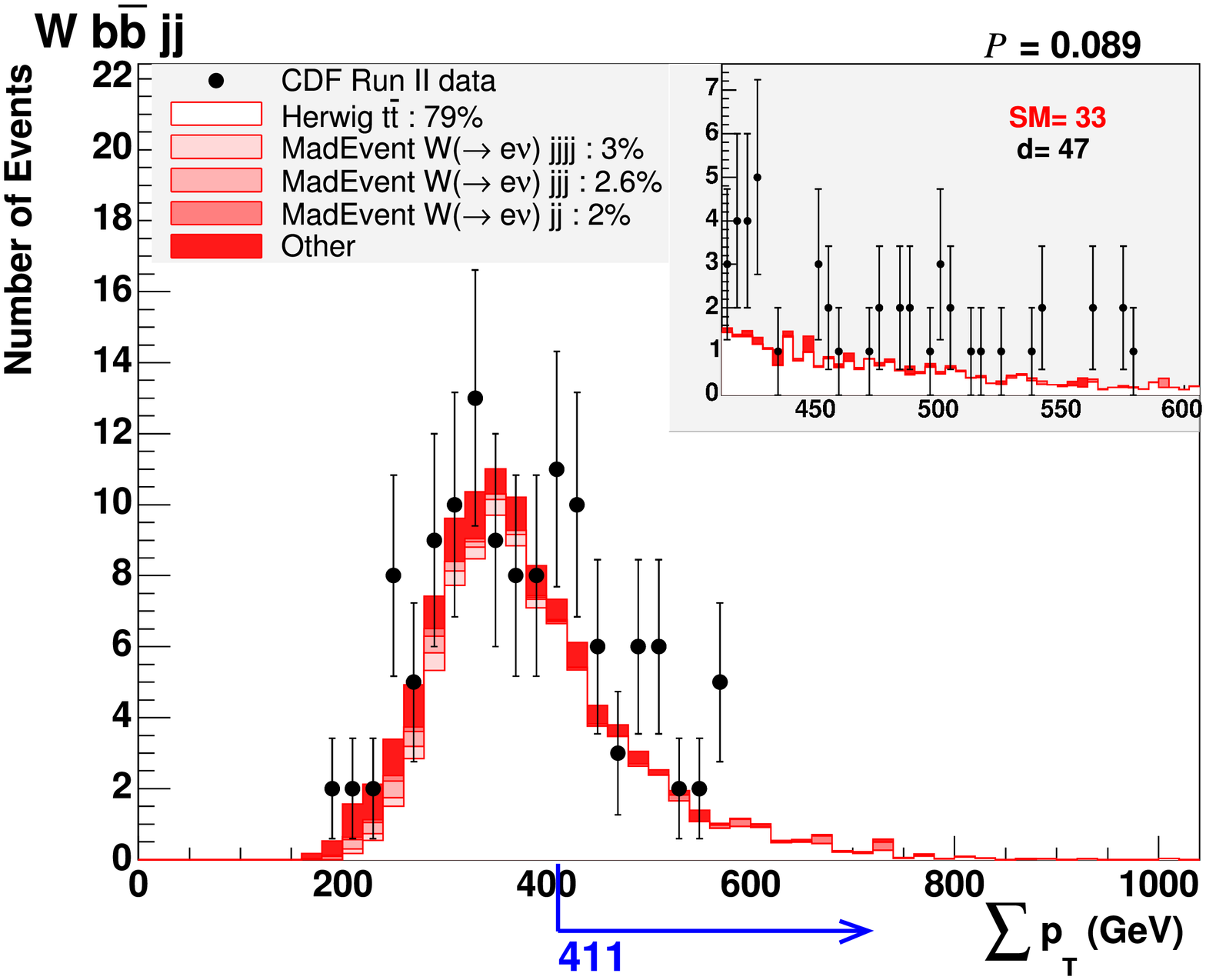} & 

\hspace*{-0.15in}\includegraphics[width=2.75in,angle=270]{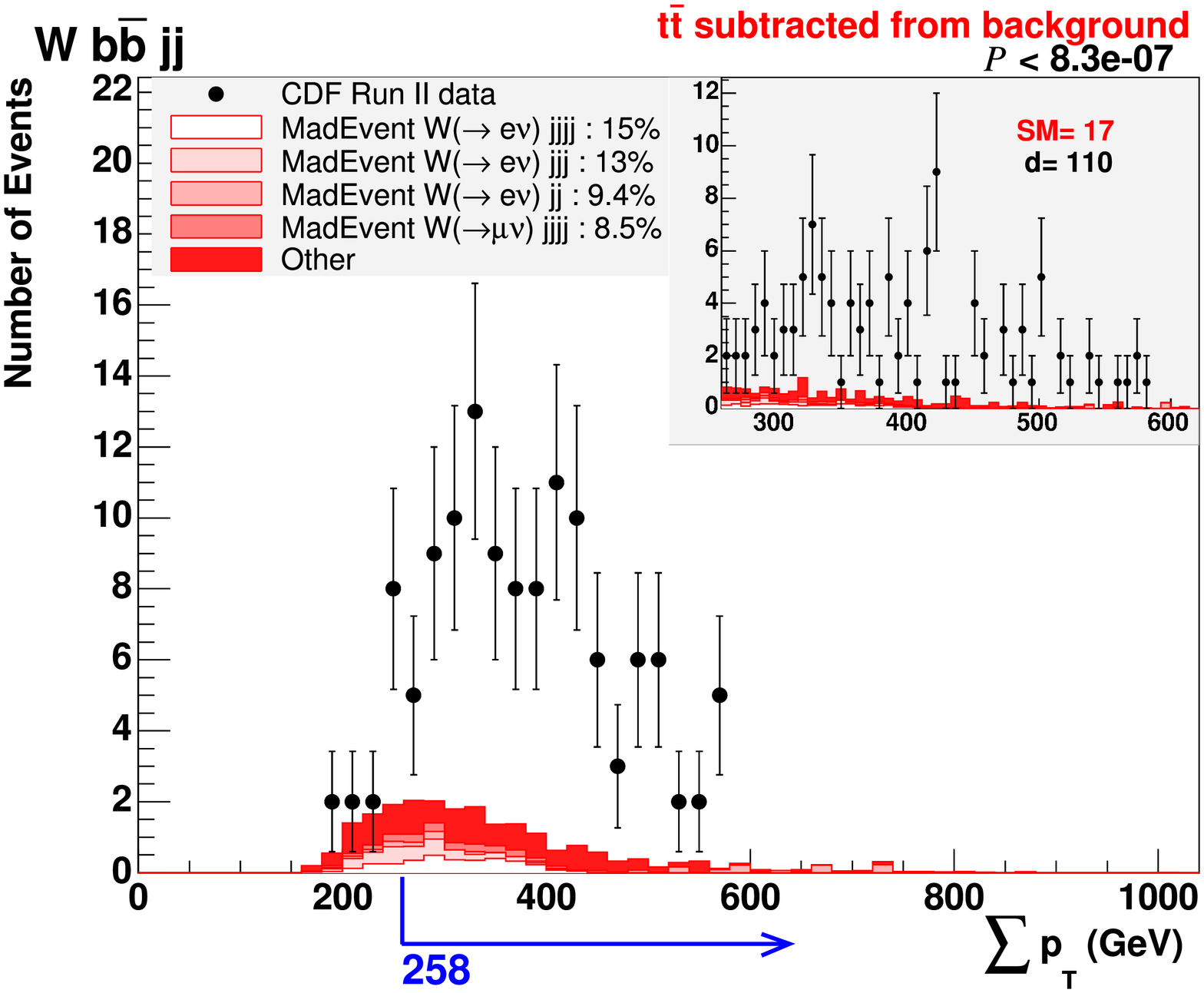} 
\end{tabular}
\caption{(Top left)  The \Sleuth\ final state $b\bar{b}\ell^+\ell'^-\pmiss$, consisting of events with one electron and one muon of opposite sign, missing momentum, and two or three jets, one or two of which are $b$-tagged.  Data \highlight{corresponding to 927~pb$^{-1}$} are shown as filled (black) circles; the standard model prediction is shown as the (red) shaded histogram.  (Top right)  The same final state with $t\bar t$ subtracted from the standard model prediction.  (Bottom row)  The \Sleuth\ final state $Wb\bar{b}jj$, with the standard model $t\bar t$ contribution included (lower left) and removed (lower right).  Significant discrepancies far surpassing \Sleuth's discovery threshold are observed in these final states with $t\bar{t}$ removed from the standard model background estimate.  If the top quark had not been predicted, \Sleuth\ would have discovered it.  
}
\label{fig:topless_SM_sensitivityTest}
\end{figure*}


\begin{figure*}
\begin{tabular}{cc}
\hspace*{-0.2in}\includegraphics[width=2.75in,angle=270]{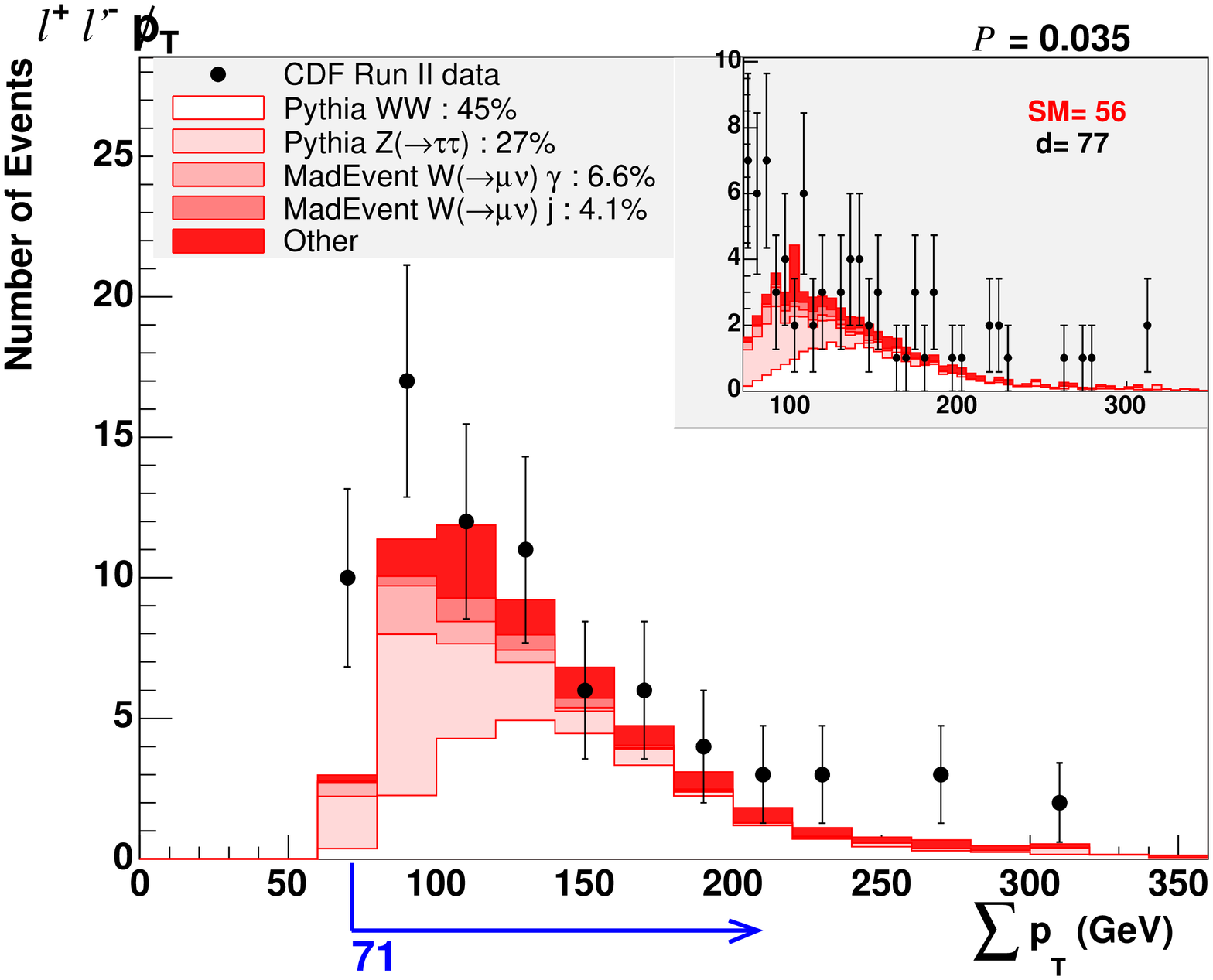} & \hspace*{0.2in}\includegraphics[width=2.75in,angle=270]{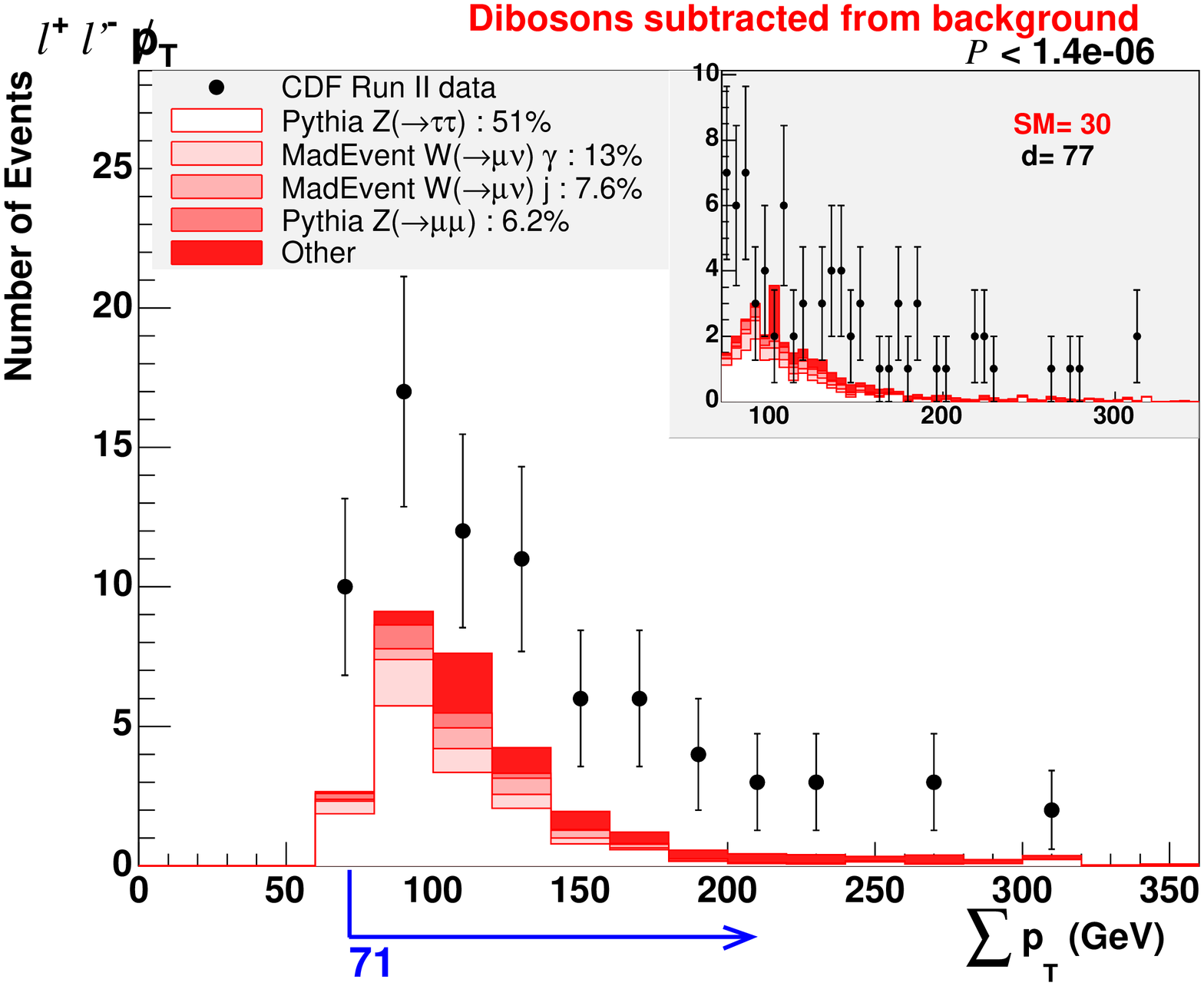} \\
\end{tabular}
\caption{(Left)  The final state $\ell^+{\ell'}^-\pmiss$, consisting of events with an electron and muon of opposite sign and missing transverse momentum, in \VistaApproximateLuminosity~pb$^{-1}$ of CDF data.  (Right) The same final state with standard model $WW$, $WZ$, and $ZZ$ contributions subtracted, and with the \Vista\ correction factors re-fit in the absence of these contributions.  
\Sleuth\ finds the final state $\ell^+{\ell'}^-\pmiss$ to contain a discrepancy surpassing the discovery threshold of $\tildeScriptP<0.001$ with the processes $WW$, $WZ$, and $ZZ$ removed from the standard model background.}
\label{fig:VVless_SM_sensitivityTest}
\end{figure*}

\begin{table*}
\begin{minipage}{7in}
\begin{tabular}{cp{5cm}c}
  {\bf Model} & \multicolumn{1}{c}{{\bf Description}} & {\bf Sensitivity} \\
  \hline
  {1}
             & GMSB, $\Lambda=82.6$~GeV, $\tan{\beta}=15$, $\mu>0$, with one messenger of $M=2\Lambda$. 
  &  \raisebox{-0.5\height}{\includegraphics[width=7.5cm]{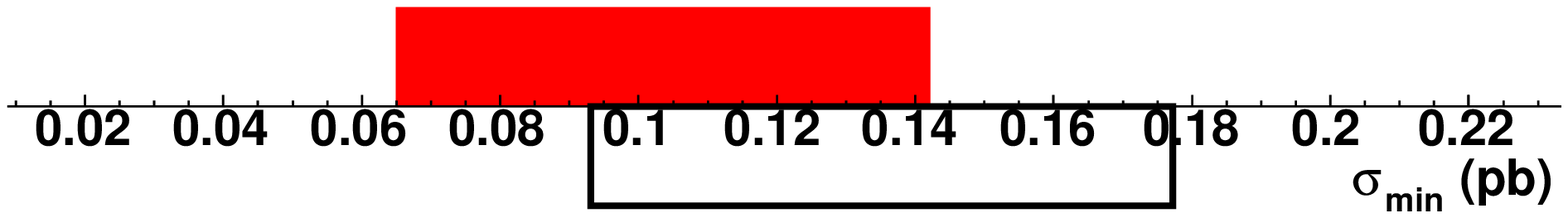}} \\ 
  {2}
             &  $Z'\to \ell^+\ell^-$, $m_{Z'}=250$~GeV, with standard model couplings to leptons.
  &   \raisebox{-0.5\height}{\includegraphics[width=7.5cm]{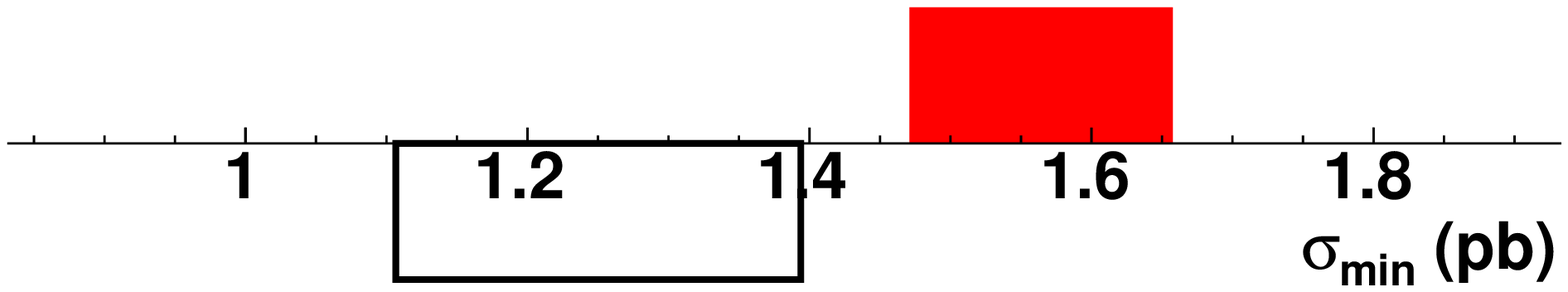}} \\ 
  {3}
             & $Z' \to q\bar{q}$, $m_{Z'}=700$~GeV, with standard model couplings to quarks.
  &   \raisebox{-0.5\height}{\includegraphics[width=7.5cm]{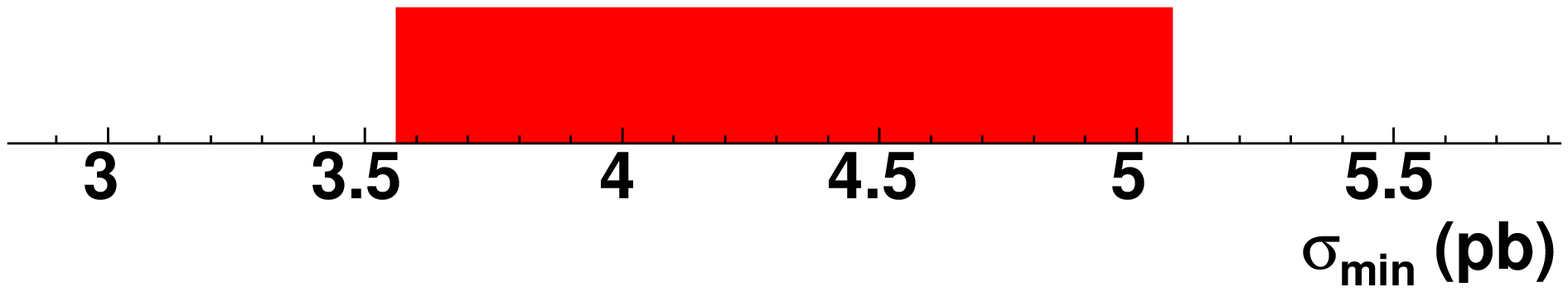}} \\ 
  {4}
             & $Z'\to q\bar{q}$, $m_{Z'}=1$~TeV, with standard model couplings to quarks.
  &   \raisebox{-0.5\height}{\includegraphics[width=7.5cm]{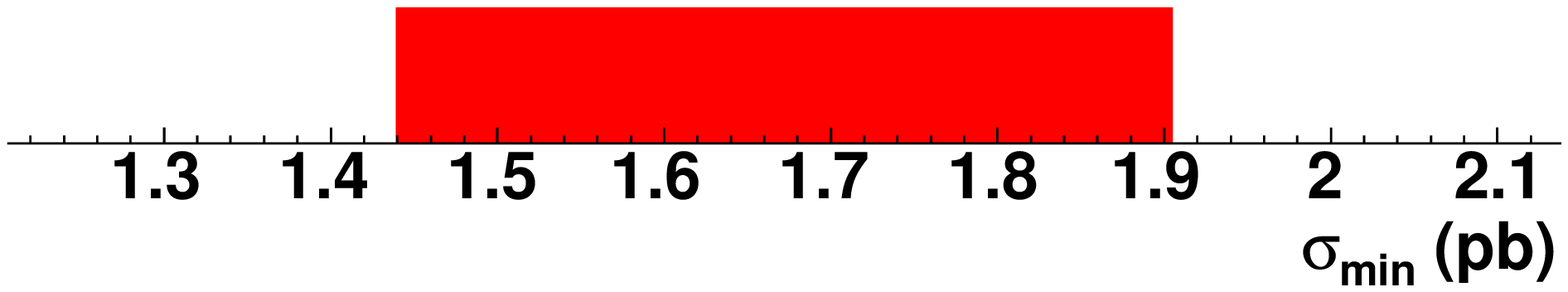}} \\ 
  {5}
             &  $Z'\to t\bar{t}$, $m_{Z'}=500$~GeV, with standard model couplings to $t\bar{t}$.
  &   \raisebox{-0.5\height}{\includegraphics[width=7.5cm]{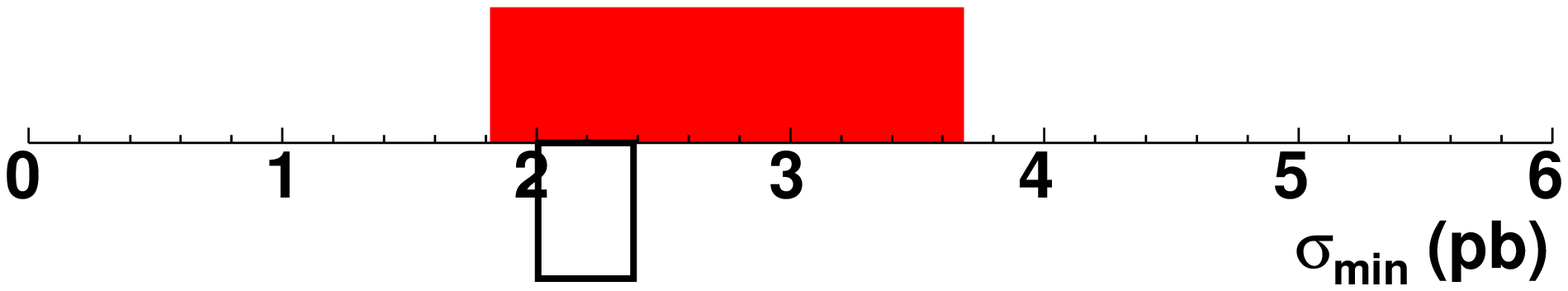}} \\ 
\end{tabular}
\end{minipage}
\caption{Summary of \Sleuth's sensitivity to several new physics models, expressed in terms of the minimum production cross section needed for discovery with \VistaApproximateLuminosity~pb$^{-1}$.  Where available, a comparison is made to the sensitivity of a dedicated search for this model.  The solid (red) box represents \Sleuth's sensitivity, and the open (white) box represents the sensitivity of the dedicated analysis. Systematic uncertainties are not included in the sensitivity calculation. The width of each box shows typical variation under fluctuation of data statistics.  In Models 3 and 4, there is no targeted analysis available for comparison.  \Sleuth\ is seen to perform comparably to the targeted analyses on models satisfying the assumptions on which \Sleuth\ is based.
}
\label{tab:sensitivitySummary}
\end{table*}

\subsection{Sensitivity}

Two important questions must be asked:
\begin{itemize}
\item Will \Sleuth\ find nothing if there is nothing to be found?
\item Will \Sleuth\ find something if there is something to be found?
\end{itemize}

If there is nothing to be found, \Sleuth\ will find nothing 999 times out of 1000, given a uniform distribution of $\twiddleScriptP$ and a discovery threshold of $\twiddleScriptP \lesssim 0.001$.  The uniform distribution of $\twiddleScriptP$ in the absence of new physics is illustrated in Fig.~\ref{fig:tildeScriptPsPlotsPseudo}, using values of $\twiddleScriptP$ obtained in pseudo experiments with pseudo data generated from the standard model prediction.  \Sleuth\ will of course return spurious signals if provided improperly modeled backgrounds.  The algorithm directly addresses the issue of whether an observed hint is due to a statistical fluctuation.  \Sleuth\ itself is unable to address systematic mismeasurement or incorrect modeling, but quite useful in bringing these to attention.

The answer to the second question depends to what degree the new physics satisfies the three assumptions on which \Sleuth\ is based:  new physics will appear predominantly in one final state, at high summed scalar transverse momentum, and as an excess of data over standard model prediction.  \Sleuth's sensitivity to any particular new phenomenon depends on the extent to which this new phenomenon satisfies these assumptions.

\subsubsection{Known standard model processes} 

Consideration of specific standard model processes can provide intuition for \Sleuth's sensitivity to new physics.  This section tests \Sleuth's sensitivity to the production of top quark pairs, $W$ boson pairs, single top, and the Higgs boson.

\paragraph{Top quark pairs.}
Top quark pair production results in two $b$ jets and two $W$ bosons, each of which may decay leptonically or hadronically.  The $W$ branching ratios are such that this signal predominantly populates the \Sleuth\ final state $Wb\bar{b}jj$, where ``$W$'' denotes an electron or muon and significant missing momentum.  Although the final states $\ell^+\ell^-\pmiss b\bar{b}$ were important in verifying the top quark pair production hypothesis in the initial observation by CDF~\cite{CDFTopDiscovery:Abe:1995hr} and D\O~\cite{D0TopDiscovery:Abachi:1995iq} in 1995, most of the statistical power came from the final state $Wb\bar{b}jj$.  The all hadronic decay final state $b\bar{b}\,4j$ has only convincingly been seen after integrating substantial Run II luminosity~\cite{TopAllHadronic:Aaltonen:2006xc}.  \Sleuth's first assumption that new physics will appear predominantly in one final state is thus reasonably well satisfied.  Since the top quark has a mass of $170.9\pm1.8$~GeV~\cite{TopQuarkMass:unknown:2007bx}, the production of two such objects leads to a signal at large \SumPt\ relative to the standard model background of $W$ bosons produced in association with jets, satisfying \Sleuth's second and third assumptions.  \Sleuth\ is expected to perform reasonably well on this example.  

To quantitatively test \Sleuth's sensitivity to top quark pair production, this process is removed from the standard model prediction, and the values of the \Vista\ correction factors are re-obtained from a global fit assuming ignorance of $t\bar{t}$ production.  \Sleuth\ easily discovers $t\bar{t}$ production in \VistaApproximateLuminosity~pb$^{-1}$ in the final states $b\bar{b}\ell^+\ell'^-\pmiss$ and $Wb\bar{b}jj$, shown in Fig.~\ref{fig:topless_SM_sensitivityTest}.  \Sleuth\ finds $\scriptP_{b\bar{b}\ell^+\ell'^-\pmiss}<1.5\times10^{-8}$ and $\scriptP_{Wb\bar{b}jj}<8.3\times10^{-7}$, far surpassing the discovery threshold of $\twiddleScriptP \lesssim 0.001$.

\highlightB{The test is repeated as a function of assumed integrated luminosity, and  \Sleuth\ is found to highlight the top quark signal at an integrated luminosity of roughly $80\pm60$~pb$^{-1}$, where the large variation arises from statistical fluctuations in the $t\bar{t}$ signal events.}  Weaker constraints on the \Vista\ correction factors at lower integrated luminosity marginally increase the integrated luminosity required to claim a discovery.

\paragraph{$W$ boson pairs.}

The sensitivity to standard model $WW$ production is tested by removing this process from the standard model background prediction and allowing the \Vista\ correction factors to be re-fit.  \highlight{In 927~pb$^{-1}$ of Tevatron Run II data, \Sleuth\ identifies an excess in the final state $\ell^+{\ell'}^-\pmiss$, consisting of an electron and muon of opposite sign and missing momentum.}  This excess corresponds to $\twiddleScriptP < 2 \times 10 ^{-4}$, sufficient for the discovery of $WW$, as shown in Fig.~\ref{fig:VVless_SM_sensitivityTest}. 

\paragraph{Single top.}
Single top quarks are produced weakly, and predominantly decay to populate the \Sleuth\ final state $Wb\bar{b}$, satisfying \Sleuth's first assumption.  Single top production will appear as an excess of events, satisfying \Sleuth's third assumption.  \Sleuth's second assumption is not well satisfied for this example, since single top production does not lie at large \SumPt\ relative to other standard model processes.  \Sleuth\ is thus expected to be outperformed by a targeted search in this example.  

\paragraph{Higgs boson.}
Assuming a standard model Higgs boson of mass $m_h=115$~GeV, the dominant observable production mechanism is $p\bar{p}\rightarrow Wh$ and $p\bar{p}\rightarrow Zh$, populating the final states $Wb\bar{b}$, $\ell^+\ell^- b\bar{b}$, and $\pmiss\,b\bar{b}$.  The signal is thus spread over three \Sleuth\ final states.  Events in the last of these ($\pmiss\,b\bar{b}$) do not pass the \Vista\ event selection, which does not use $\pmiss$ as a trigger object.  \Sleuth's first assumption is thus poorly satisfied for this example.
The standard model Higgs boson signal will appear as an excess, but as in the case of single top production it does not appear at particularly large \SumPt\ relative to other standard model processes.  Since the standard model Higgs boson poorly satisfies \Sleuth's first and second assumptions, a targeted search for this specific signal is expected to outperform \Sleuth.  

\subsubsection{Specific models of new physics}
\label{sec:SleuthSensitivity:SpecificModels}

To build intuition for \Sleuth's sensitivity to new physics signals, several sensitivity tests are conducted for a variety of new physics possibilities.  Some of the new physics models chosen have already been considered by more specialized analyses within CDF, making possible a comparison between \Sleuth's sensitivity and the sensitivity of these previous analyses.  

\Sleuth's sensitivity can be compared to that of a dedicated search by determining the minimum new physics cross section $\sigma_\text{min}$ required for a discovery by each.  The discovery for \Sleuth\ occurs when $\tildeScriptP < 0.001$.  \highlight{In most \Sleuth\ regions satisfying the discovery threshold of $\tildeScriptP < 0.001$, the probability for the predicted number of events to fluctuate up to or above the number of events observed corresponds to greater than $5\sigma$.}  The discovery for the dedicated search occurs when the observed excess of data corresponds to a $5\sigma$ effect.  Smaller $\sigma_\text{min}$ corresponds to greater sensitivity.

The sensitivity tests are performed by first generating pseudo data from the standard model background prediction.  Signal events for the new physics model are generated, passed through the chain of CDF detector simulation and event reconstruction, and consecutively added to the pseudo data until \Sleuth\ finds $\tildeScriptP<0.001$.  The number of signal events needed to trigger discovery is used to calculate $\sigma_\text{min}$.

For each dedicated analysis to which \Sleuth\ is compared, the number of standard model events expected in \VistaApproximateLuminosity~pb$^{-1}$ within the region targeted is used to calculate the number of signal events required in that region to produce a discrepancy corresponding to $5\sigma$.  Using the signal efficiency determined in the dedicated analysis, $\sigma_\text{min}$ is calculated.  The effect of systematic uncertainties are removed from the dedicated analyses, and are not included for \Sleuth.  \highlightB{The inclusion of systematic uncertainties will reduce the sensitivity of both \Sleuth\ and the dedicated analysis to the extent that the systematic parameters are allowed to vary.  \Vista\ and \Sleuth\ have the advantage of using a large data set to constrain them.}

The results of five such sensitivity tests are summarized in Table~\ref{tab:sensitivitySummary}.  \Sleuth\ is seen to perform comparably to targeted analyses on models satisfying the assumptions on which \Sleuth\ is based.  For models in which \Sleuth's simple use of $\SumPt$ can be improved upon by optimizing for a specific feature, a targeted search may be expected to achieve greater sensitivity.  One of the important features of \Sleuth\ is that it not only performs reasonably well, but that it does so broadly.  In Model 1, a search for a particular model point in a gauge mediated supersymmetry breaking (GMSB) scenario, \Sleuth\ gains an advantage by exploiting a final state not considered in the targeted analysis~\cite{Acosta:2004sb}.  In Model 2, a search for a $Z'$ decaying to lepton pairs, the targeted analysis~\cite{SamHarper:Abulencia:2006iv} exploits the narrow resonance in the $e^+e^-$ invariant mass.  In Models 3 and 4, which are searches for a hadronically decaying $Z'$ of different masses, there is no targeted analysis against which to compare.  In Model 5, a search for a $Z'\to t\bar{t}$ resonance, the signal appears at large summed scalar transverse momentum in a particular final state, resulting in comparable sensitivity between \Sleuth\ and the targeted analysis~\cite{JacoZprimettbar2}.

\begin{figure}
\begin{tabular}{c}
\includegraphics[width=2.1in,angle=270]{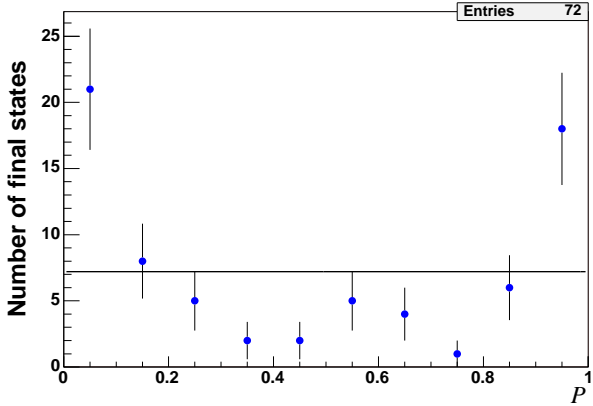} \\
\end{tabular}
\caption{The distribution of \scriptP\ in the data, with one entry for each final state considered by \Sleuth.}
\label{fig:scriptPsPlots}
\end{figure}

\begin{figure*}
\begin{tabular}{cc}
\includegraphics[width=2.5in,angle=270]{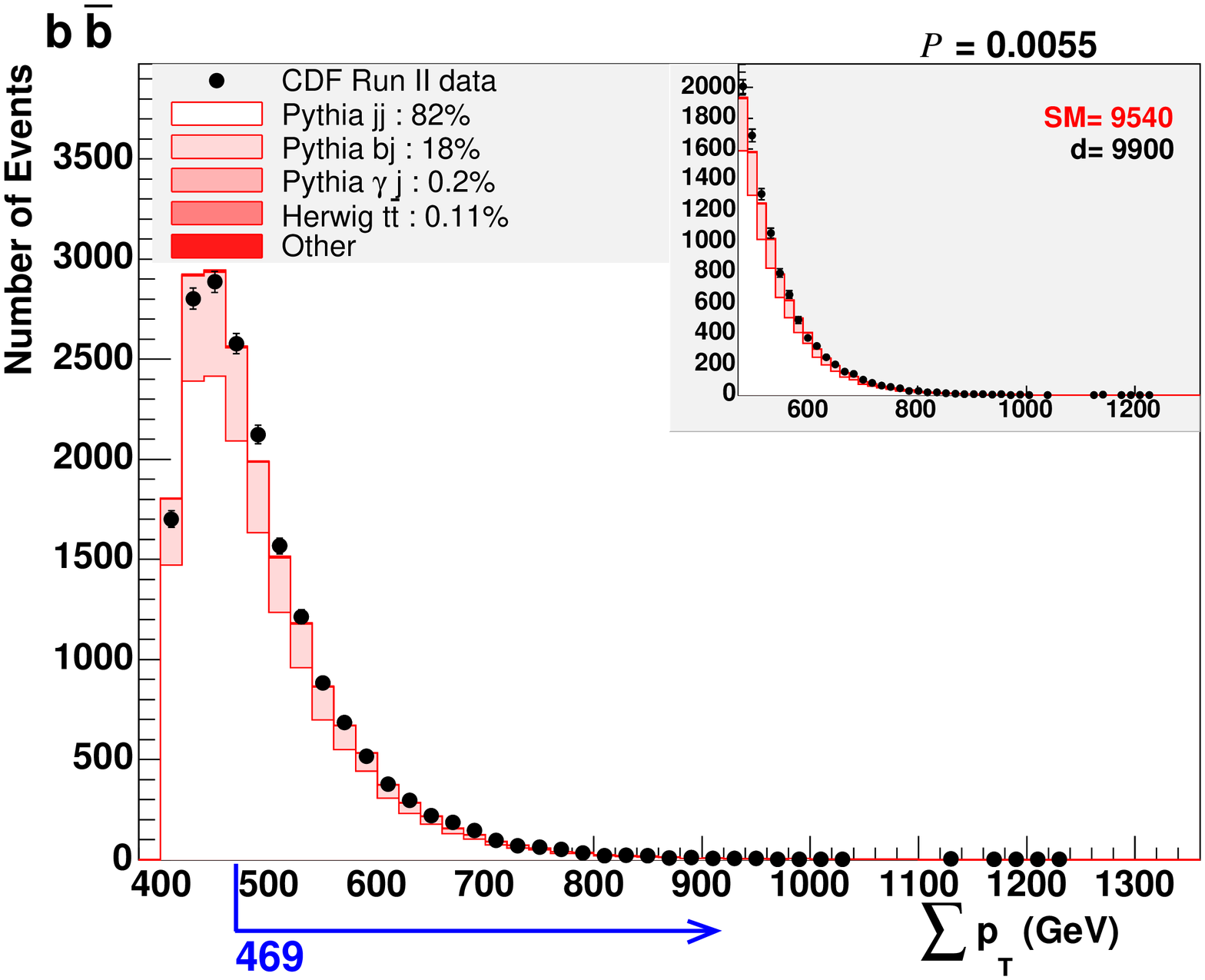} &
\includegraphics[width=2.5in,angle=270]{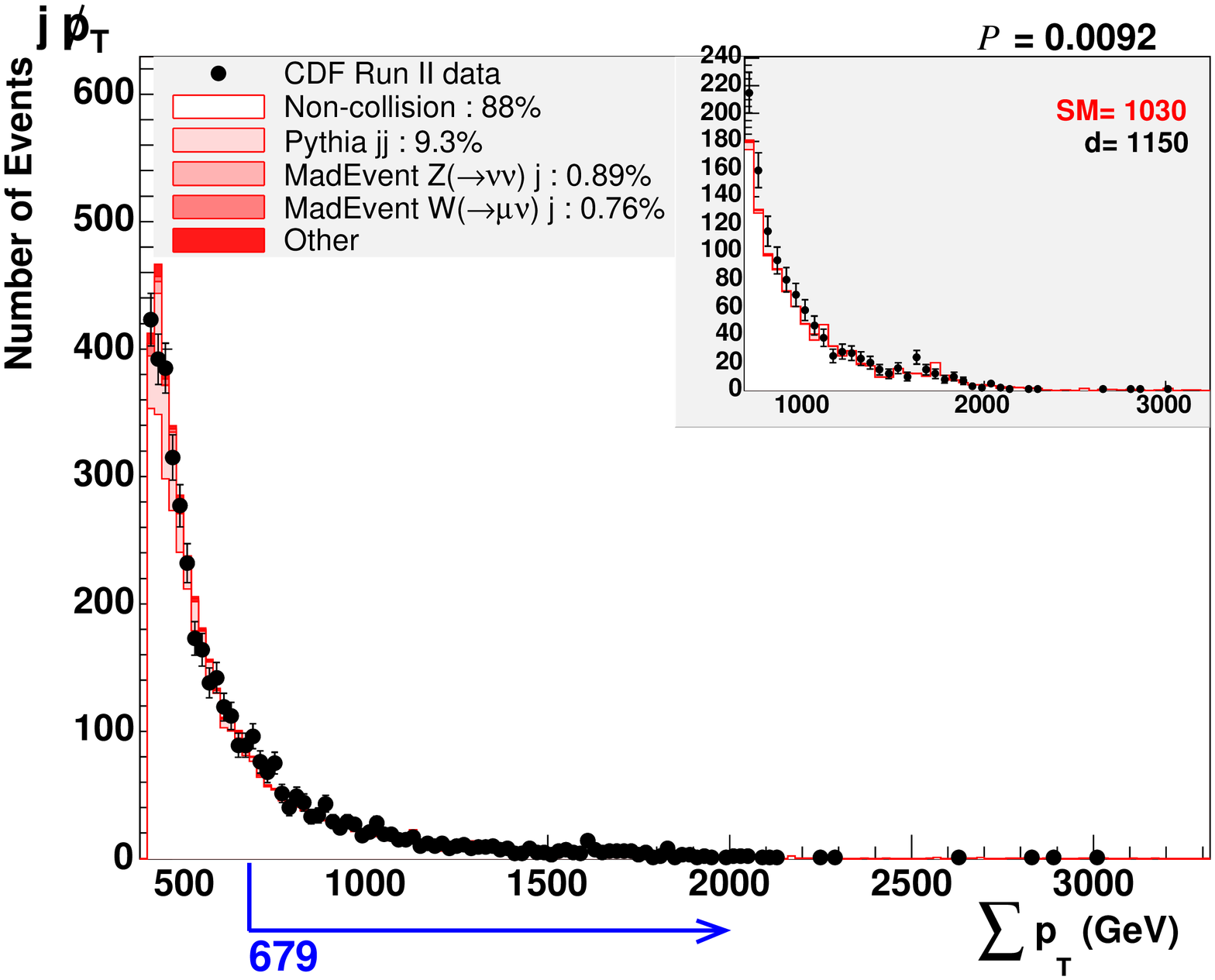} \\
\includegraphics[width=2.5in,angle=270]{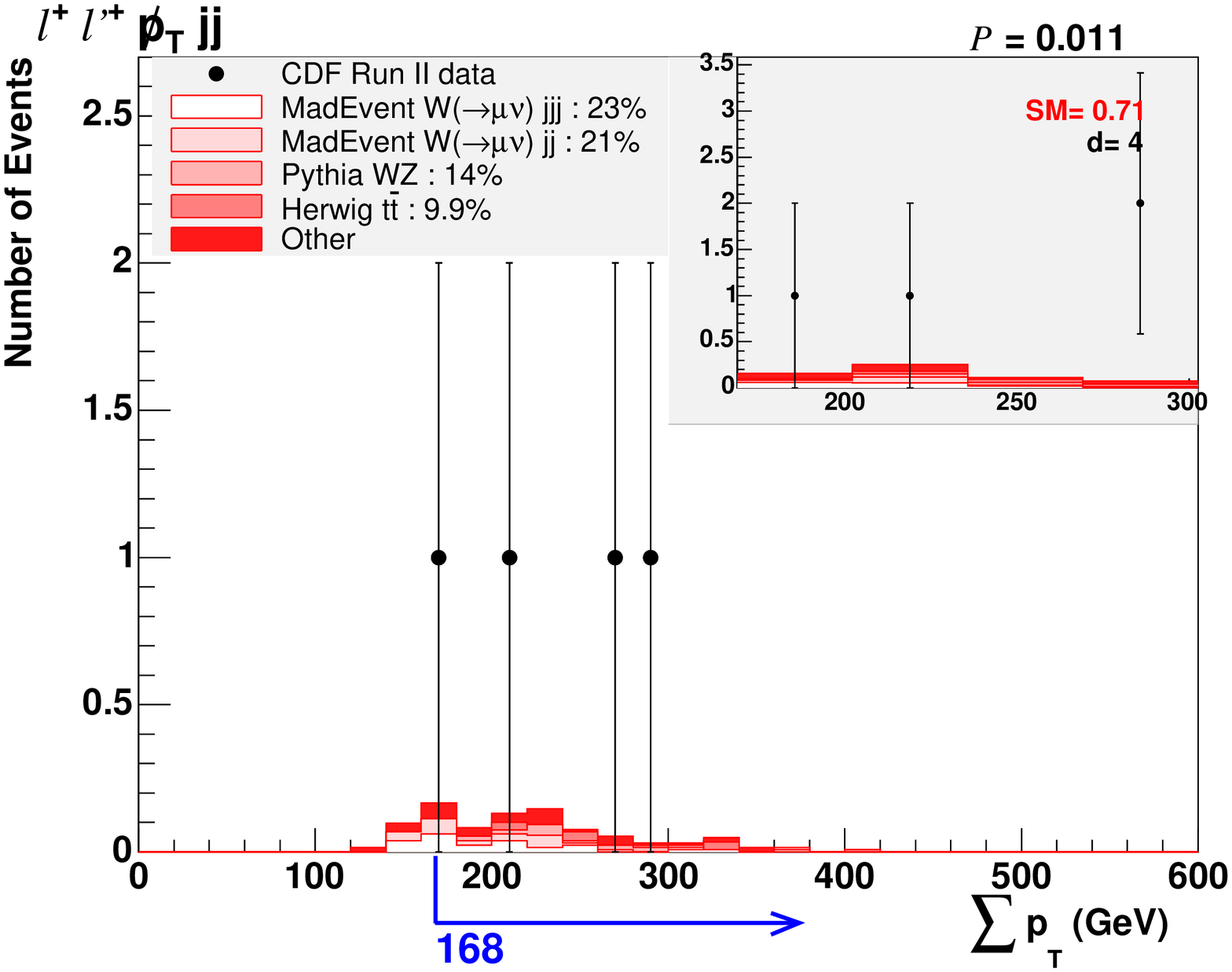} &
\includegraphics[width=2.5in,angle=270]{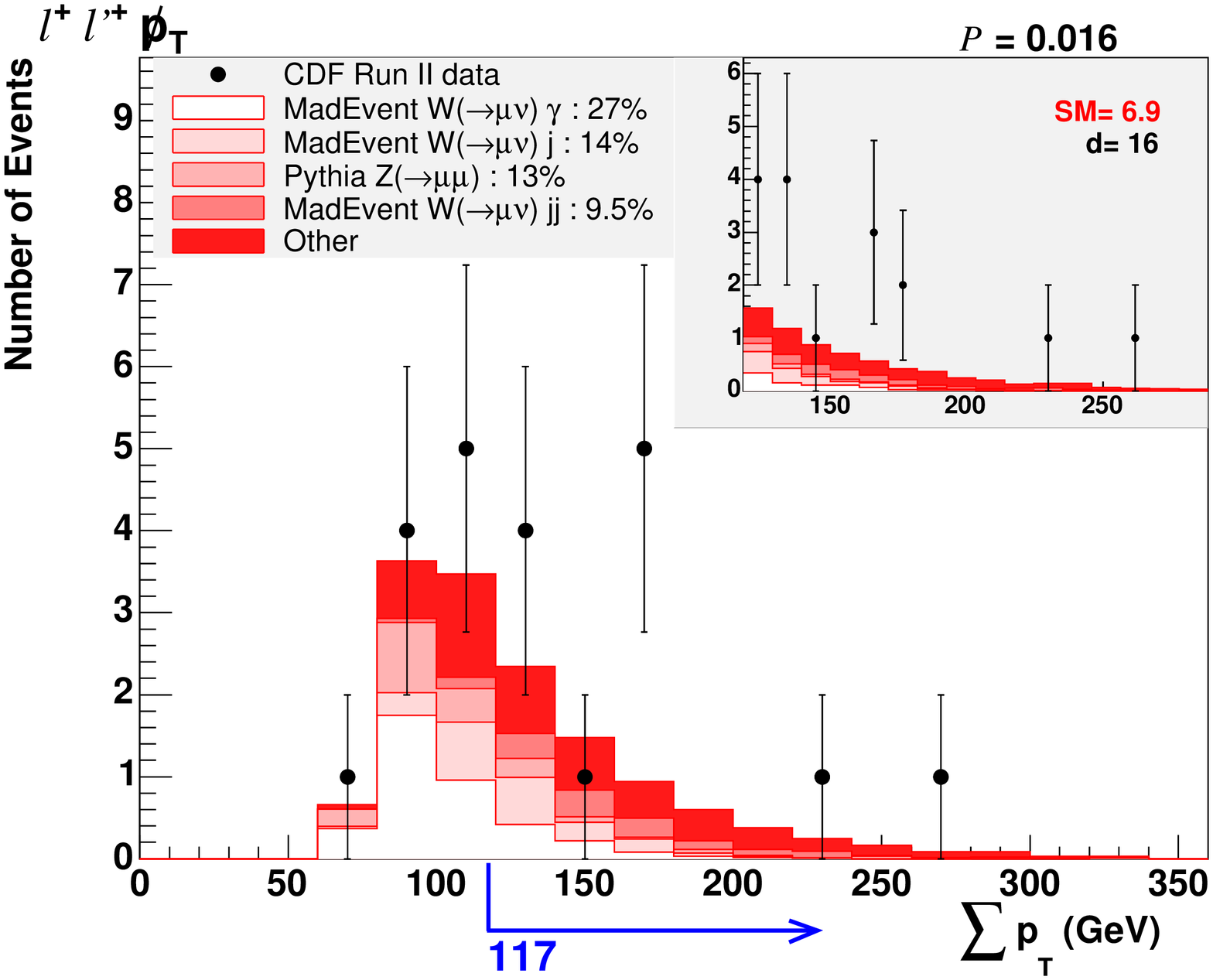} \\
\includegraphics[width=2.5in,angle=270]{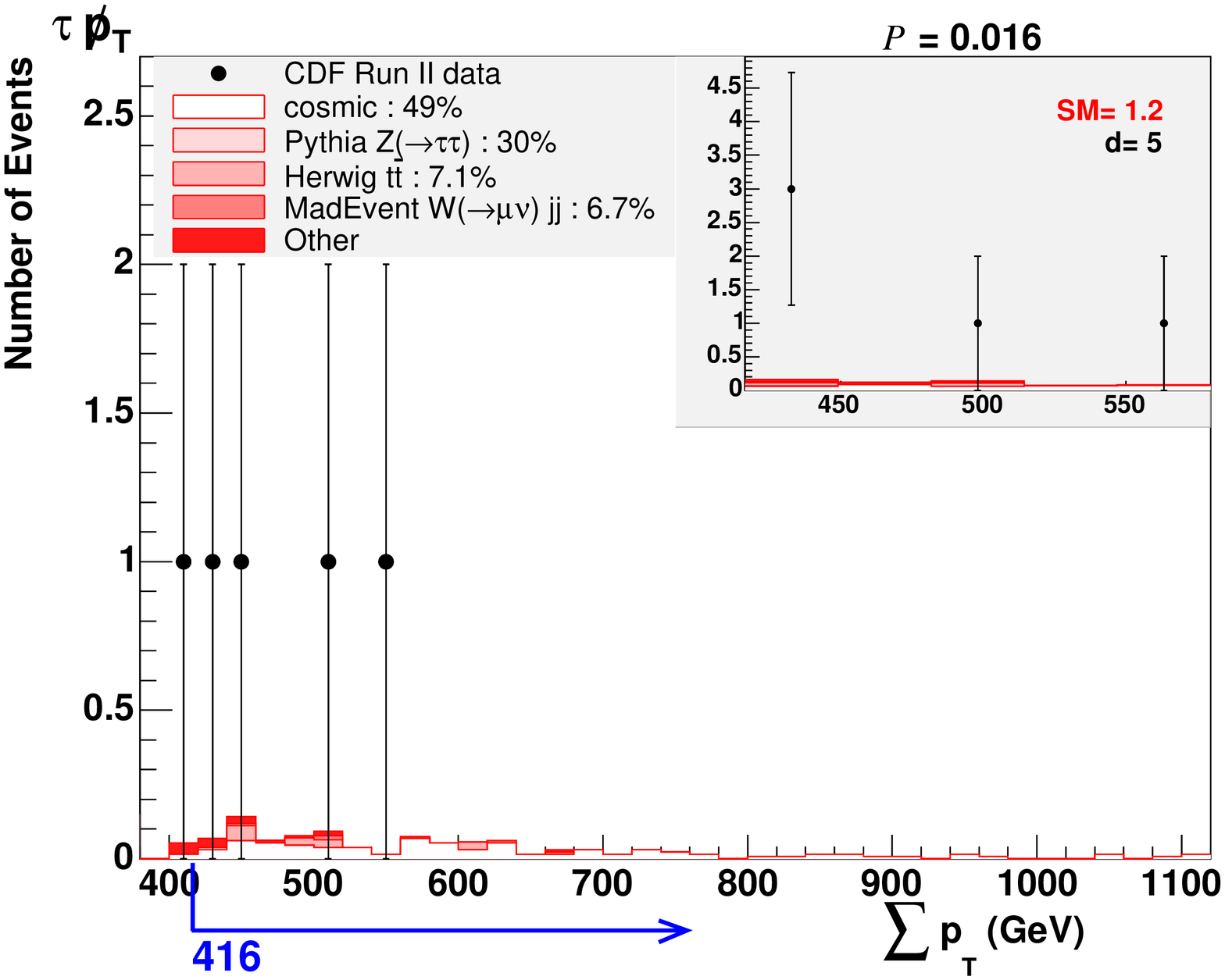} &
\includegraphics[width=2.5in,angle=270]{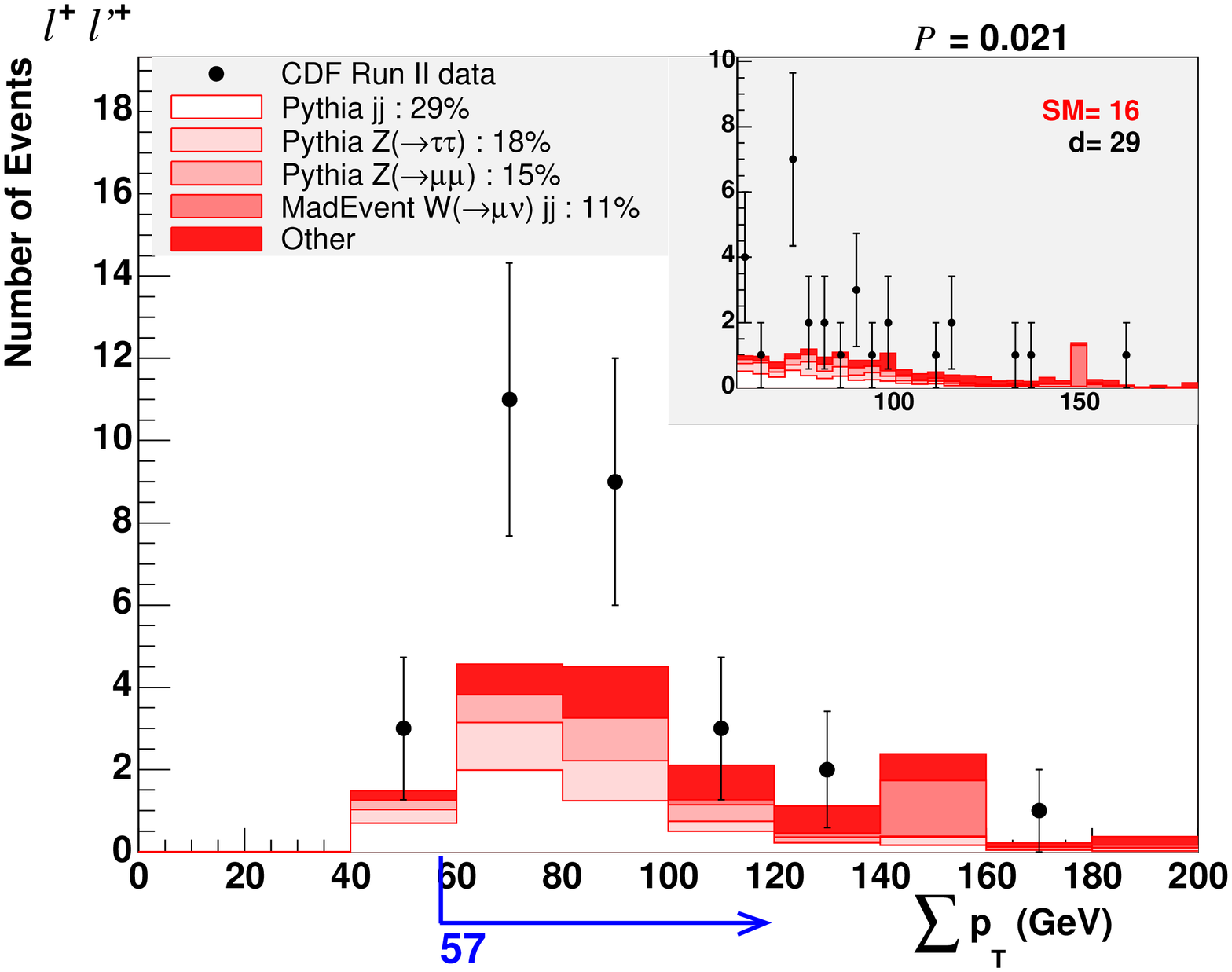} \\
\end{tabular}
\caption{The most interesting final states identified by \Sleuth.  The region chosen by \Sleuth, extending up to infinity, is shown by the (blue) arrow just below the horizontal axis.  Data are shown as filled (black) circles, and the standard model prediction is shown as the shaded (red) histogram.  The \Sleuth\ final state is labeled in the upper left corner of each panel, with $\ell$ denoting $e$ or $\mu$, and $\ell^+\ell'^+$ denoting an electron and muon with the same electric charge.  The number at upper right in each panel shows ${\cal P}$, the fraction of hypothetical similar experiments in which something at least as interesting as the region shown would be seen in this final state.  The inset in each panel shows an enlargement of the region selected by \Sleuth, together with the number of events (${\text{SM}}$) predicted by the standard model in this region, and the number of data events ($d$) observed in this region.
\label{fig:SleuthPlots}}
\end{figure*}


\subsection{Results}
\label{sec:Sleuth:Results}

The distribution of \scriptP\ for the final states considered by \Sleuth\ in the data is shown in Fig.~\ref{fig:scriptPsPlots}.  The concavity of this distribution reflects the degree to which the correction model described in Sec.~\ref{sec:Vista:CorrectionModel} has been tuned.  A crude correction model tends to produce a distribution that is concave upwards, as seen in this figure, while an overly tuned correction model produces a distribution that is concave downwards, with more final states than expected having $\scriptP$ near the midpoint of the unit interval.  

The most interesting final states identified by \Sleuth\ are shown in Fig.~\ref{fig:SleuthPlots}, together with a quantitative measure (\scriptP) of the interest of the most interesting region in each final state, determined as described in Sec.~\ref{sec:Sleuth:Regions}.  The legends of Fig.~\ref{fig:SleuthPlots} show the primary contributing standard model processes in each of these final states, together with the fractional contribution of each.  The top six final states, which correspond to entries in the leftmost bin in Fig.~\ref{fig:scriptPsPlots}. span a range of populations, relevant physics objects, and important background contributions.  

The final state $b\bar{b}$, consisting of two or three reconstructed jets, one or two of which are $b$-tagged, heads the list.  These events enter the analysis by satisfying the \Vista\ offline selection requiring one or more jets or $b$-jets with $p_T>200$~GeV.  The definition of \Sleuth's \SumPt\ variable is such that all events in this final state consequently have $\SumPt>400$~GeV.  \Sleuth\ chooses the region $\SumPt>469$~GeV, which includes nearly $10^4$ data events.  \highlight{The standard model prediction in this region is sensitive to the $b$-tagging efficiency $\poo{b}{b}$ and the fake rate $\poo{j}{b}$, which have few strong constraints on their values for jets with $p_T>200$~GeV other than those imposed by other \Vista\ kinematic distributions within this and a few other related final states.}  For this region \Sleuth\ finds $\scriptP_{b\bar{b}}=0.0055$, which is unfortunately not statistically significant after accounting for the trials factor associated with looking in many different final states, as discussed below.

The final state $j\pmiss$, consisting of events with one reconstructed jet and significant missing transverse momentum, is the second final state identified by \Sleuth.  The primary background is due to non-collision processes, including cosmic rays and beam halo backgrounds, whose estimation is discussed in Appendix~\ref{sec:CorrectionModelDetails:CosmicRays}.  Since the hadronic energy is not required to be deposited in time with the beam crossing, \Sleuth's analysis of this final state is sensitive to particles with a lifetime between 1~ns and 1~$\mu$s that lodge temporarily in the hadronic calorimeter, complementing Ref.~\cite{Hugo:Abulencia:2006kk}.

The final states $\ell^+ {\ell'}^{+} \pmiss jj$, $\ell^+ {\ell'}^{+} \pmiss$, and $\ell^+ {\ell'}^{+}$ all contain an electron ($\ell$) and muon ($\ell'$) with identical reconstructed charge (either both positive or both negative).  The final states with and without missing transverse momentum are qualitatively different in terms of the standard model processes contributing to the background estimate, with the final state $\ell^+ {\ell'}^{-}$ composed mostly of dijets where one jet is misreconstructed as an electron and a second jet is misreconstructed as a muon; $Z\rightarrow \tau^+\tau^-$, where one tau decays to a muon and the other to a leading $\pi^0$, one of the two photons from which converts while traveling through the silicon support structure to result in an electron reconstructed with the same sign as the muon, as described in Appendix~\ref{sec:MisidentificationMatrix}; and $Z\rightarrow\mu^+\mu^-$, in which a photon is produced, converts, and is misreconstructed as an electron.  The final states containing missing transverse momentum are dominated by the production of $W(\rightarrow\mu\nu)$ in association with one or more jets, with one of the jets misreconstructed as an electron.  The muon is significantly more likely than the electron to have been produced in the hard interaction, since the fake rate $\poo{j}{\mu}$ is roughly an order of magnitude smaller than the fake rate $\poo{j}{e}$, as observed in Table~\ref{tbl:CorrectionFactorDescriptionValuesSigmas}.  The final state $\ell^+ {\ell'}^{-} \pmiss jj$, which contains two or three reconstructed jets in addition to the electron, muon, and missing transverse momentum, also has some contribution from $WZ$ and top quark pair production.

The final state $\tau\pmiss$ contains one reconstructed tau, significant missing transverse momentum, and one reconstructed jet with $p_T>200$~GeV.  This final state in principle also contains events with one reconstructed tau, significant missing transverse momentum, and zero reconstructed jets, but such events do not satisfy the offline selection criteria described in Sec.~\ref{sec:Vista:OfflineTrigger}.  Roughly half of the background is non-collision, in which two different cosmic ray muons (presumably from the same cosmic ray shower) leave two distinct energy deposits in the CDF hadronic calorimeter, one with $p_T>200$~GeV, and one with a single associated track from a $p\bar{p}$ collision occurring during the same bunch crossing.  Less than a single event is predicted from this non-collision source (using techniques described in Appendix~\ref{sec:CorrectionModelDetails:CosmicRays}) over the past five years of Tevatron running.

In these CDF data, \Sleuth\ finds $\twiddleScriptP = 0.46$.  The fraction of hypothetical similar CDF experiments (assuming a fixed standard model prediction, detector simulation, and correction model) that would exhibit a final state with \scriptP\ smaller than the smallest \scriptP\ observed in the CDF Run II data is approximately 46\%.  The actual value obtained for $\tildeScriptP$ is not of particular interest, except to note that this value is significantly greater than the threshold of $\lesssim 0.001$ required to claim an effect of statistical significance.  \Sleuth\ has not revealed a discrepancy of sufficient statistical significance to justify a new physics claim.

Systematics are incorporated into \Sleuth\ in the form of the flexibility in the \Vista\ correction model, as described previously.  This flexibility is significantly more important in practice than the uncertainties on particular correction factor values obtained from the fit, although the latter are easier to discuss.  The relative importance of correction factor value uncertainties on \Sleuth's result depends on the number of predicted standard model events ($b$) in \Sleuth's high \SumPt\ tail.  The uncertainties on the correction factors of Table~\ref{tbl:CorrectionFactorDescriptionValuesSigmas} are such that the appropriate addition in quadrature gives a typical uncertainty of $\approx 10\%$ on the total background prediction in each final state.  Using $\sigma_{\text{sys}}\approx 10\% \times b$ and $\sigma_{\text{stat}}\approx \sqrt{b}$, the relative importance of systematic uncertainty and statistical uncertainty is estimated to be $\sigma_{\text{sys}}/\sigma_{\text{stat}}=10\% \times b/\sqrt{b}$.  The importance of systematic and statistical uncertainties are thus comparable for high \SumPt\ tails containing $b\sim 100$ predicted events.  The effect of systematic uncertainties is provided in this approximation rather than through a rigorous integration over these uncertainties as nuisance parameters due to the high computational cost of performing the integration.  This estimate of systematic uncertainty is valid only within the particular correction model resulting in the list of correction factors shown in Table~\ref{tbl:CorrectionFactorDescriptionValuesSigmas}; additional changes to the correction model may result in larger variation.  The inclusion of additional systematic uncertainties does not qualitatively change the conclusion that \Sleuth\ has not revealed a discrepancy of sufficient statistical significance to justify a new physics claim.

Due to the large number of final states considered, there are regions (such as those shown in Fig.~\ref{fig:SleuthPlots}) in which the probability for the standard model prediction to fluctuate up to or above the number of events observed in the data corresponds to a significance exceeding $3\sigma$ if the appropriate trials factor is not accounted for.  A doubling of data may therefore result in discovery.  In particular, although the excesses in Fig.~\ref{fig:SleuthPlots} are currently consistent with simple statistical fluctuations, if any of them are genuinely due to new physics, \Sleuth\ will find they pass the discovery threshold of $\twiddleScriptP<0.001$ with roughly a doubling of data.


\section{Conclusions}
\label{sec:Conclusions}

A broad search for new physics (\Vista) has been performed in \VistaApproximateLuminosity~pb$^{-1}$ of CDF Run II data.  A complete standard model background estimate has been obtained and compared with data in 344 populated exclusive final states and 16,486 relevant kinematic distributions, most of which have not been previously considered.  Consideration of exclusive final state populations yields no statistically significant ($>3\sigma$) discrepancy after the trials factor is accounted for.  Quantifying the difference in shape of kinematic distributions using the Kolmogorov-Smirnov statistic, significant discrepancies are observed between data and standard model prediction.  These discrepancies are believed to arise from mismodeling of the parton shower and intrinsic $k_T$, and represent observables for which a QCD-based understanding is highly motivated.  None of the shape discrepancies highlighted motivates a new physics claim.

A further systematic search (\Sleuth) for regions of excess on the high-$\SumPt$ tails of exclusive final states has been performed, representing a quasi-model-independent search for new electroweak scale physics.  Most of the exclusive final states searched with \Sleuth\ have not been considered by previous Tevatron analyses.  A measure of interest rigorously accounting for the trials factor associated with looking in many regions with few events is defined, and used to quantify the most interesting region observed in the CDF Run II data.  No region of excess on the high-$\SumPt$ tail of any of the \Sleuth\ exclusive final states surpasses the discovery threshold.


Although this global analysis of course cannot prove that no new physics is hiding in these data, this broad search of the Tevatron Run II data represents one of the single most encompassing tests of the particle physics standard model at the energy frontier.


\acknowledgments

Tim Stelzer and Fabio Maltoni have provided particularly valuable assistance in the estimation of standard model backgrounds.  Sergey Alekhin provided helpful correspondence in understanding the theoretical uncertainties on $W$ and $Z$ production imposed as constraints on the \Vista\ correction factor fit.  Sascha Caron provided insight gained from the general search at H1.  Torbj\"orn Sj\"ostrand assisted in the understanding of \Pythia's treatment of parton showering applied to the \Vista\ $3j$ $\Delta R(j_2,j_3)$ discrepancy.

We thank the Fermilab staff and the technical staffs of the participating institutions for their vital contributions. This work was supported by the U.S. Department of Energy and National Science Foundation; the Italian Istituto Nazionale di Fisica Nucleare; the Ministry of Education, Culture, Sports, Science and Technology of Japan; the Natural Sciences and Engineering Research Council of Canada; the National Science Council of the Republic of China; the Swiss National Science Foundation; the A.P. Sloan Foundation; the Bundesministerium f\"ur Bildung und Forschung, Germany; the Korean Science and Engineering Foundation and the Korean Research Foundation; the Science and Technology Facilities Council and the Royal Society, UK; the Institut National de Physique Nucleaire et Physique des Particules/CNRS; the Russian Foundation for Basic Research; the Comisi\'on Interministerial de Ciencia y Tecnolog\'{\i}a, Spain; the European Community's Human Potential Programme; the Slovak R\&D Agency; and the Academy of Finland.


\appendix

\begin{table}
\hspace*{-0.3in}\mbox{
\begin{tabular}{c|rrrrrrrrr}
 & $e^+$ & $e^-$ & $\mu^+$ & $\mu^-$ & $\tau^+$ & $\tau^-$ & $\gamma$ & $j$ & $b$ \\ \hline 
$e^+$  & 62228 & 33 & 0 & 0 & 182 & 0 & 2435 & 28140 & 0  \\
$e^-$  & 24 & 62324 & 0 & 0 & 0 & 192 & 2455 & 28023 & 1  \\
$\mu^+$  & 0 & 0 & 50491 & 0 & 6 & 0 & 0 & 606 & 0  \\
$\mu^-$  & 0 & 1 & 0 & 50294 & 0 & 6 & 0 & 577 & 0  \\
$\gamma$  & 1393 & 1327 & 0 & 0 & 1 & 1 & 67679 & 21468 & 0  \\
$\pi^0$  & 1204 & 1228 & 0 & 0 & 5 & 8 & 58010 & 33370 & 0  \\
$\pi^+$  & 266 & 0 & 115 & 0 & 41887 & 6 & 95 & 54189 & 37  \\
$\pi^-$  & 1 & 361 & 0 & 88 & 13 & 41355 & 148 & 54692 & 44  \\
$K^+$  & 156 & 1 & 273 & 0 & 42725 & 7 & 37 & 52317 & 24  \\
$K^-$  & 1 & 248 & 0 & 165 & 28 & 41562 & 115 & 53917 & 22  \\
$B^+$  & 100 & 0 & 77 & 1 & 100 & 10 & 40 & 66062 & 25861  \\
$B^-$  & 2 & 85 & 3 & 68 & 11 & 99 & 45 & 66414 & 25621  \\
$B^0$  & 88 & 27 & 87 & 17 & 77 & 32 & 21 & 65866 & 25046  \\
$\bar{B^0}$  & 17 & 79 & 11 & 71 & 41 & 77 & 21 & 66034 & 25103  \\
$D^+$  & 126 & 6 & 62 & 0 & 1485 & 67 & 207 & 79596 & 11620  \\
$D^-$  & 4 & 134 & 3 & 74 & 64 & 1400 & 234 & 79977 & 11554  \\
$D^0$  & 60 & 13 & 27 & 2 & 312 & 1053 & 248 & 88821 & 5487  \\
$\bar{D^0}$  & 15 & 46 & 5 & 28 & 1027 & 253 & 237 & 89025 & 5480  \\
$K^0_L$  & 1 & 4 & 0 & 0 & 71 & 60 & 202 & 96089 & 26  \\
$K^0_S$  & 26 & 31 & 2 & 1 & 170 & 525 & 9715 & 76196 & 0  \\
$\tau^+$  & 1711 & 13 & 1449 & 0 & 4167 & 2 & 673 & 50866 & 607  \\
$\tau^-$  & 12 & 1716 & 0 & 1474 & 6 & 3940 & 621 & 51125 & 580  \\
$u$  & 8 & 10 & 1 & 0 & 446 & 31 & 247 & 94074 & 26  \\
$d$  & 3 & 4 & 0 & 0 & 64 & 308 & 191 & 94322 & 22  \\
$g$  & 2 & 0 & 0 & 0 & 17 & 14 & 12 & 81865 & 99  \\
\end{tabular}}
\caption{Central single particle misidentification matrix.  Using a single particle gun, $10^5$ particles of each type shown at the left of the table are shot with $p_T=25$~GeV into the central CDF detector, uniformly distributed in $\theta$ and in $\phi$.  The resulting reconstructed object types are shown at the top of the table, labeling the table columns.  Thus the rightmost element of this matrix in the fourth row from the bottom shows $\poo{\tau^-}{b}$, the number of negatively charged tau leptons (out of $10^5$) reconstructed as a $b$-tagged jet.
\label{tbl:misId_cdfSim_central}}
\end{table}


\section{\Vista\ correction model details}
\label{sec:CorrectionModelDetails}

This appendix contains details of the \Vista\ correction model.  Appendix~\ref{sec:MisidentificationMatrix} covers the physical mechanisms underlying fake rates.  Appendix~\ref{sec:AdditionalBackgroundSources} contains information about additional background sources, including backgrounds from cosmic rays and beam halo, multiple interactions, and the effects of intrinsic $k_T$.  Appendix~\ref{sec:CorrectionFactorFitDetails} contains details of the \Vista\ correction factor fit, including the construction of the $\chi^2$ function that is minimized and the resulting covariance matrix.  Appendix~\ref{sec:VistaCorrectionModel:CorrectionFactorValues} discusses the values of the correction factors that are obtained. 

\subsection{Fake rate physics}
\label{sec:MisidentificationMatrix}

The following facts begin to build a unified understanding of fake rates for electrons, muons, taus, and photons.  This understanding is woven throughout the \Vista\ correction model, and significantly informs and constrains the \Vista\ correction process.  Explicit constraints derived from these studies are provided in Appendix~\ref{sec:CorrectionFactorFitDetails}.  The underlying physical mechanisms for these fakes lead to simple and well justified relations among them.

\begin{figure}
\includegraphics[width=3.0in]{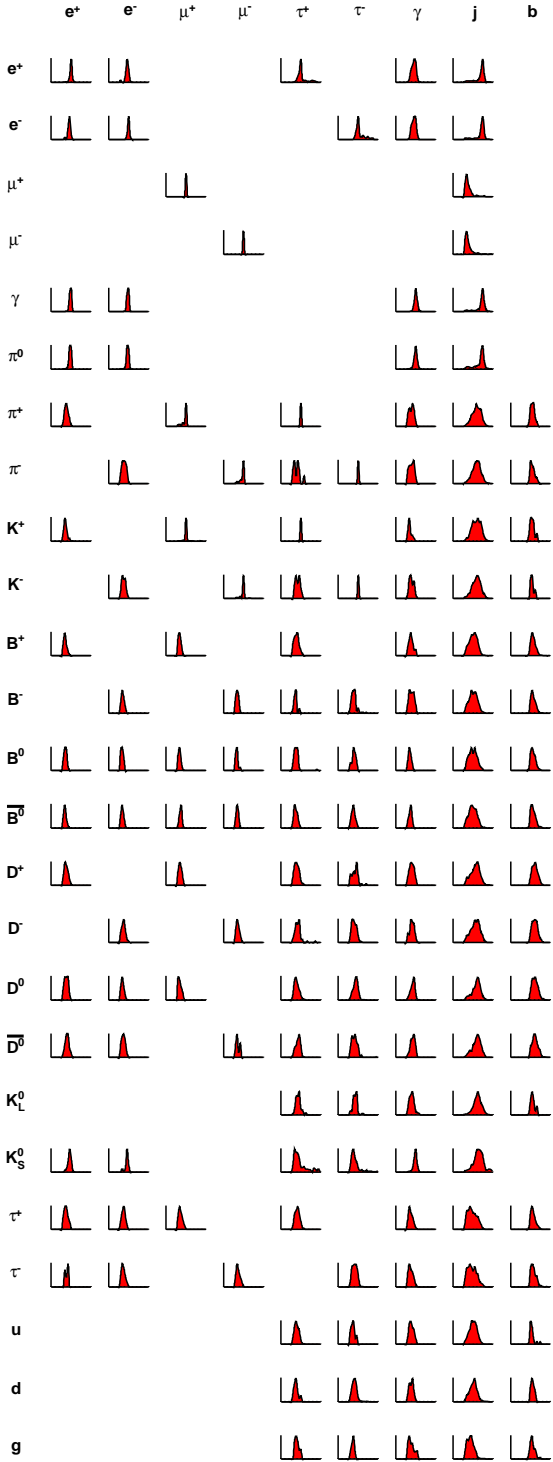}
\caption{Transverse momentum distribution of reconstructed objects (labeling columns) arising from single particles (labeling rows) with $p_T=25$~GeV shot from a single particle gun into the central CDF detector.  The area under each histogram is equal to the number of events in the corresponding misidentification matrix element of Table~\ref{tbl:misId_cdfSim_central}, with the vertical axis of each histogram scaled to the peak of each distribution.  A different vertical scale is used for each histogram, and histograms with fewer than ten events are not shown.  The horizontal axis ranges from 0 to 50~GeV.
\label{fig:misId_cdfSim_central_pt}}
\end{figure}


Table~\ref{tbl:misId_cdfSim_central} shows the response of the CDF detector simulation, reconstruction, and object identification algorithms to single particles.  Using a single particle gun, $10^5$ particles of each type shown at the left of the table are shot with $p_T=25$~GeV into the CDF detector, uniformly distributed in $\theta$ and in $\phi$.  The resulting reconstructed object types are shown at the top of the table, labeling the columns.  The first four entries on the diagonal at upper left show the efficiency for reconstructing electrons and muons~\footnote{The electron and muon efficiencies shown in this table are different from the correction factors {\tt 0025} and {\tt 0027} in Table~\ref{tbl:CorrectionFactorDescriptionValuesSigmas}, which show the ratio of the object efficiencies in the data to the object identification efficiencies in \CdfSim.}.  The fraction of electrons misidentified as photons, shown in the top row, seventh column, is seen to be roughly equal to the fraction of photons identified as electrons or positrons, shown in the fifth row, first and second columns, and measures the number of radiation lengths in the innermost regions of the CDF tracker.  The fraction of $B$ mesons identified as electrons or muons, primarily through semileptonic decay, are shown in the four left columns, eleventh through fourteenth rows.  Other entries provide similarly useful information, most easily comprehensible from simple physics.

The transverse momenta of the objects reconstructed from single particles are displayed in Fig.~\ref{fig:misId_cdfSim_central_pt}.  \cdfSpecific{Table~\ref{tbl:misId_cdfSim_50GeV_central} shows a similar study with $10^4$ particles at $p_T=50$~GeV.}  The relative resolutions for the measurement of electron and muon momenta are shown in the first four histograms on the diagonal at upper left.  The histograms in the left column, sixth through eighth rows, show that single neutral pions misreconstructed as electrons have their momenta well measured, while single charged pions misreconstructed as electrons have their momenta systematically undermeasured, as discussed below.  The histogram in the top row, second column from the right, shows that electrons misreconstructed as jets have their energies systematically overmeasured.  Other histograms in Fig.~\ref{fig:misId_cdfSim_central_pt} contain similarly relevant information, easily overlooked without the benefit of this study, but understandable from basic physics considerations once the effect has been brought to attention.


\begin{figure*}
\begin{tabular}{ccc}
\includegraphics[width=2.5in,angle=270]{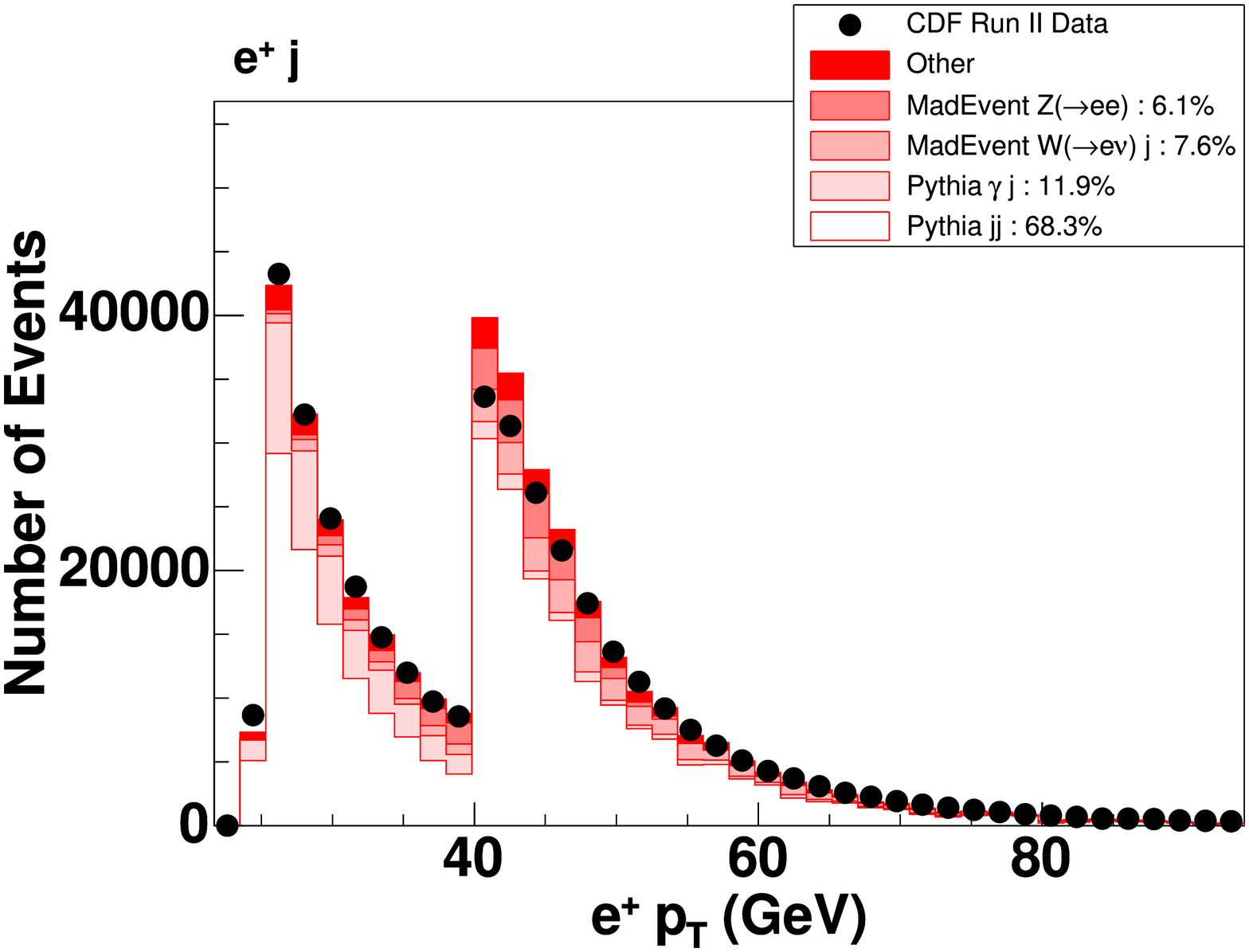} & \ \ \ \ \ \ \ & \includegraphics[width=2.5in,angle=270]{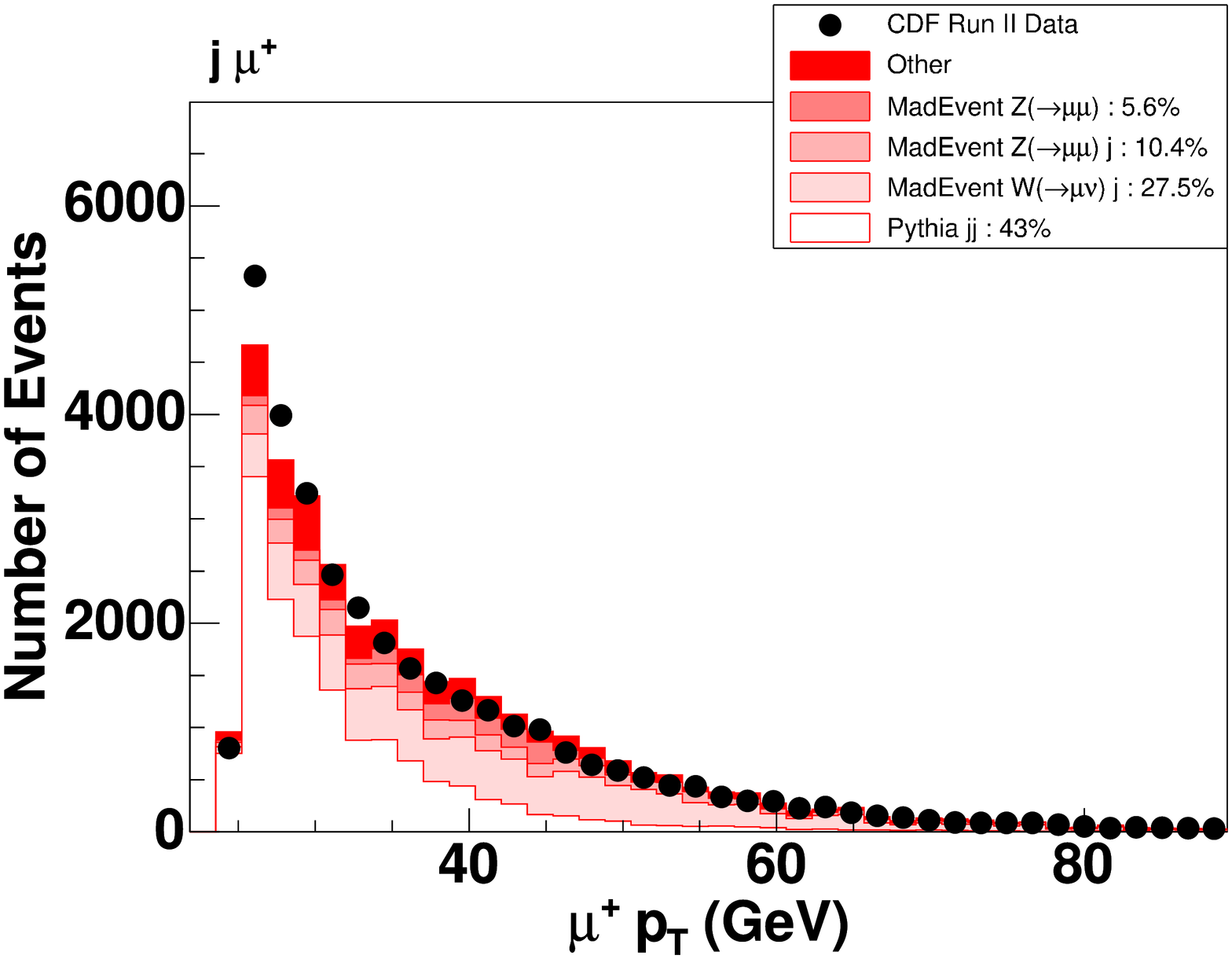} \\
\includegraphics[width=2.5in,angle=270]{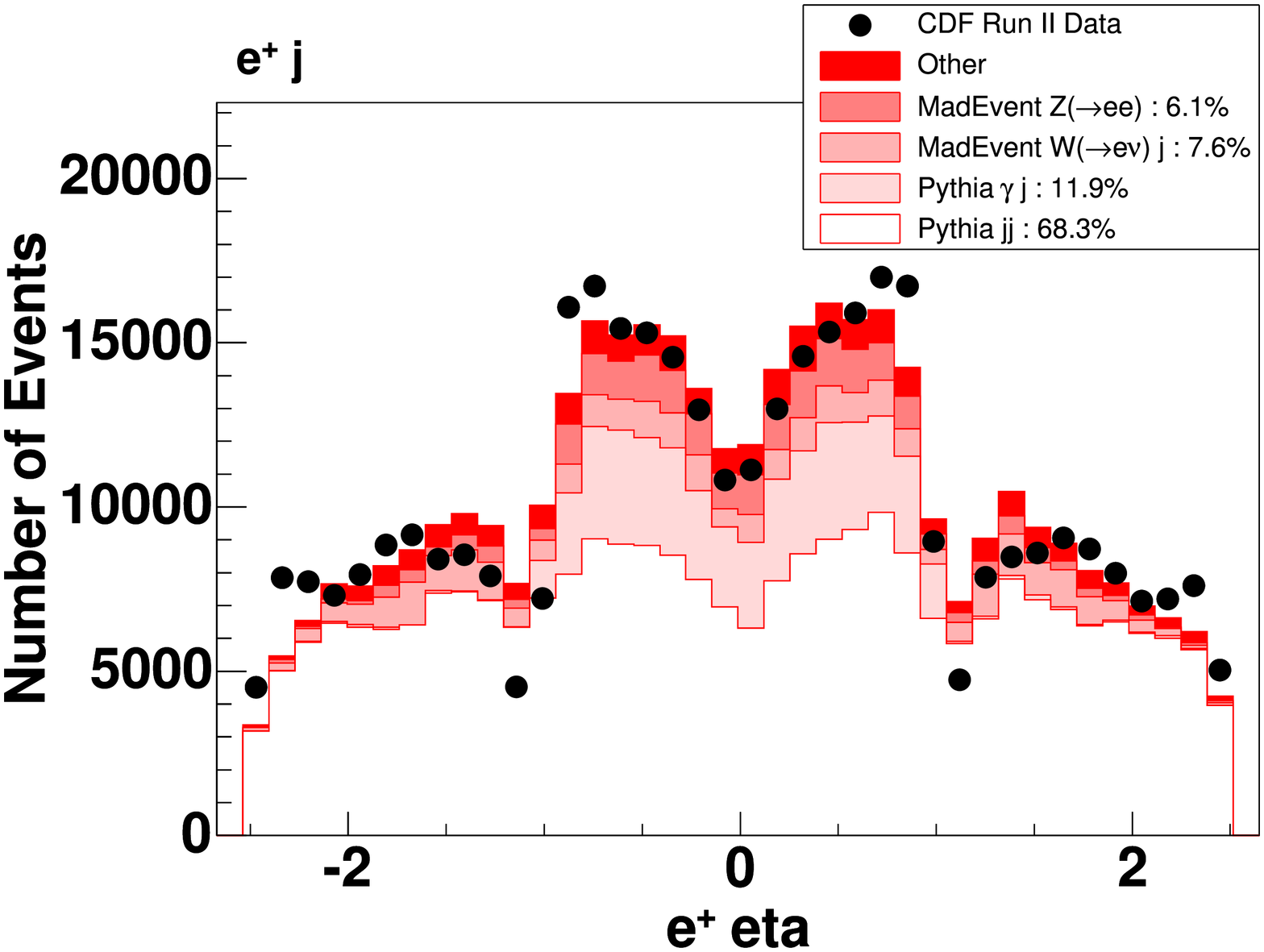} & \ \ \ \ \ \ \ & \includegraphics[width=2.5in,angle=270]{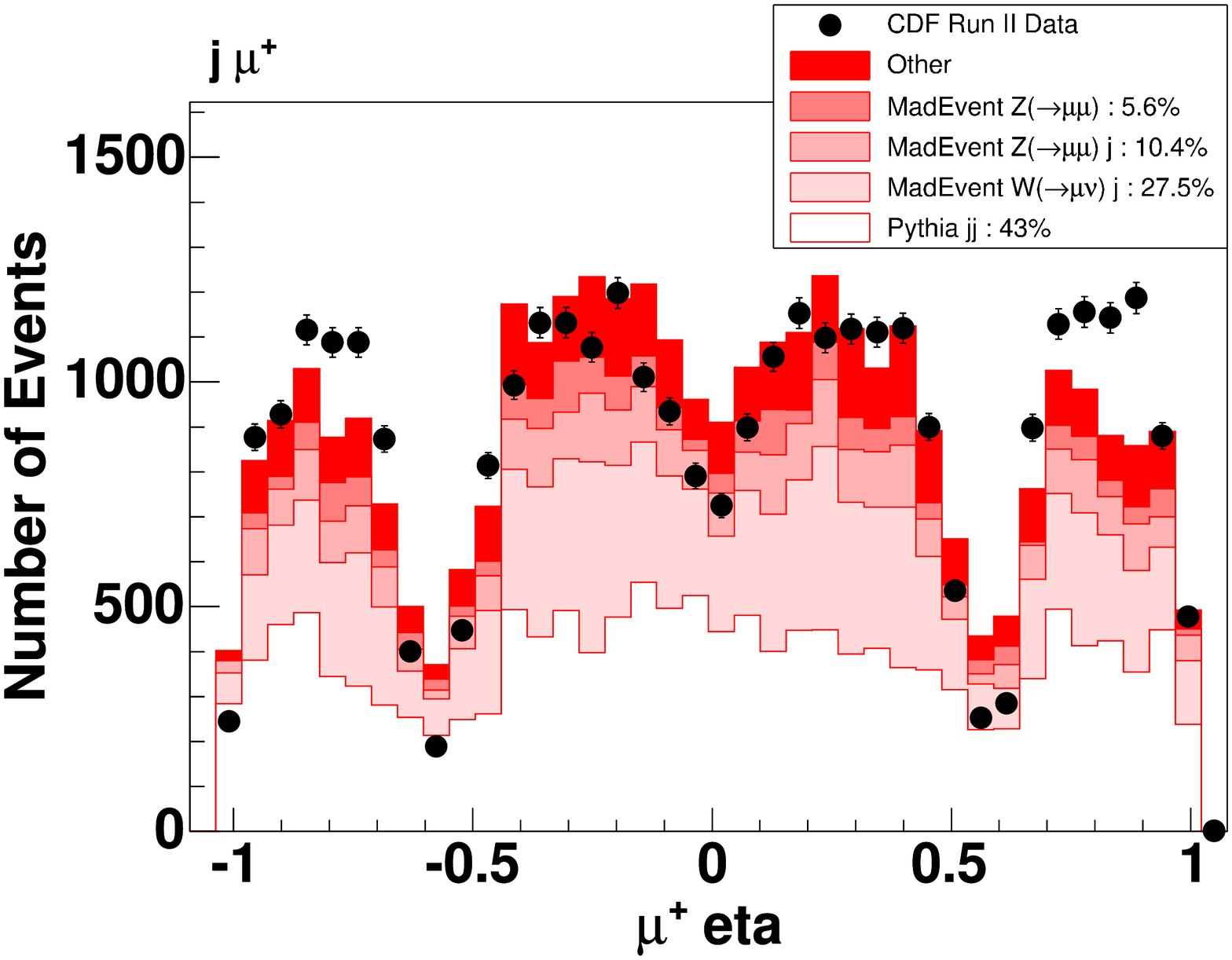} \\
\end{tabular}
\caption{ A few of the most discrepant distributions in the final states $ej$ and $j\mu$, which are greatly affected by the fake rates $\poo{j}{e}$ and $\poo{j}{\mu}$, respectively.  These distributions are among the 13 significantly discrepant distributions identified as resulting from coarseness of the correction model employed.  The vertical axis shows the number of events; the horizontal axes show the transverse momentum and pseudorapidity of the lepton.  Filled (black) circles show CDF data, and the shaded (red) histogram shows the standard model prediction.  Events enter the $ej$ final state either on a central electron trigger with $p_T>25$~GeV, or on a plug electron trigger with $p_T>40$~GeV.  The fake rate $\poo{j}{e}$ is significantly larger in the plug region than in the central region of the CDF detector.  Muons are identified with separate detectors covering the regions $\abs{\eta}<0.6$ and $0.6<\abs{\eta}<1.0$.}
\label{fig:1j1fakeLepton_1}
\end{figure*}

\begin{figure*}
\begin{tabular}{ccc}
\includegraphics[width=2.5in,angle=270]{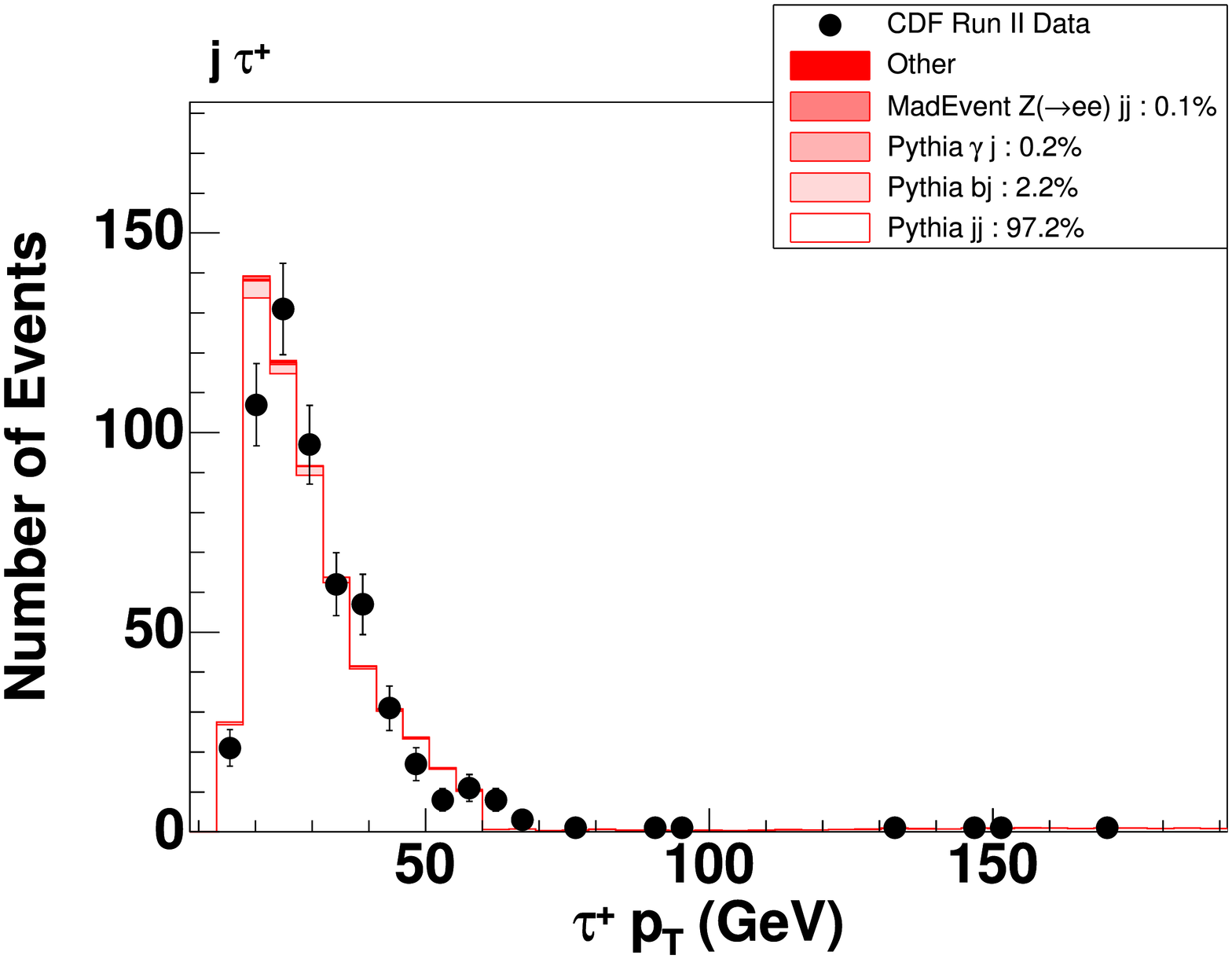} & \ \ \ \ \ \ \ & \includegraphics[width=2.5in,angle=270]{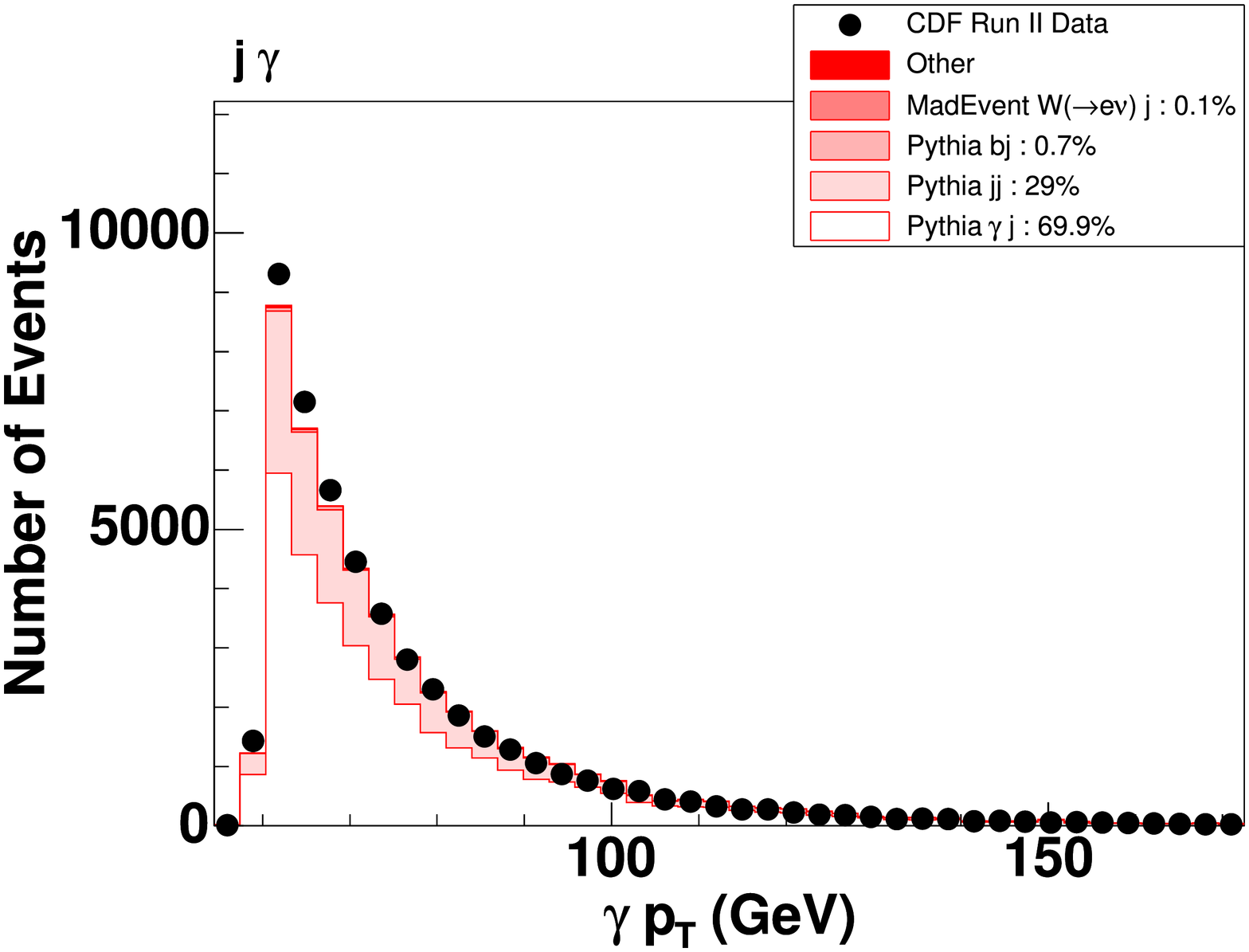} \\
\includegraphics[width=2.5in,angle=270]{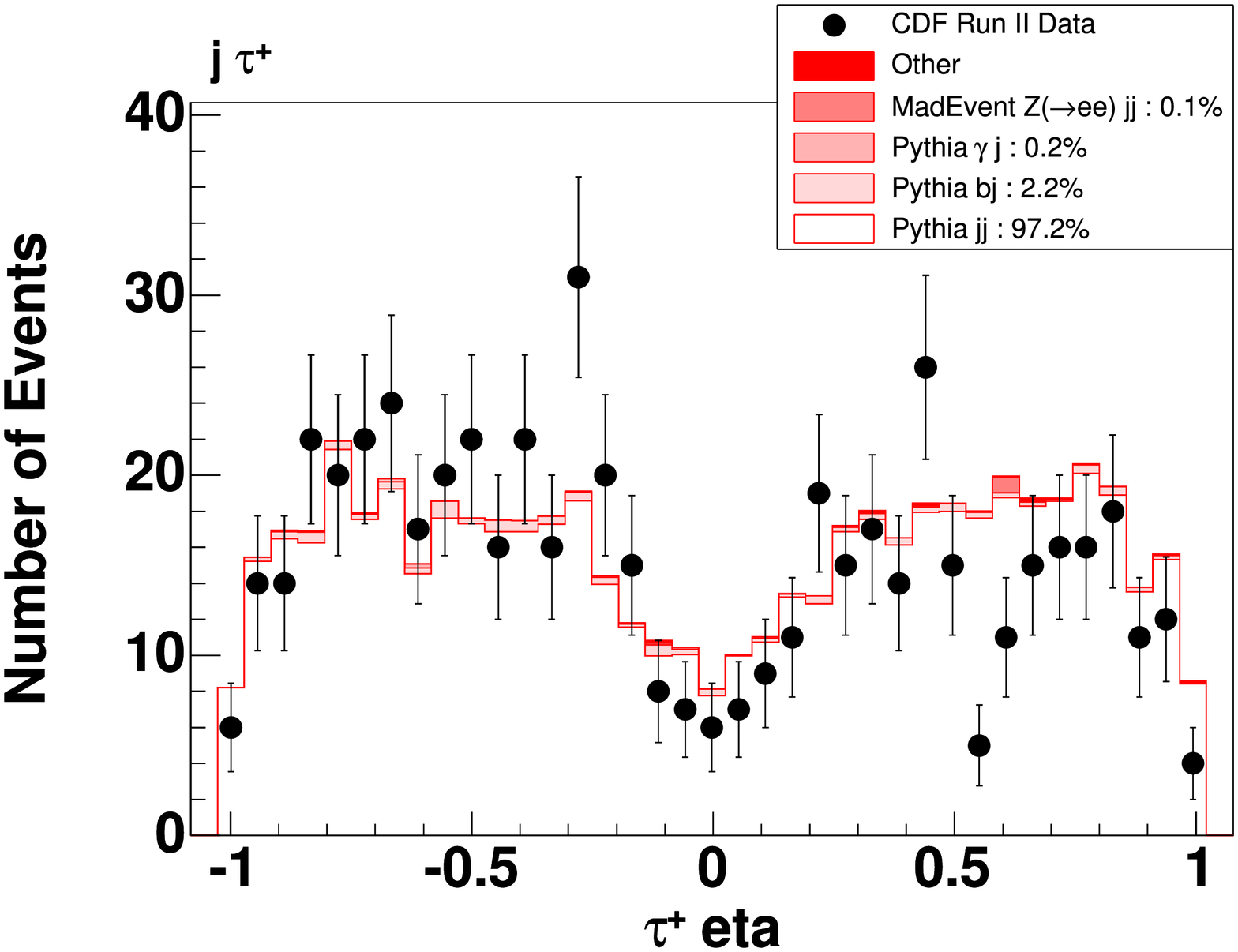} & \ \ \ \ \ \ \ & \includegraphics[width=2.5in,angle=270]{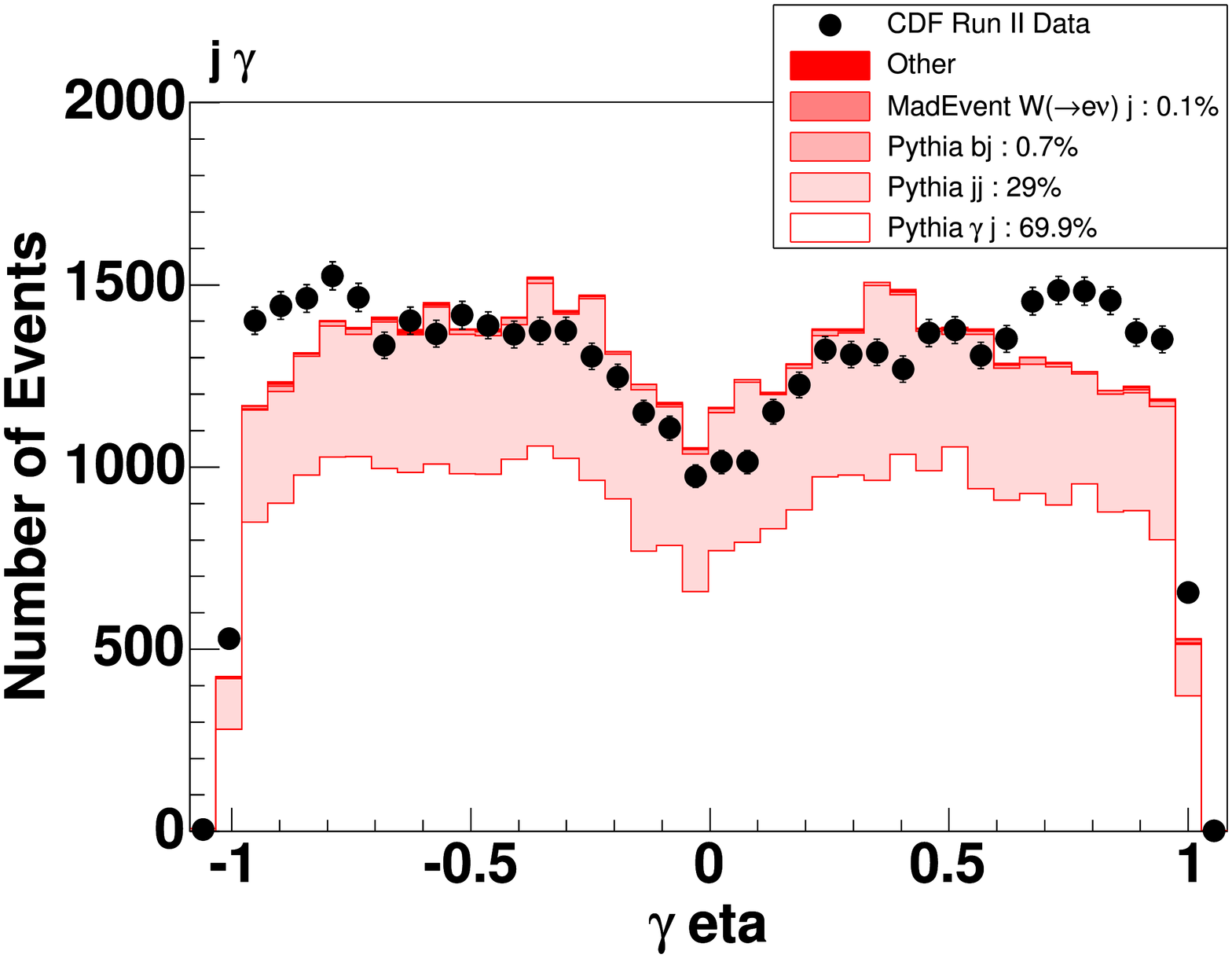}
\end{tabular}
\caption{A few of the most discrepant distributions in the final states $j\tau$ and $j\gamma$, which are greatly affected by the fake rates $\poo{j}{\tau}$ and $\poo{j}{\gamma}$, respectively.  The vertical axis shows the number of events; the horizontal axes show the transverse momentum and pseudorapidity of the tau lepton and photon.  Filled (black) circles show CDF data, and the shaded (red) histogram shows the standard model prediction.  The distributions in the $j\gamma$ final state are among the 13 significantly discrepant distributions identified as resulting from coarseness of the correction model employed.}
\label{fig:1j1fakeLepton_2}
\end{figure*}

Here and below $\poo{q}{X}$ denotes a quark fragmenting to $X$ carrying nearly all of the parent quark's energy, and $\poo{j}{X}$ denotes a parent quark or gluon being misreconstructed in the detector as $X$.

The probability for a light quark jet to be misreconstructed as an $e^+$ can be written
\begin{eqnarray}
            \poo{j}{e^+} = & \poo{q}{\gamma} \, \poo{\gamma}{e^+} + \nonumber \\
                                   & \poo{q}{\pi^0} \, \poo{\pi^0}{e^+} + \nonumber \\
                                   & \poo{q}{\pi^+} \, \poo{\pi^+ }{e^+} + \nonumber \\ 
                                   & \poo{q}{K^+} \, \poo{K^+ }{e^+}. 
\label{eqn:poo_j_e+}
\end{eqnarray}
A similar equation holds for a light quark jet faking an $e^-$.  

The probability for a light quark jet to be misreconstructed as a $\mu^+$ can be written
\begin{eqnarray}
            \poo{j}{\mu^+} = & \poo{q}{\pi^+} \, \poo{\pi^+}{\mu^+} + \nonumber \\
                             & \poo{q}{K^+} \, \poo{K^+}{\mu^+}.
\label{eqn:poo_j_mu+}
\end{eqnarray}
Here $\poo{\pi}{\mu}$ denotes pion decay-in-flight, and $\poo{K}{\mu}$ denotes kaon decay-in-flight; other processes contribute negligibly.  A similar equation holds for a light quark jet faking a $\mu^-$. 

The only non-negligible underlying physical mechanisms for a jet to fake a photon are for the parent quark or gluon to fragment into a photon or a neutral pion, carrying nearly all the energy of the parent quark or gluon.  Thus
\begin{eqnarray}
            \poo{j}{\gamma} = & \poo{q}{\pi^0} \, \poo{\pi^0}{\gamma} + \nonumber \\
                              & \poo{q}{\gamma} \, \poo{\gamma}{\gamma}.
\label{eqn:poo_j_ph}
\end{eqnarray}


Up and down quarks and gluons fragment nearly equally to each species of pion; hence
\begin{eqnarray}
          \frac{1}{3} \, \poo{q}{\pi} & = \poo{q}{\pi^+} & =  \poo{q}{\pi^-} \nonumber \\
                                      & = \poo{q}{\pi^0},
\label{eqn:poo_q_pi}
\end{eqnarray}
where $\poo{q}{\pi}$ denotes fragmentation into any pion carrying nearly all of the parent quark's energy.
Fragmentation into each type of kaon also occurs with equal probability; hence
\begin{eqnarray}
     \frac{1}{4} \poo{q}{K} & = \poo{q}{K^+} = \poo{q}{K^-}       \nonumber \\
                            & =    \poo{q}{K^0} = \poo{q}{\bar{K^0}},
\label{eqn:poo_q_K}
\end{eqnarray}
where $\poo{q}{K}$ denotes fragmentation into any kaon carrying nearly all of the parent quark's energy.

\Pythia\ contains a parameter that sets the number of string fragmentation kaons relative to the number of fragmentation pions.  The default value of this parameter, which has been tuned to LEP I data, is 0.3; for every 1 up quark and every 1 down quark, 0.3 strange quarks are produced.  Strange particles are produced perturbatively in the hard interaction itself, and in perturbative radiation, at a ratio larger than 0.3:1:1.  This leads to the inequality
\begin{eqnarray}
   0.3 \lesssim \frac{\poo{q}{K}}{\poo{q}{\pi}} < 1,
\label{eqn:poo_q_Kpi}
\end{eqnarray}
where $\poo{q}{K}$ and $\poo{q}{\pi}$ are as defined above.

The probability for a jet to be misreconstructed as a tau lepton can be written
\begin{equation}
    \poo{j}{\tau^+} = \poo{j}{\tau^+_1} + \poo{j}{\tau^+_3},
\label{eqn:poo_j_tau13}
\end{equation}
where $\poo{j}{\tau^+_1}$ denotes the probability for a jet to fake a 1-prong tau, and $\poo{j}{\tau^+_3}$ denotes the probability for a jet to fake a 3-prong tau.  For 1-prong taus,
\begin{eqnarray}
    \poo{j}{\tau^+_1} = & \poo{q}{\pi^+} \, \poo{\pi^+}{\tau^+} + \nonumber \\
                        & \poo{q}{K^+} \, \poo{K^+}{\tau^+}.
\label{eqn:poo_j_tau1}
\end{eqnarray}
Similar equations hold for negatively charged taus.

\begin{figure}
\hspace*{-0.12in}\includegraphics[width=2.75in,angle=270]{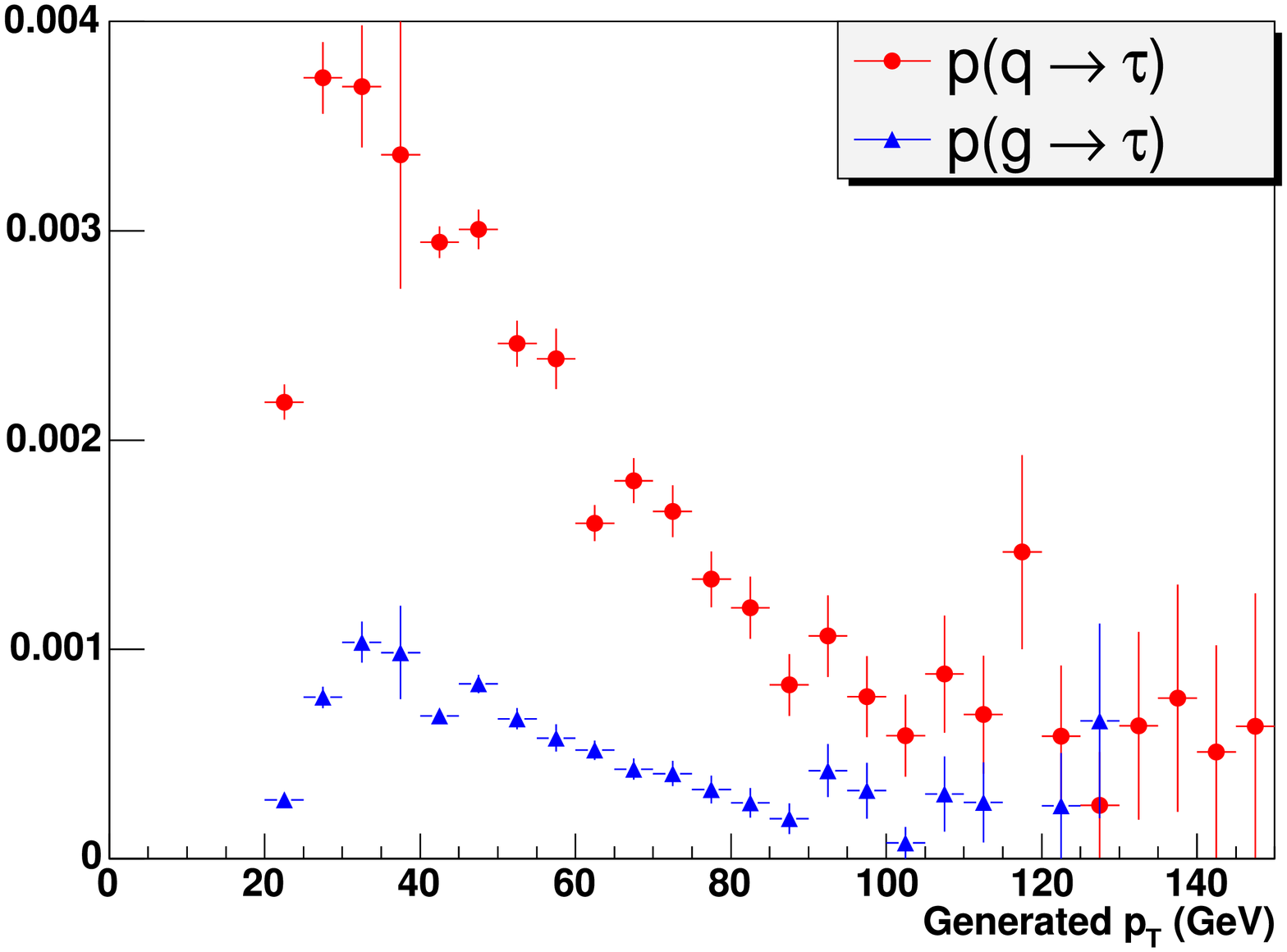}
\caption{The probability for a generated parton to be misreconstructed as a one-prong $\tau$, as a function of the parton's generated $p_T$.  Filled (red) circles show the probability for a jet arising from a parent quark to be misreconstructed as a one-prong tau.  Filled (blue) triangles show the probability for a jet arising from a parent gluon to be misreconstructed as a one-prong tau.}
\label{fig:j2tauProbabilities}
\end{figure}

Figure~\ref{fig:j2tauProbabilities} shows the probability for a quark (or gluon) to fake a one-prong tau, as a function of transverse momentum.  Using fragmentation functions tuned on LEP\,1 data, \Pythia\ predicts the probability for a quark jet to fake a one-prong tau to be roughly four times the probability for a gluon jet to fake a one-prong tau.  This difference in fragmentation is incorporated into \Vista's treatment of jets faking electrons, muons, taus, and photons.  The \Vista\ correction model includes such correction factors as the probability for a jet with a parent quark to fake an electron ({\tt 0033} and {\tt 0034}) and the probability for a jet with a parent quark to fake a muon ({\tt 0035}); the probability for a jet with a parent gluon to fake an electron or muon is then obtained by dividing the values of these fitted correction factors by four.

\cdfSpecific{
This effect is investigated using fake one-prong taus reconstructed in \Pythia\ dijet samples\cdfSpecific{ ({\tt Pythia\_jj\_018}, {\tt Pythia\_jj\_040} and {\tt Pythia\_jj\_060}, corresponding to lower cuts on $\hat{p}_T$ of 18, 40 and 60~GeV, respectively)}.

Figure~\ref{fig:j2tauPt} shows that the reconstructed fake tau has about $75\pm18\%$ of the $p_T$ of the prominent generated particle, defined to be the generated particle carrying the greatest $p_T$ and being within a cone of $\Delta R < 0.4$ centered on the reconstructed tau.  The $p_T$ of the misreconstructed tau is on average more undermeasured if the generated parton is a gluon than if it is a quark.  This reduction in the $p_T$ of the fake tau is implemented in \Vista\ when a jet is made to fake a $\tau$ during the misreconstruction process.

Figure~\ref{fig:j2tauNeutrals} shows the remaining generated $p_T$ to be carried by neutral particles:  mostly $\pi^0$'s, followed by $K^0_L$'s and $\eta$'s decaying to photons or to three neutral pions.  The $p_T$ of the fake tau is determined by the track and reconstructed $\pi^0$'s.

\begin{figure}
\includegraphics[width=3.0in]{pTofFakeTauOverPtOfGeneratedParton}
\caption{Distribution of the $p_T$ of the fake $\tau$ over the $p_T$ of the prominent generated particle (pgp), which is defined as the generated particle within $\Delta R < 0.4$ from the reconstructed $\tau$ with the greatest $p_T$. The pgp is almost always a quark or a gluon, and more likely to be a quark by a factor of four.}
\label{fig:j2tauPt}
\end{figure}

\begin{figure}
\begin{tabular}{c}
\includegraphics[width=3.0in]{PNGPoverPGP}\\
\includegraphics[width=3.0in]{fakeTauDifferenceIsInNeutrals} \\
\end{tabular}
\caption{Upper: The distribution of the $p_T$ of the prominent neutral generated particle (pngp), which is the neutral generated particle with the greatest $p_T$ within a cone of $\Delta R < 0.4$ from the fake one-prong $\tau$, divided by the $p_T$ of the prominent generated particle (pgp), which happens to be either a quark or a gluon. Lower: $p_T$ of the pngp plus the $p_T$ of the reconstructed $\tau$, divided by the $p_T$ of the pgp. The fact that this distribution peaks around 1 shows that the generated $p_T$ that is missing from the fake $\tau$ was carried by the pngp. Most of the times the pngp is a $\pi^0$.}
\label{fig:j2tauNeutrals}
\end{figure}
}

The physical mechanism underlying the process whereby an incident photon or neutral pion is misreconstructed as an electron is a conversion in the material serving as the support structure of the silicon vertex detector.  This process produces exactly as many $e^+$ as $e^-$, leading to 
\begin{eqnarray}
            \frac{1}{2} \, \poo{\gamma}{e} = \poo{\gamma}{e^+} = \poo{\gamma}{e^-} \nonumber \\
            \frac{1}{2} \, \poo{\pi^0}{e}  = \poo{\pi^0}{e^+}  = \poo{\pi^0}{e^-},
\label{eqn:poo_pi0ph_e}
\end{eqnarray}
where $e$ is an electron or positron.
 
From Fig.~\ref{fig:misId_cdfSim_central_pt}, the average $p_T$ of electrons reconstructed from 25~GeV incident photons is $23.9\pm1.4$~GeV.  The average $p_T$ of electrons reconstructed from incident 25~GeV neutral pions is $23.7\pm1.3$~GeV.

 The charge asymmetry between $\poo{K^+}{e^+}$ and $\poo{K^-}{e^-}$ in Table~\ref{tbl:misId_cdfSim_central} arises because $K^-$ can capture on a nucleon, producing a single hyperon.  Conservation of baryon number and strangeness prevents $K^+$ from capturing on a nucleon, reducing the $K^+$ cross section relative to the $K^-$ cross section by roughly a factor of two.

The physical process primarily responsible for $\pi^\pm \rightarrow e^\pm$ is inelastic charge exchange
\begin{eqnarray}
            \pi^- p  \rightarrow  \pi^0 n  \nonumber \\
            \pi^+ n  \rightarrow  \pi^0 p
\label{eqn:chargeExchangeProcesses}
\end{eqnarray}
occurring within the electromagnetic calorimeter.  The charged pion leaves the ``electron's'' track in the CDF tracking chamber, and the $\pi^0$ produces the ``electron's'' electromagnetic shower.  No true electron appears at all in this process, except as secondaries in the electromagnetic shower originating from the $\pi^0$.

The average $p_T$ of reconstructed ``electrons'' originating from a single charged pion is $18.8\pm2.2$~GeV, indicating that the misreconstructed ``electron'' in this case is measured to have on average only 75\% of the total energy of the parent quark or gluon.  This is expected, since the recoiling nucleon from the charge exchange process carries some of the incident pion's momentum.

 An additional small loss in energy for a jet misreconstructed as an electron, photon, or muon is expected since the leading $\pi^+$, $K^+$, $\pi^0$, or $\gamma$ takes only some fraction of the parent quark's energy.

The cross sections for $\pi^- p  \rightarrow  \pi^0 n$ and $\pi^+ n  \rightarrow  \pi^0 p$, proceeding through the isospin $I$ conserving and $I_3$ independent strong interaction, are roughly equal.  The corresponding particles in the two reactions are related by interchanging the signs of their $z$-components of isospin.  
\cdfSpecific{
\begin{eqnarray}
  \sqrt{1/3} \, \ket{3/2, -1/2 } - \sqrt{2/3} \, \ket{ 1/2, -1/2 }  \nonumber \\
  \rightarrow   \sqrt{2/3} \, \ket{ 3/2, -1/2 } + \sqrt{1/3} \, \ket{ 1/2, -1/2 },
\end{eqnarray}
the cross section is proportional to ${2/9} \abs{M_3 - M_1}^2$.  Kets $\ket{I,I_3}$ are in isospin space; $M_3$ denotes $\bra{3/2} H \ket{3/2}$ for the Hamiltonian $H$ responsible for the strong interaction, which is assumed to conserve total isospin $I$ and to be independent of $I_3$; and $M_1$ similarly denotes $\bra{1/2} H \ket{1/2}$.
  For $\pi^+ n \rightarrow  \pi^0 p$,
\begin{eqnarray}
  \sqrt{1/3} \, \ket{ 3/2, 1/2 } + \sqrt{2/3} \, \ket{ 1/2, 1/2 }  \nonumber  \\
   \rightarrow   \sqrt{2/3} \, \ket{ 3/2, 1/2 } - \sqrt{1/3} \, \ket{ 1/2, 1/2 },
\end{eqnarray}
the cross section is again proportional to ${2/9} \abs{M_3 - M_1}^2$.  
}

The probability for a 25 GeV $\pi^+$ to decay to a $\mu^+$ can be written
\begin{eqnarray}
            \poo{\pi^+}{\mu^+} = & p(\text{decays within tracker}) +    \nonumber \\
                                & p(\text{decays within calorimeter}).
\label{eqn:poo_pi_mu_decayInFlight}
\end{eqnarray}
The probability for the pion to decay within the tracking volume is
\begin{equation}
            p(\text{decays within tracker}) = 1-e^{-R_{\text{tracker}}/\gamma (c \tau)},
\label{eqn:poo_pi_mu_decayInFlight_cot}
\end{equation}
where $\gamma=25$~GeV~/~140~MeV~$=180$ is the pion's Lorentz boost, the proper decay length of the charged pion is $(c \tau) = 7.8$~meters, and the radius of the CDF tracking volume is $R_{\text{tracker}} = 1.5$~meters, giving $p(\text{decays within tracker}) = 0.001$.  The probability for the pion to decay within the calorimeter volume is
\begin{equation}
            p(\text{decays within calorimeter}) \approx \lambda_I / \gamma (c \tau),
\label{eqn:poo_pi_mu_decayInFlight_calorimeter}
\end{equation}
where $\lambda_I \approx 0.4$~meters is the nuclear interaction length for charged pions on lead or iron and the path length through the calorimeter is $L_{\text{cal}} \approx 2$~meters, leading to $p(\text{decays within calorimeter}) \approx 0.00025$.  Summing the contributions from decay within the tracking volume and decay within the calorimeter volume, $\poo{\pi^+ }{ \mu^+} \approx 0.00125$.

The primary physical mechanism by which a jet fakes a photon is for the parent quark or gluon to fragment into a leading $\pi^0$ carrying nearly all the momentum.  The highly boosted $\pi^0$ decays within the beam pipe to two photons that are sufficiently collinear to appear in the preshower, electromagnetic calorimeter, and shower maximum detector as a single photon.  Thus
\begin{equation}
\poo{j}{\gamma} = \poo{q}{\pi^0} \poo{\pi^0}{\gamma}.
\end{equation}
An immediate corollary is that the misreconstructed ``photon'' carries the energy of the parent quark or gluon, and is well measured.

 Typical jets are measured with poorer energy resolution than jets that have faked electrons, muons, or photons.

 Since $\poo{q}{\pi^0} \gg \poo{q}{\gamma}$, it follows from Eq.~\ref{eqn:poo_q_pi} and Table~\ref{tbl:misId_cdfSim_central} that the conversion contribution to $\poo{j}{e}$ is $\approx 75\%$, and the charge exchange contribution is $\approx 25\%$:
\begin{eqnarray}
\frac{0.75}{0.25} =  ( & \poo{q}{\gamma} \, \poo{\gamma}{e^+} + & \, \nonumber \\
                       & \poo{q}{\pi^0} \, \poo{\pi^0}{e^+}     & ) \, / \nonumber \\
                     ( & \poo{q}{\pi^+} \, \poo{\pi^+ }{e^+} +  & \, \nonumber \\
                       & \poo{q}{K^+} \, \poo{K^+ }{e^+}        & ). 
\label{eqn:poo_conversion_e}
\end{eqnarray}
\cdfSpecific{This is consistent with the conclusion obtained in Ref.~\cite{BackgroundToElectronsCdfNote}.}

The number of $e^+\,j$ events in data is 0.9 times the number of $e^-\,j$ events.  This charge asymmetry arises from $\poo{K^+}{e^+}$ and $\poo{K^-}{e^-}$ in Table~\ref{tbl:misId_cdfSim_central}.  Quantitatively,
\begin{eqnarray}
    \frac{\poo{j}{e^+}}{\poo{j}{e^-}} = \frac{0.9+ 0.2 \, \poo{K^+}{e^+} / \poo{K}{e}}{0.9+0.2 \, \poo{K^-}{e^-} / \poo{K}{e}},
\label{eqn:poo_j_e_chargeAsymmetry}
\end{eqnarray}
where 0.9 is the sum of 0.75 from Eq.~\ref{eqn:poo_conversion_e} and $0.15\approx0.25\times 0.6$ from Eq.~\ref{eqn:poo_q_Kpi}, and 0.2 is twice $1-0.9$.  From $\poo{K^+}{e^+}$ and $\poo{K^-}{e^-}$ in Table~\ref{tbl:misId_cdfSim_central}, $\poo{K^+}{e^+} / \poo{K}{e} = 1/3$ and $\poo{K^-}{e^-} / \poo{K}{e} = 2/3$, predicting $\poo{j}{e^+}/\poo{j}{e^-}=0.935$, in reasonable agreement with the ratio of the observed number of events in the $e^+\,j$ and $e^-\,j$ final states.

The number of $j\,\mu^+$ events observed in CDF Run II is 1.1 times the number of $j\,\mu^-$ events observed.  This charge asymmetry arises from $\poo{K^+}{\mu^+}$ and $\poo{K^-}{\mu^-}$ in Table~\ref{tbl:misId_cdfSim_central}.

The physical mechanism by which a prompt photon fakes a tau lepton is for the photon to convert, producing an electron or positron carrying most of the photon's energy, which is then misreconstructed as a tau.  The probability for this to occur is equal for positively and negatively charged taus,
\begin{equation}
       \frac{1}{2} \poo{\gamma}{\tau} = \poo{\gamma}{\tau^+} = \poo{\gamma}{\tau^-},
\end{equation}
and is related to previously defined quantities by 
\begin{equation}
       \poo{\gamma}{\tau} = \poo{\gamma}{e} \, \frac{1}{\poo{e}{e}} \, \poo{e}{\tau},
\end{equation}
where $\poo{\gamma}{e}$ denotes the fraction of produced photons that are reconstructed as electrons, $\poo{e}{e}$ denotes the fraction of produced electrons that are reconstructed as electrons, and hence $\poo{\gamma}{e} / \poo{e}{e}$ is the fraction of produced photons that pair produce a single leading electron.

 Note $\poo{e}{\gamma}\approx\poo{\gamma}{e}$ from Table~\ref{tbl:misId_cdfSim_central}, as expected, with value of $\approx 0.03$ determined by the amount of material in the inner detectors and the tightness of isolation criteria.  A hard bremsstrahlung followed by a conversion is responsible for electrons to be reconstructed with opposite sign; hence
\begin{eqnarray}
            \poo{e^\pm}{e^\mp} =       & \poo{e^+}{e^-} = \poo{e^-}{e^+} \nonumber \\
                               \approx & \frac{1}{2} \, \poo{e^\pm}{\gamma}\poo{\gamma}{e^\mp},
\end{eqnarray}
where the factor of $1/2$ comes because the material already traversed by the $e^\pm$ will not be traversed again by the $\gamma$. In particular, track curvature mismeasurement is not responsible for erroneous sign determination in the central region of the CDF detector.

From knowledge of the underlying physical mechanisms by which jets fake electrons, muons, taus, and photons, the simple use of a reconstructed jet as a lepton or photon with an appropriate fake rate applied to the weight of the event needs slight modification to correctly handle the fact that a jet that has faked a lepton or photon generally is measured more accurately than a hadronic jet.  Rather than using the momentum of the reconstructed jet, the momentum of the parent quark or gluon is determined by adding up all Monte Carlo particle level objects within a cone of $\Delta R = 0.4$ about the reconstructed jet.  In misreconstructing a jet in an event, the momentum of the corresponding parent quark or gluon is used rather than the momentum of the reconstructed jet.  A jet that fakes a photon then has momentum equal to the momentum of the parent quark or gluon plus a fractional correction equal to $0.01\times(\text{parent}\,p_T - 25~\text{GeV})/(25~\text{GeV})$ to account for leakage out of the cone of $\Delta R = 0.4$, and a further smearing of $0.2~\sqrt{\text{GeV}} \times \sqrt{\text{parent}\,p_T}$, reflecting the electromagnetic resolution of the CDF detector.  The momenta of jets that fake photons are multiplied by an overall factor of 1.12, and jets that fake electrons, muons, or taus are multiplied by an overall factor of $0.95$.  These numbers are determined by the $\ell\pmiss$, $\ell j$, and $\gamma j$ final states.  The distributions most sensitive to these numbers are the missing energy and the jet $p_T$.

A $b$ quark fragmenting into a leading $b$ hadron that then decays leptonically or semileptonically results in an electron or muon that shares the $p_T$ of the parent $b$ quark with the associated neutrino.  If all hadronic decay products are soft, the distribution of the momentum fraction carried by the charged lepton can be obtained by considering the decay of a scalar to two massless fermions.  Isolated and energetic electrons and muons arising from parent $b$ quarks in this way are modeled as having $p_T$ equal to the parent $b$ quark $p_T$, multiplied by a random number uniformly distributed between 0 and 1.  

\subsection{Additional background sources}
\label{sec:AdditionalBackgroundSources}

This appendix provides additional details on the estimation of the standard model prediction.

\subsubsection{Cosmic ray and beam halo muons}
\label{sec:CorrectionModelDetails:CosmicRays}

\begin{figure*}
\begin{tabular}{ccc}
\includegraphics[width=2.5in,angle=270]{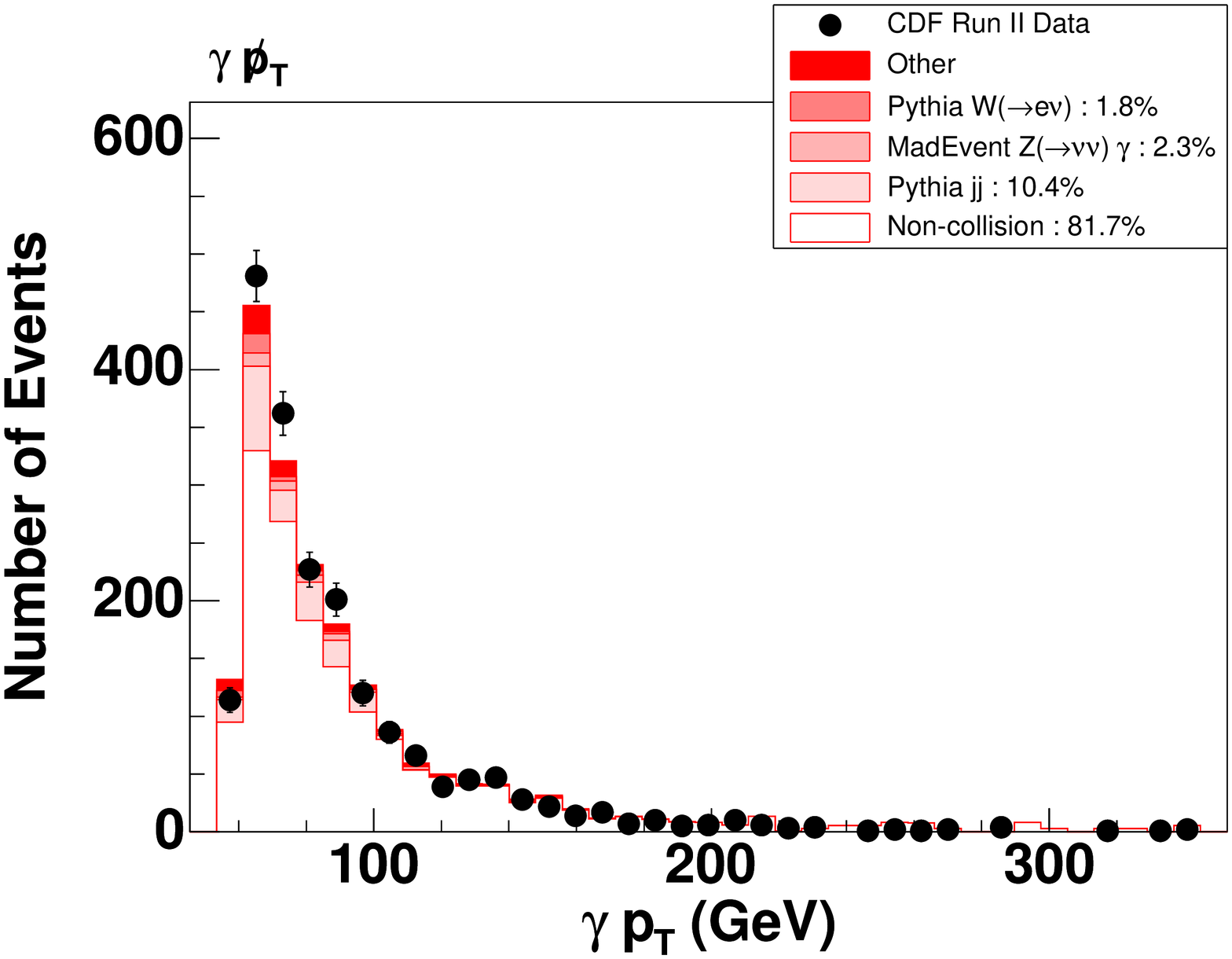} & \ \ \ \ \ \ \ & \includegraphics[width=2.5in,angle=270]{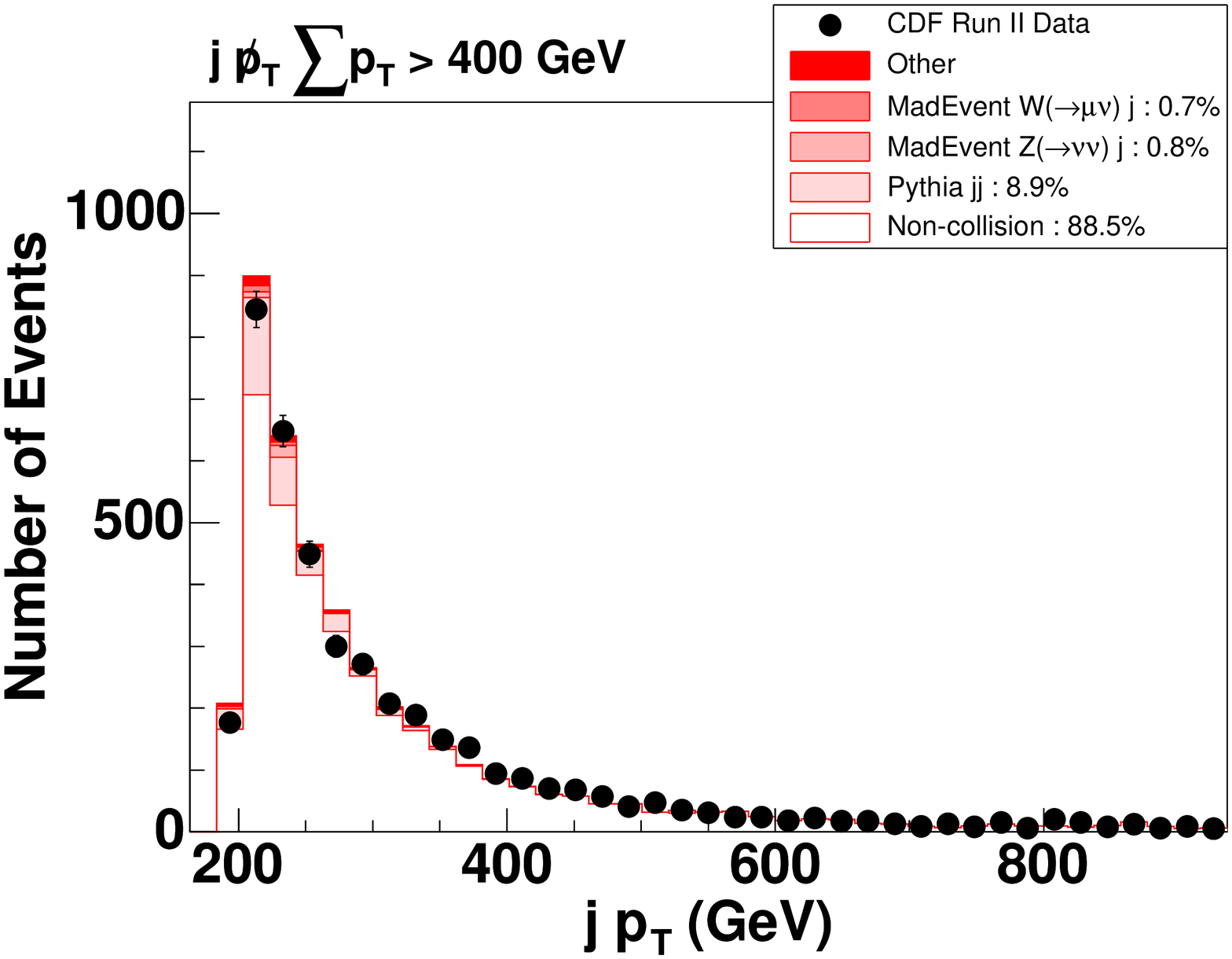} \\
\includegraphics[width=2.5in,angle=270]{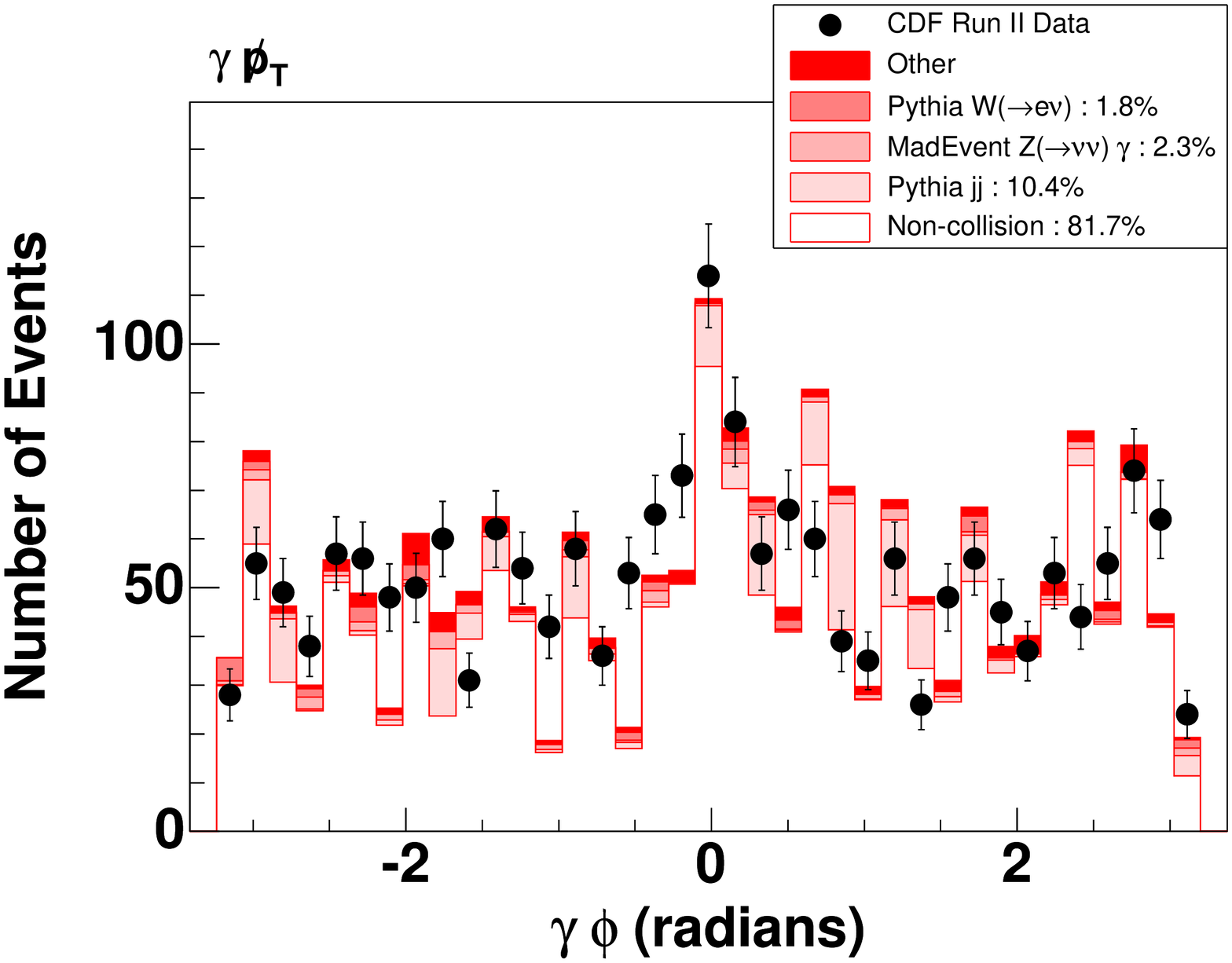} & \ \ \ \ \ \ \ & \includegraphics[width=2.5in,angle=270]{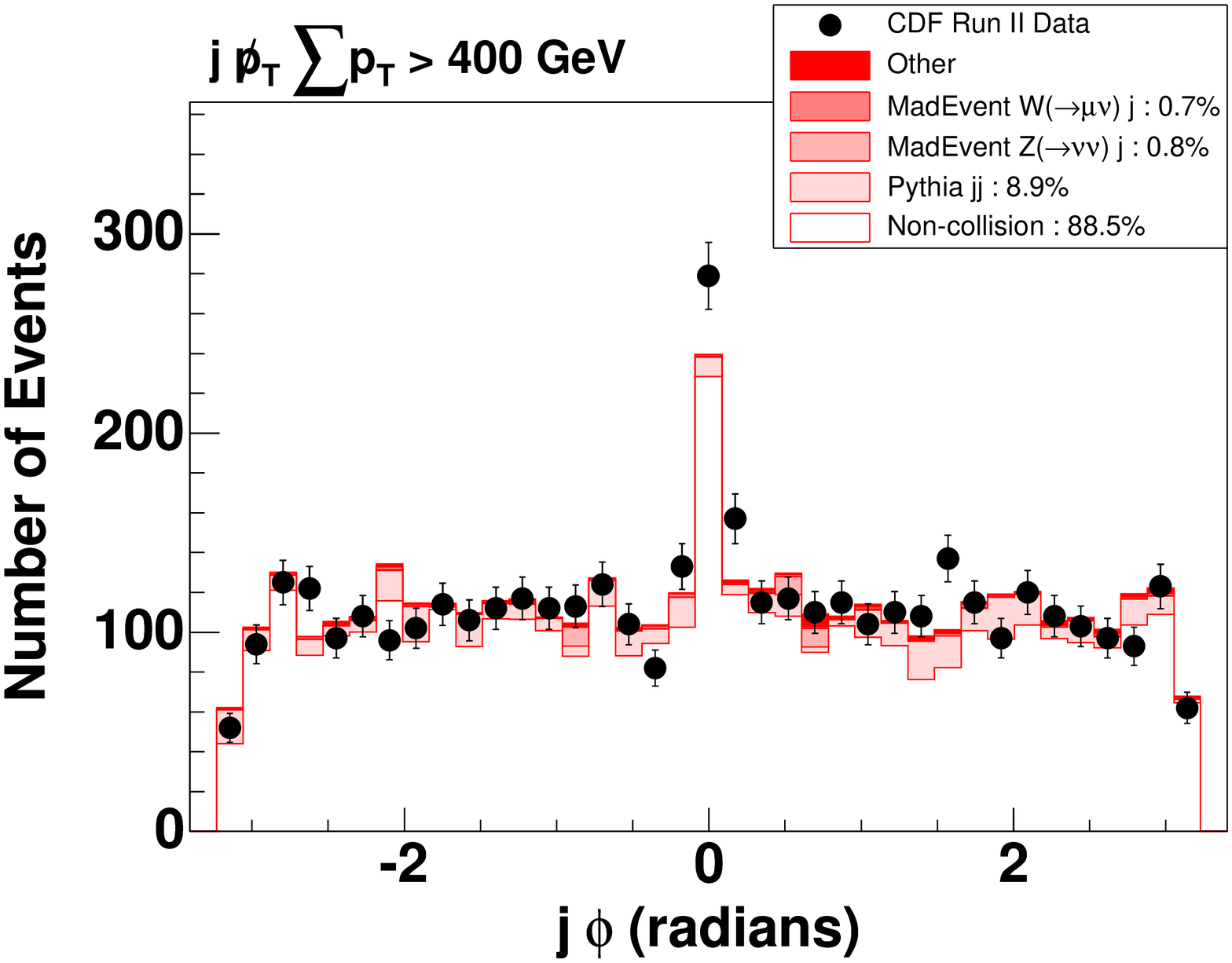} \\
\end{tabular}
\caption{The distribution of transverse momentum and azimuthal angle for photons and jets in the $\gamma\pmiss$ and $j\pmiss$ final states, dominated by cosmic ray and beam halo muons.  The vertical axis shows the number of events in each bin.  Data are shown as filled (black) circles; the standard model prediction is shown as the shaded (red) histogram.  Here the ``standard model'' prediction includes contributions from cosmic ray and beam halo muons, estimated using events containing fewer than three reconstructed tracks.  The contribution from cosmic ray muons is flat in $\phi$, while the contribution from beam halo is localized to $\phi=0$.  The only degrees of freedom for the background to these final states are the {\tt{cosmic}}~$\gamma$ and {\tt{cosmic}}~$j$ correction factors, whose values are determined from the global \Vista\ fit and provided in Table~\ref{tbl:CorrectionFactorDescriptionValuesSigmas}.}
\label{fig:1ph1j}
\end{figure*}

There are four dominant categories of events caused by cosmic ray muons penetrating the detector: $\mu\pmiss$, $\mu^+\mu^-$, $\gamma\pmiss$, and $j\pmiss$.  There is negligible contribution from cosmic ray secondaries of any particle type other than muons.

A cosmic ray muon penetrating the CDF detector whose trajectory passes within 1~mm of the beam line and within $-60<z<60$~cm of the origin may be reconstructed as two outgoing muons.  In this case the cosmic ray event is partitioned into the final state $\mu^+\mu^-$.  If one of the tracks is missed, the cosmic ray event is partitioned into the final state $\mu\pmiss$.  The standard CDF cosmic ray filter, which makes use of drift time information in the central tracking chamber, is used to reduce these two categories of cosmic ray events.

CDF data events with exactly one track (corresponding to one muon) and events with exactly two tracks (corresponding to two muons) are used to estimate the cosmic ray muon contribution to the final states $\mu\pmiss$ and $\mu^+\mu^-$ after the cosmic ray filter.  This sample of events is used as the standard model background process {\tt{cosmic}}~$\mu$.  The {\tt{cosmic}}~$\mu$ sample does not contribute to the events passing the analysis offline trigger, whose cleanup cuts require the presence of three or more tracks.  Roughly 100 events are expected from cosmic ray muons in the categories $\mu^+\pmiss$ and $\mu^+\mu^-$.  The sample {\tt{cosmic}}~$\mu$ is neglected from the background estimate, since there is no discrepancy that demands its inclusion.

The remaining two categories are $\gamma\pmiss$ and $j\pmiss$, resulting from a cosmic ray muon that penetrates the CDF electromagnetic or hadronic calorimeter and undergoes a hard bremsstrahlung in one calorimeter cell.  Such an interaction can mimic a single photon or a single jet, respectively.  The reconstruction algorithm infers the presence of significant missing energy balancing the ``photon'' or ``jet.''  If this cosmic ray interaction occurs during a bunch crossing in which there is a $p\bar{p}$ interaction producing three or more tracks, the event will be partitioned into the final state $\gamma\pmiss$ or $j\pmiss$.

CDF data events with fewer than three tracks are used to estimate the cosmic ray muon contribution to the final states $\gamma\pmiss$ and $j\pmiss$.  These samples of events are used as standard model background processes {\tt{cosmic}}~$\gamma$ and {\tt{cosmic}}~$j$ for the modeling of this background, corresponding to offline triggers requiring a photon with $p_T>60$~GeV, or a jet with $p_T>40$~GeV (prescaled) or $p_T>200$~GeV (unprescaled), respectively.  These samples do not contribute to the events passing the analysis offline trigger, whose cleanup cuts require three or more tracks.  The contribution of these events is adjusted with correction factors that are listed as {\tt{cosmic}}~$\gamma$ and {\tt{cosmic}}~$j$ ``$k$-factors'' in Table~\ref{tbl:CorrectionFactorDescriptionValuesSigmas}, but which are more properly understood as reflecting the number of bunch crossings with zero $p\bar{p}$ interactions (resulting in zero reconstructed tracks) relative to the number of bunch crossings with one or more interactions (resulting in three or more reconstructed tracks).  Since the number of bunch crossings with no inelastic $p\bar{p}$ interactions is used to determine the CDF instantaneous luminosity, these cosmic correction factors can be viewed as containing direct information about the luminosity-averaged instantaneous luminosity.

The cosmic ray muon contribution to the final states $\gamma\pmiss$ and $j\pmiss$ is uniform as a function of the CDF azimuthal angle $\phi$.  Consider the CDF detector to be a thick cylindrical shell, and consider two arbitrary infinitesimal volume elements at different locations in the material of the shell.  Since the two volume elements have similar overburdens, the number of cosmic ray muons with $E\gtrsim20$~GeV penetrating the first volume element is very nearly the same as the number of cosmic ray muons with $E\gtrsim20$~GeV penetrating the second volume element.  Since the material of the CDF calorimeters is uniform as a function of CDF azimuthal angle $\phi$, it follows that the cosmic ray muon contribution to the final states $\gamma\pmiss$ and $j\pmiss$ should also be uniform as a function of $\phi$.  In particular, it is noted that the $\phi$ dependence of this contribution depends solely on the material distribution of CDF calorimeter, which is uniform in $\phi$, and has no dependence on the distribution of the horizon angle of the muons from cosmic rays streaking through the atmosphere.

The final states $\gamma\pmiss$ and $j\pmiss$ are also populated by beam halo muons, traveling horizontally through the CDF detector in time with a bunch.  A beam halo muon can undergo a hard bremsstrahlung in the electromagnetic or hadronic calorimeters, producing an energy deposit that can be reconstructed as a photon or jet, respectively.  These beam halo muons tend to lie in the plane and outside of the Tevatron ring, thus horizontally penetrating the CDF detector along $\hat{z}$ at $y=0$, $x>0$, and hence $\phi=0$.  

Figure~\ref{fig:1ph1j} shows the $\gamma\pmiss$ and $j\pmiss$ final states, in which events come primarily from cosmic ray and beam halo muons.

\subsubsection{Multiple interactions}
\label{sec:Overlaps}

In order to estimate event overlaps, consider an interesting event observed in final state C, which looks like an overlap of two events in the final states A and B.  An example is C={\tt e+e-4j}, A={\tt e+e-} and B={\tt 4j}.  It is desired to estimate how many C events are expected from the overlap of A and B events, given the observed frequencies of A and B.

Let ${\cal L}(t)$ be the instantaneous luminosity as a function of time $t$; let 
\begin{equation}
L = \int_{\text{RunII}}{{\cal L}(t) dt} = \VistaApproximateDefiniteLuminosity~\text{pb}^{-1}
\end{equation}
denote the total integrated luminosity; and let
\begin{equation}
\bar{{\cal L}} = \frac{\int_{\text{RunII}}{{\cal L}(t){\cal L}(t) dt}}{\int_{\text{RunII}}{{\cal L}(t) dt}} \approx 10^{32}~\text{cm}^{-2}\text{s}^{-1}
\end{equation}
be the luminosity-averaged instantaneous luminosity.  Denote by $t_0$ the time interval of 396~ns between successive bunch crossings.  The total number of effective bunch crossings $X$ is then 
\begin{equation}
X = \frac{L}{\bar{{\cal L}} t_0} \approx 2.3\times 10^{13}.
\end{equation}
Letting $A$ and $B$ denote the number of observed events in final states A and B, it follows that the number of events in the final state C expected from overlap of A and B is
\begin{equation}
C=\frac{AB}{X}.
\end{equation}
Overlap events are included in the \Vista\ background estimate, although their contribution is generally negligible.

\subsubsection{Intrinsic $k_T$}
\label{sec:CorrectionModelDetails:IntrinsicKt}

Significant discrepancy is observed in many final states containing two objects {\tt o1} and {\tt o2} in the variables $\Delta\phi${\tt{(o1,o2)}}, {\tt uncl}~$p_T$, and $\pmiss_T$.  These discrepancies are ascribed to the sum of two effects:  (1) an intrinsic Fermi motion of the colliding partons within the proton and anti-proton, and (2) soft radiation along the beam axis.  The sum of these two effects appears to be larger in Nature than predicted by \Pythia\ with the parameter tunes used for the generation of the samples employed in this analysis.  This discrepancy is well known from previous studies at the Tevatron and elsewhere, and affects this analysis similarly to other Tevatron analyses.  

The $W$ and $Z$ electroweak samples used in this analysis have been generated with an adjusted \Pythia\ parameter that increases the intrinsic $k_T$.  For all other generated standard model events, the net effect of the Fermi motion of the colliding partons and the soft non-perturbative radiation is hypothesized to be described by an overall ``effective intrinsic $k_T$,'' and the center of mass of each event is given a transverse kick.  Specifically, for every event of invariant mass $m$ and generated summed transverse momentum \SumPt, a random number $k_T$ is pulled from the probability distribution 
\begin{eqnarray}
p(k_T) \varpropto (k_T < m/5) \times & [\frac{4}{5} g(k_T;\mu=0,\sigma_1) + \nonumber \\
                                     &  \frac{1}{5} g(k_T;\mu=0,\sigma_2)],
\end{eqnarray}
where $(k_T < m/5)$ evaluates to unity if true and zero if false; $g(k_T; \mu,\sigma)$ is a Gaussian function of $k_T$ with center at $\mu$ and width $\sigma$;  $\sigma_1=2.55\,\text{GeV}+0.0085\, \SumPt$ is the width of the core of the double Gaussian; and $\sigma_2=5.25\,\text{GeV}+0.0175 \, \SumPt$ is the width of the second, wider Gaussian. 
The event is then boosted to an inertial frame traveling with speed $\abs{\vec{\beta}}=k_T/m$ with respect to the lab frame, in a direction transverse to the beam axis, where $m$ is the invariant mass of all reconstructed objects in the event, along an azimuthal angle pulled randomly from a uniform distribution between 0 and $2\pi$.  The momenta of identified objects are recalculated in the lab frame.  Sixty percent of the recoil kick is assigned to unclustered momentum in the event.  The remaining forty percent of the recoil kick is assumed to disappear down the beam pipe, and contributes to the missing transverse momentum in the event.  This picture, and the particular parameter values that accompany this story, are determined primarily by the {\tt{uncl}}~$p_T$ and $\pmiss_T$ distributions in highly populated two-object final states, including the low-$p_T$ $2j$ final state, the high-$p_T$ $2j$ final state, and the final states $j\gamma$, $e^+e^-$, and $\mu^+\mu^-$.

Under the hypothesis described, reasonable although imperfect agreement with observation is obtained.  The result of this analysis supports the conclusions of previous studies indicating that the effective intrinsic $k_T$ needed to match observation is quite large relative to naive expectation.  That the data appear to require such a large effective intrinsic $k_T$ may be pointing out the need for some basic improvement to our understanding of this physics.

\subsection{Global fit}
\label{sec:CorrectionFactorFitDetails}

This section describes the construction of the global $\chi^2$ used in the \Vista\ global fit.

\subsubsection{$\chi^2_k$}
\label{sec:CorrectionFactorFitDetails:chi_k}
The bins in the CDF high-$p_T$ data sample are labeled by the index $k=(k_1,k_2)$, where each value of $k_1$ represents a phrase such as ``this bin contains events with three objects: one with $17<p_T<25$~GeV and $\abs{\eta}<0.6$, one with $40<p_T<60$~GeV and $0.6<\abs{\eta}<1.0$, and one with $25<p_T<40$~GeV and $1.0<\abs{\eta}$,'' and each value of $k_2$ represents a phrase such as ``this bin contains events with three objects: an electron, muon, and jet, respectively.''  The reason for splitting $k$ into $k_1$ and $k_2$ is that a jet can fake an electron (mixing the contents of $k_2$), but an object with $\abs{\eta}<0.6$ cannot fake an object with $0.6<\abs{\eta}<1.0$ (no mixing of $k_1$).  The term corresponding to the $k^\text{th}$ bin takes the form of Eq.~\ref{eq:chi_k}, where $\text{Data}[k]$ is the number of data events observed in the $k^\text{th}$ bin, $\text{SM}[k]$ is the number of events predicted by the standard model in the $k^\text{th}$ bin, $\delta\text{SM}[k]$ is the Monte Carlo statistical uncertainty on the standard model prediction in the $k^\text{th}$ bin, and $\sqrt{\text{SM}[k]}$ is the statistical uncertainty on the prediction in the $k^\text{th}$ bin. To legitimize the use of Gaussian errors, only bins containing eight or more data events are considered.  The standard model prediction $\text{SM}[k]$ for the $k^\text{th}$ bin can be written in terms of the introduced correction factors as
\begin{eqnarray}
\lefteqn{\text{SM}[k] = \text{SM}[(k_1,k_2)] =} & \nonumber \\
                & \sum_{{k_2}'\in\text{objectLists}}\sum_{l\in\text{processes}} \nonumber \\
                &  (\int{{\cal L}\,dt}) \cdot (\text{kFactor}[l])\cdot (\text{SM}_0[(k_1,{k_2}')][l])\cdot \nonumber \\
                & (\text{probabilityToBeSoMisreconstructed}[(k_1,{k_2}')][k_2])\cdot   \nonumber \\
                &  (\text{probabilityPassesTrigger}[(k_1,k_2)]),
\end{eqnarray}
where $\text{SM}[k]$ is the standard model prediction for the $k^\text{th}$ bin; the index $k$ is the Cartesian product of the two indices $k_1$ and $k_2$ introduced above, labeling the regions of the detector in which there are energy clusters and the identified objects corresponding to those clusters, respectively; the index ${k_2}'$ is a dummy summation index; the index $l$ labels standard model background processes, such as dijet production or $W$+1~jet production; $\text{SM}_0[(k_1,{k_2}')][l]$ is the initial number of standard model events predicted in bin $(k_1,{k_2}')$ from the process labeled by the index $l$; $\text{probabilityToBeMisreconstructedThus}[(k_1,{k_2}')][k_2]$ is the probability that an event produced with energy clusters in the detector regions labeled by $k_1$ that are identified as objects labeled by ${k_2}'$ would be mistaken as having objects labeled by $k_2$; and $\text{probabilityPassesTrigger}[(k_1,k_2)]$ represents the probability that an event produced with energy clusters in the detector regions labeled by $k_1$ that are identified as objects labeled by $k_2$ would pass the trigger.

The quantity $\text{SM}_0[(k_1,{k_2}')][l]$ is obtained by generating some number $n_l$ (say $10^4$) of Monte Carlo events corresponding to the process $l$.  The event generator provides a cross section $\sigma_l$ for this process $l$.  The weight of each of these Monte Carlo events is equal to $\sigma_l /n_l$.  Passing these events through the CDF simulation and reconstruction, the sum of the weights of these events falling into the bin $(k_1,{k_2}')$ is $\text{SM}_0[(k_1,{k_2}')][l]$.

\subsubsection{$\chi^2_{\text{constraints}}$}
\label{sec:CorrectionFactorFitDetails:chi_constraints}

The term $\chi^2_{\text{constraints}}(\vec{s})$ in Eq.~\ref{eqn:chiSqd} reflects constraints on the values of the correction factors determined by data other than those in the global high-$p_T$ sample.  These constraints include $k$-factors taken from theoretical calculations and numbers from the CDF literature when use is made of CDF data external to the \Vista\ high-$p_T$ sample.  The constraints imposed are:
\begin{itemize}
\item The luminosity ({\tt 0001}) is constrained to be within 6\% of the value measured by the CDF \v{C}erenkov luminosity counters.
\item The fake rate $\poo{q}{\gamma}$  ({\tt 0039}) is constrained to be $2.6\times10^{-4} \pm 1.5\times10^{-5}$, from the single particle gun study of Appendix~\ref{sec:MisidentificationMatrix}.
\item The fake rate $\poo{e}{\gamma}$ ({\tt 0032}) plus the efficiency $\poo{e}{e}$ ({\tt 0026}) for electrons in the plug is constrained to be within 1\% of unity.
\item Noting $\poo{q}{\gamma}$ corresponds to correction factor {\tt 0039}, $\poo{q}{\pi^\pm}=2\, \poo{q}{\pi^0}$, and $\poo{q}{\pi^0}=\poo{q}{\gamma} / \poo{\pi^0}{\gamma}$, and taking $\poo{\pi^0}{\gamma}=0.6$ and $\poo{\pi^\pm}{\tau}=0.415$ from the single particle gun study of Appendix~\ref{sec:MisidentificationMatrix}, the fake rate $\poo{q}{\tau}$ ({\tt 0038}) is constrained to $\poo{q}{\tau}=\poo{q}{\pi^\pm}\poo{\pi^\pm}{\tau} \, \pm 10\%$.
\item The $k$-factors for dijet production ({\tt 0018} and {\tt 0019}) are constrained to $1.10 \pm 0.05$ and $1.33\pm 0.05$ in the kinematic regions $\hat{p}_T<150$~GeV and $\hat{p}_T>150$~GeV, respectively, where $\hat{p}_T$ is the transverse momentum of the scattered partons in the $2\rightarrow2$ process in the colliding parton center of momentum frame\cdfSpecific{~\cite{dijetNLO}}.
\item The inclusive $k$-factor for $\gamma + N$jets ({\tt 0004}--{\tt 0007}) is constrained to $1.25\pm 0.15$~\cite{NLOJET:Nagy:2003tz,NLOJET:Nagy:2001xb}.
\item The inclusive $k$-factor for $\gamma\gamma + N$jets ({\tt 0008}--{\tt 0010}) is constrained to $2.0 \pm 0.15$~\cite{Diphox:Binoth:1999qq}.
\item The inclusive $k$-factors for $W$ and $Z$ production ({\tt 0011}--{\tt 0014} and {\tt 0015}--{\tt 0017}) are subject to a 2-dimensional Gaussian constraint, with mean at the NNLO/LO theoretical values~\cite{Stirling:Sutton:1991ay}, and a covariance matrix that encapsulates the highly correlated theoretical uncertainties, as discussed in Appendix~\ref{sec:VistaCorrectionModel:CorrectionFactorValues}.
\item Trigger efficiency correction factors are constrained to be less than unity.
\item All correction factors are constrained to be positive.
\end{itemize}

\subsubsection{Covariance matrix}
\label{sec:CorrectionFactorCovarianceMatrix}

\begin{sidewaystable*}
\hspace*{-0.5in}
\tiny\begin{minipage}{8.1in}\begin{verbatim}
       0001 0002 0003 0004 0005 0006 0007 0008 0009 0010 0011 0012 0013 0014 0015 0016 0017 0018 0019 0020 0021 0022 0023 0024 0025 0026 0027 0028 0029 0030 0031 0032 0033 0034 0035 0036 0037 0038 0039 0040 0041 0042 0043 0044 
0001   1    -.32 -.7  -.56 -.53 -.45 -.26 -.36 -.21 -.14 -.87 -.77 -.51 -.28 -.82 -.55 -.31 -.95 -.96 -.94 -.94 -.88 -.88 -.62 -.54 -.17 -.46 -.37 -.09 -.1  0    -.24 +.08 +.17 +.08 -.01 -.04 +.01 +.02 -.02 -.22 -.21 -.13 -.11 
0002   -.32 1    +.21 +.37 +.38 +.33 +.2  +.34 +.18 +.12 +.28 +.25 +.17 +.09 +.27 +.18 +.1  +.3  +.31 +.31 +.3  +.28 +.28 +.2  +.18 +.06 +.15 +.12 -.31 +.02 0    +.08 -.14 -.06 -.03 0    +.01 -.03 -.07 -.07 +.07 +.07 +.04 +.04 
0003   -.7  +.21 1    +.39 +.37 +.31 +.18 +.25 +.14 +.1  +.61 +.53 +.35 +.2  +.57 +.38 +.21 +.66 +.66 +.66 +.65 +.61 +.61 +.43 +.38 +.12 +.32 +.26 +.06 +.07 -.01 +.17 -.05 -.12 -.06 +.01 +.03 -.01 -.01 +.01 +.15 +.14 +.09 +.08 
0004   -.56 +.37 +.39 1    +.9  +.77 +.48 +.61 +.33 +.23 +.49 +.43 +.29 +.16 +.46 +.32 +.18 +.5  +.53 +.53 +.52 +.49 +.49 +.35 +.3  +.1  +.25 +.2  -.46 +.03 0    +.13 -.44 -.09 -.03 -.01 +.02 -.32 -.62 -.17 +.11 +.11 +.07 +.06 
0005   -.53 +.38 +.37 +.9  1    +.75 +.46 +.62 +.31 +.21 +.46 +.41 +.27 +.15 +.44 +.3  +.17 +.5  +.51 +.48 +.5  +.47 +.46 +.33 +.29 +.1  +.24 +.19 -.49 +.03 -.02 +.12 -.43 -.09 -.04 0    +.02 -.29 -.57 -.16 +.11 +.1  +.07 +.06 
0006   -.45 +.33 +.31 +.77 +.75 1    +.4  +.54 +.29 +.13 +.4  +.35 +.24 +.13 +.38 +.26 +.14 +.43 +.44 +.42 +.42 +.35 +.4  +.28 +.25 +.09 +.2  +.16 -.45 +.02 -.01 +.1  -.36 -.07 -.04 0    +.01 -.24 -.46 -.14 +.09 +.09 +.06 +.05 
0007   -.26 +.2  +.18 +.48 +.46 +.4  1    +.34 +.18 +.09 +.23 +.2  +.13 +.08 +.22 +.15 +.08 +.24 +.25 +.25 +.24 +.21 +.22 +.02 +.14 +.05 +.12 +.09 -.29 +.01 -.02 +.06 -.23 -.04 -.02 +.01 +.01 -.15 -.3  -.09 +.05 +.05 +.03 +.03 
0008   -.36 +.34 +.25 +.61 +.62 +.54 +.34 1    +.37 +.28 +.32 +.28 +.19 +.1  +.3  +.22 +.12 +.34 +.34 +.34 +.33 +.31 +.31 +.22 +.18 +.06 +.16 +.12 -.61 -.03 0    +.11 -.29 -.06 -.03 0    +.01 -.09 -.17 -.28 +.07 +.07 +.04 +.04 
0009   -.21 +.18 +.14 +.33 +.31 +.29 +.18 +.37 1    +.06 +.19 +.17 +.11 +.06 +.2  +.06 +.11 +.2  +.2  +.19 +.19 +.18 +.18 +.13 +.08 +.03 +.07 +.06 -.31 +.05 0    +.06 -.15 -.03 -.01 0    +.01 -.04 -.08 -.29 +.04 +.04 +.03 +.02 
0010   -.14 +.12 +.1  +.23 +.21 +.13 +.09 +.28 +.06 1    +.13 +.11 +.08 +.06 +.13 +.11 -.03 +.13 +.14 +.13 +.13 +.12 +.12 +.09 +.05 -.01 +.05 +.04 -.19 +.06 0    +.07 -.1  -.03 -.01 0    +.01 -.04 -.07 -.26 +.03 +.04 +.01 +.01 
0011   -.87 +.28 +.61 +.49 +.46 +.4  +.23 +.32 +.19 +.13 1    +.85 +.58 +.32 +.89 +.61 +.33 +.83 +.84 +.82 +.82 +.76 +.77 +.54 +.25 +.09 +.16 +.15 +.07 +.07 0    +.12 +.1  +.04 +.02 0    +.01 -.01 -.02 +.01 -.13 -.04 -.11 -.09 
0012   -.77 +.25 +.53 +.43 +.41 +.35 +.2  +.28 +.17 +.11 +.85 1    +.33 +.35 +.79 +.49 +.33 +.72 +.74 +.74 +.72 +.68 +.67 +.47 +.21 +.08 +.15 +.13 +.06 +.06 +.01 +.11 +.1  -.02 -.09 -.01 +.01 -.01 -.01 +.01 -.14 +.01 -.06 -.05 
0013   -.51 +.17 +.35 +.29 +.27 +.24 +.13 +.19 +.11 +.08 +.58 +.33 1    -.21 +.52 +.35 +.15 +.5  +.49 +.46 +.48 +.46 +.45 +.36 +.15 +.06 +.1  +.09 +.04 +.04 -.01 +.07 +.05 +.07 -.07 0    0    -.01 -.01 +.01 -.1  -.07 -.06 -.05 
0014   -.28 +.09 +.2  +.16 +.15 +.13 +.08 +.1  +.06 +.06 +.32 +.35 -.21 1    +.29 +.26 -.04 +.28 +.27 +.28 +.26 +.21 +.26 +.09 +.08 +.03 +.05 +.05 +.02 +.02 0    +.03 +.01 0    -.07 0    0    -.01 -.01 +.01 -.05 -.01 -.03 -.02 
0015   -.82 +.27 +.57 +.46 +.44 +.38 +.22 +.3  +.2  +.13 +.89 +.79 +.52 +.29 1    +.58 +.35 +.77 +.78 +.77 +.76 +.71 +.71 +.5  +.09 +.04 +.06 +.05 +.05 +.02 0    +.03 -.02 -.03 -.06 0    0    -.01 -.02 +.03 +.04 +.03 +.04 +.03 
0016   -.55 +.18 +.38 +.32 +.3  +.26 +.15 +.22 +.06 +.11 +.61 +.49 +.35 +.26 +.58 1    -.09 +.52 +.53 +.52 +.52 +.49 +.48 +.35 +.1  +.03 +.08 +.07 +.03 +.02 0    +.04 -.03 -.01 -.1  0    +.01 -.01 -.02 +.02 0    +.01 -.02 -.01 
0017   -.31 +.1  +.21 +.18 +.17 +.14 +.08 +.12 +.11 -.03 +.33 +.33 +.15 -.04 +.35 -.09 1    +.3  +.3  +.29 +.29 +.25 +.28 +.16 +.03 -.02 +.04 +.04 +.02 +.04 0    +.04 -.03 -.06 -.06 0    +.01 -.01 -.01 -.02 +.03 +.05 +.01 +.01 
0018   -.95 +.3  +.66 +.5  +.5  +.43 +.24 +.34 +.2  +.13 +.83 +.72 +.5  +.28 +.77 +.52 +.3  1    +.91 +.92 +.89 +.85 +.83 +.6  +.51 +.16 +.43 +.35 +.09 +.1  -.07 +.23 -.16 -.23 -.16 +.02 0    -.01 -.03 +.01 +.21 +.18 +.12 +.1  
0019   -.96 +.31 +.66 +.53 +.51 +.44 +.25 +.34 +.2  +.14 +.84 +.74 +.49 +.27 +.78 +.53 +.3  +.91 1    +.91 +.91 +.84 +.85 +.59 +.52 +.16 +.44 +.36 +.09 +.1  +.03 +.23 -.07 -.17 -.08 -.06 +.04 -.01 -.02 +.02 +.21 +.2  +.12 +.11 
0020   -.94 +.31 +.66 +.53 +.48 +.42 +.25 +.34 +.19 +.13 +.82 +.74 +.46 +.28 +.77 +.52 +.29 +.92 +.91 1    +.87 +.84 +.83 +.6  +.51 +.16 +.43 +.35 +.08 +.1  -.05 +.23 -.13 -.24 -.13 +.01 +.01 -.02 -.03 +.01 +.21 +.2  +.12 +.11 
0021   -.94 +.3  +.65 +.52 +.5  +.42 +.24 +.33 +.19 +.13 +.82 +.72 +.48 +.26 +.76 +.52 +.29 +.89 +.91 +.87 1    +.82 +.83 +.57 +.51 +.16 +.43 +.35 +.08 +.1  +.04 +.23 -.07 -.16 -.07 -.08 +.04 -.01 -.02 +.02 +.2  +.19 +.12 +.1  
0022   -.88 +.28 +.61 +.49 +.47 +.35 +.21 +.31 +.18 +.12 +.76 +.68 +.46 +.21 +.71 +.49 +.25 +.85 +.84 +.84 +.82 1    +.73 +.55 +.47 +.15 +.4  +.33 +.08 +.09 -.04 +.21 -.1  -.21 -.1  +.01 +.02 -.01 -.03 +.02 +.19 +.18 +.11 +.1  
0023   -.88 +.28 +.61 +.49 +.46 +.4  +.22 +.31 +.18 +.12 +.77 +.67 +.45 +.26 +.71 +.48 +.28 +.83 +.85 +.83 +.83 +.73 1    +.53 +.48 +.15 +.4  +.33 +.08 +.09 +.01 +.21 -.06 -.15 -.07 -.04 +.03 -.01 -.02 +.02 +.19 +.18 +.11 +.1  
0024   -.62 +.2  +.43 +.35 +.33 +.28 +.02 +.22 +.13 +.09 +.54 +.47 +.36 +.09 +.5  +.35 +.16 +.6  +.59 +.6  +.57 +.55 +.53 1    +.33 +.11 +.28 +.23 +.05 +.06 -.01 +.15 -.09 -.16 -.07 +.01 +.02 -.01 -.02 +.01 +.13 +.13 +.08 +.07 
0025   -.54 +.18 +.38 +.3  +.29 +.25 +.14 +.18 +.08 +.05 +.25 +.21 +.15 +.08 +.09 +.1  +.03 +.51 +.52 +.51 +.51 +.47 +.48 +.33 1    +.23 +.6  +.49 +.05 +.04 -.01 +.25 -.03 -.23 -.05 +.01 +.04 -.01 -.02 +.09 +.12 +.28 +.19 +.17 
0026   -.17 +.06 +.12 +.1  +.1  +.09 +.05 +.06 +.03 -.01 +.09 +.08 +.06 +.03 +.04 +.03 -.02 +.16 +.16 +.16 +.16 +.15 +.15 +.11 +.23 1    +.18 +.15 +.01 +.01 0    -.66 -.03 +.37 -.01 0    -.02 0    -.01 +.19 +.07 -.44 +.05 +.04 
0027   -.46 +.15 +.32 +.25 +.24 +.2  +.12 +.16 +.07 +.05 +.16 +.15 +.1  +.05 +.06 +.08 +.04 +.43 +.44 +.43 +.43 +.4  +.4  +.28 +.6  +.18 1    +.29 +.05 +.1  0    +.27 -.15 -.25 0    0    +.05 -.01 -.01 0    +.35 +.3  -.33 +.33 
0028   -.37 +.12 +.26 +.2  +.19 +.16 +.09 +.12 +.06 +.04 +.15 +.13 +.09 +.05 +.05 +.07 +.04 +.35 +.36 +.35 +.35 +.33 +.33 +.23 +.49 +.15 +.29 1    +.05 +.08 0    +.23 -.1  -.19 +.03 0    +.04 -.01 -.01 0    +.26 +.23 +.32 -.54 
0029   -.09 -.31 +.06 -.46 -.49 -.45 -.29 -.61 -.31 -.19 +.07 +.06 +.04 +.02 +.05 +.03 +.02 +.09 +.09 +.08 +.08 +.08 +.08 +.05 +.05 +.01 +.05 +.05 1    +.06 0    +.03 +.31 -.02 0    +.01 +.01 +.01 +.01 +.21 +.03 +.03 +.01 +.01 
0030   -.1  +.02 +.07 +.03 +.03 +.02 +.01 -.03 +.05 +.06 +.07 +.06 +.04 +.02 +.02 +.02 +.04 +.1  +.1  +.1  +.1  +.09 +.09 +.06 +.04 +.01 +.1  +.08 +.06 1    0    -.13 -.02 -.03 0    0    +.03 0    0    -.76 +.08 +.05 +.01 +.01 
0031   0    0    -.01 0    -.02 -.01 -.02 0    0    0    0    +.01 -.01 0    0    0    0    -.07 +.03 -.05 +.04 -.04 +.01 -.01 -.01 0    0    0    0    0    1    0    +.07 +.04 +.07 -.83 +.03 +.01 +.03 +.01 -.01 0    -.01 -.01 
0032   -.24 +.08 +.17 +.13 +.12 +.1  +.06 +.11 +.06 +.07 +.12 +.11 +.07 +.03 +.03 +.04 +.04 +.23 +.23 +.23 +.23 +.21 +.21 +.15 +.25 -.66 +.27 +.23 +.03 -.13 0    1    -.06 -.48 -.02 0    +.05 0    0    -.08 +.17 +.57 +.06 +.05 
0033   +.08 -.14 -.05 -.44 -.43 -.36 -.23 -.29 -.15 -.1  +.1  +.1  +.05 +.01 -.02 -.03 -.03 -.16 -.07 -.13 -.07 -.1  -.06 -.09 -.03 -.03 -.15 -.1  +.31 -.02 +.07 -.06 1    +.23 +.17 -.02 -.01 +.2  +.39 +.14 -.55 -.18 -.21 -.18 
0034   +.17 -.06 -.12 -.09 -.09 -.07 -.04 -.06 -.03 -.03 +.04 -.02 +.07 0    -.03 -.01 -.06 -.23 -.17 -.24 -.16 -.21 -.15 -.16 -.23 +.37 -.25 -.19 -.02 -.03 +.04 -.48 +.23 1    +.16 -.01 -.04 +.01 +.02 +.09 -.31 -.89 -.22 -.19 
0035   +.08 -.03 -.06 -.03 -.04 -.04 -.02 -.03 -.01 -.01 +.02 -.09 -.07 -.07 -.06 -.1  -.06 -.16 -.08 -.13 -.07 -.1  -.07 -.07 -.05 -.01 0    +.03 0    0    +.07 -.02 +.17 +.16 1    -.02 +.02 +.01 +.01 0    -.12 -.1  -.26 -.23 
0036   -.01 0    +.01 -.01 0    0    +.01 0    0    0    0    -.01 0    0    0    0    0    +.02 -.06 +.01 -.08 +.01 -.04 +.01 +.01 0    0    0    +.01 0    -.83 0    -.02 -.01 -.02 1    0    0    +.01 0    +.01 +.01 +.02 +.01 
0037   -.04 +.01 +.03 +.02 +.02 +.01 +.01 +.01 +.01 +.01 +.01 +.01 0    0    0    +.01 +.01 0    +.04 +.01 +.04 +.02 +.03 +.02 +.04 -.02 +.05 +.04 +.01 +.03 +.03 +.05 -.01 -.04 +.02 0    1    +.01 +.01 -.03 +.06 +.07 +.03 +.02 
0038   +.01 -.03 -.01 -.32 -.29 -.24 -.15 -.09 -.04 -.04 -.01 -.01 -.01 -.01 -.01 -.01 -.01 -.01 -.01 -.02 -.01 -.01 -.01 -.01 -.01 0    -.01 -.01 +.01 0    +.01 0    +.2  +.01 +.01 0    +.01 1    +.51 +.06 0    0    0    0    
0039   +.02 -.07 -.01 -.62 -.57 -.46 -.3  -.17 -.08 -.07 -.02 -.01 -.01 -.01 -.02 -.02 -.01 -.03 -.02 -.03 -.02 -.03 -.02 -.02 -.02 -.01 -.01 -.01 +.01 0    +.03 0    +.39 +.02 +.01 +.01 +.01 +.51 1    +.12 0    -.01 0    0    
0040   -.02 -.07 +.01 -.17 -.16 -.14 -.09 -.28 -.29 -.26 +.01 +.01 +.01 +.01 +.03 +.02 -.02 +.01 +.02 +.01 +.02 +.02 +.02 +.01 +.09 +.19 0    0    +.21 -.76 +.01 -.08 +.14 +.09 0    0    -.03 +.06 +.12 1    -.04 -.11 +.01 +.01 
0041   -.22 +.07 +.15 +.11 +.11 +.09 +.05 +.07 +.04 +.03 -.13 -.14 -.1  -.05 +.04 0    +.03 +.21 +.21 +.21 +.2  +.19 +.19 +.13 +.12 +.07 +.35 +.26 +.03 +.08 -.01 +.17 -.55 -.31 -.12 +.01 +.06 0    0    -.04 1    +.37 +.39 +.33 
0042   -.21 +.07 +.14 +.11 +.1  +.09 +.05 +.07 +.04 +.04 -.04 +.01 -.07 -.01 +.03 +.01 +.05 +.18 +.2  +.2  +.19 +.18 +.18 +.13 +.28 -.44 +.3  +.23 +.03 +.05 0    +.57 -.18 -.89 -.1  +.01 +.07 0    -.01 -.11 +.37 1    +.25 +.22 
0043   -.13 +.04 +.09 +.07 +.07 +.06 +.03 +.04 +.03 +.01 -.11 -.06 -.06 -.03 +.04 -.02 +.01 +.12 +.12 +.12 +.12 +.11 +.11 +.08 +.19 +.05 -.33 +.32 +.01 +.01 -.01 +.06 -.21 -.22 -.26 +.02 +.03 0    0    +.01 +.39 +.25 1    +.07 
0044   -.11 +.04 +.08 +.06 +.06 +.05 +.03 +.04 +.02 +.01 -.09 -.05 -.05 -.02 +.03 -.01 +.01 +.1  +.11 +.11 +.1  +.1  +.1  +.07 +.17 +.04 +.33 -.54 +.01 +.01 -.01 +.05 -.18 -.19 -.23 +.01 +.02 0    0    +.01 +.33 +.22 +.07 1    
\end{verbatim}
\end{minipage}
\caption{Correction factor correlation matrix.  The top row and left column show correction factor codes.  Each element of the matrix shows the correlation between the correction factors corresponding to the column and row.  Each matrix element is dimensionless; the elements along the diagonal are unity; the matrix is symmetric; positive elements indicate positive correlation, and negative elements anti-correlation.}
\label{tbl:CorrectionFactorCorrelationMatrix}
\end{sidewaystable*}

This section describes the correction factor covariance matrix $\Sigma$.  The inverse of the covariance matrix is obtained from 
\begin{equation}
\Sigma^{-1}_{ij} = \frac{1}{2} \frac{\partial^2 \chi^2(\vec s)}{\partial s_i \partial s_j} \bigg| _{\vec s_0},
\label{eqn:logL4}
\end{equation}
where $\chi^2(\vec{s})$ is defined by Eq.~\ref{eqn:chiSqd} as a function of the correction factor vector $\vec{s}$, vector elements $s_i$ and $s_j$ are the $i^{\text{th}}$ and $j^{\text{th}}$ correction factors, and $\vec{s}_0$ is the vector of correction factors that minimizes $\chi^2(\vec{s})$.  Numerical estimation of the right hand side of Eq.~\ref{eqn:logL4} is achieved by calculating $\chi^2$ at $\vec s_0$ and at positions slightly displaced from $\vec s_0$ in the direction of the $i^{\text{th}}$ and $j^{\text{th}}$ correction factors, denoted by the unit vectors $\hat i$ and $\hat j$.  Approximating the second partial derivative
\begin{eqnarray}
  \label{eq:secondDerivative}
  \frac{\partial^2 \chi^2}{\partial s_j \partial s_i} \bigg|_{\vec s_0} &=& \frac{\chi^2(\vec s_0 + \hat i \, \delta s_i + \hat j \, \delta s_j) - \chi^2(\vec s_0 + \hat j \, \delta s_j)}{\delta s_j \delta s_i}  - \nonumber \\
                                                                        & & \frac{\chi^2(\vec s_0 + \hat i \, \delta s_i) - \chi^2(\vec s_0)}{\delta s_j \delta s_i} \nonumber
\end{eqnarray}
leads to
\begin{eqnarray}
\Sigma^{-1}_{ij} & = & [ \chi^2(\vec{s}_0+\delta s_i \, \hat{i}+\delta s_j \, \hat{j}) \nonumber \\
            &   &  - \chi^2(\vec{s}_0+\delta s_i \, \hat{i}) \nonumber \\
            &   &  - \chi^2(\vec{s}_0+\delta s_j \, \hat{j}) \nonumber \\
            &   & + \chi^2(\vec{s}_0) ] /  (2 \delta s_i \, \delta s_j),
\label{eqn:CorrectionFactorSigmaInverse}
\end{eqnarray}
for appropriately small steps $\delta s_i$ and $\delta s_j$ away from the minimum.  The covariance matrix $\Sigma$ is calculated by inverting $\Sigma^{-1}$.  The diagonal element $\Sigma_{ii}$ is the variance $\sigma_i^2$ of the $i^{\text{th}}$ correction factor, and the correlation $\rho_{ij}$ between the $i^{\text{th}}$ and $j^{\text{th}}$ correction factors is $\rho_{ij}=\Sigma_{ij}/\sigma_i\sigma_j$.  The variances of each correction factor, corresponding to the diagonal elements of the covariance matrix, are shown in Table~\ref{tbl:CorrectionFactorDescriptionValuesSigmas}.  The correlation matrix obtained is shown in Table~\ref{tbl:CorrectionFactorCorrelationMatrix}.

\subsection{Correction factor values}
\label{sec:VistaCorrectionModel:CorrectionFactorValues}

This section provides notes on the values of the \Vista\ correction factors obtained from a global fit of standard model prediction to data.  The correction factors considered are numbers that can in principle be calculated {\em{a priori}}, but whose calculation is in practice not feasible.  These correction factors divide naturally into two classes, the first of which reflects the difficulty of calculating the standard model prediction to all orders, and the second of which reflects the difficulty of understanding from first principles the response of the experimental apparatus.

The theoretical correction factors considered are of two types.  The difficulty of calculating the standard model prediction for many processes to all orders in perturbation theory is handled through the introduction of $k$-factors, representing the ratio of the true all orders prediction to the prediction at lowest order in perturbation theory.  Uncertainties in the distribution of partons inside the colliding proton and anti-proton as a function of parton momentum are in principle handled through the introduction of correction factors associated with parton distribution functions, but there are currently no discrepancies to motivate this.

Experimental correction factors correspond to numbers describing the response of the CDF detector that are precisely calculable in principle, but that are in practice best constrained by the high-$p_T$ data themselves.  These correction factors take the form of the integrated luminosity, object identification efficiencies, object misidentification probabilities, trigger efficiencies, and energy scales.  

\subsubsection{$k$-factors}

For nearly all standard model processes, $k$-factors are used as an overall multiplicative constant, rather than being considered to be a function of one or more kinematic variables.  The spirit of the approach is to introduce as few correction factors as possible, and to only introduce correction factors motivated by specific discrepancies.

\begin{figure}
\vspace*{-0.2in}
\includegraphics[width=3.5in]{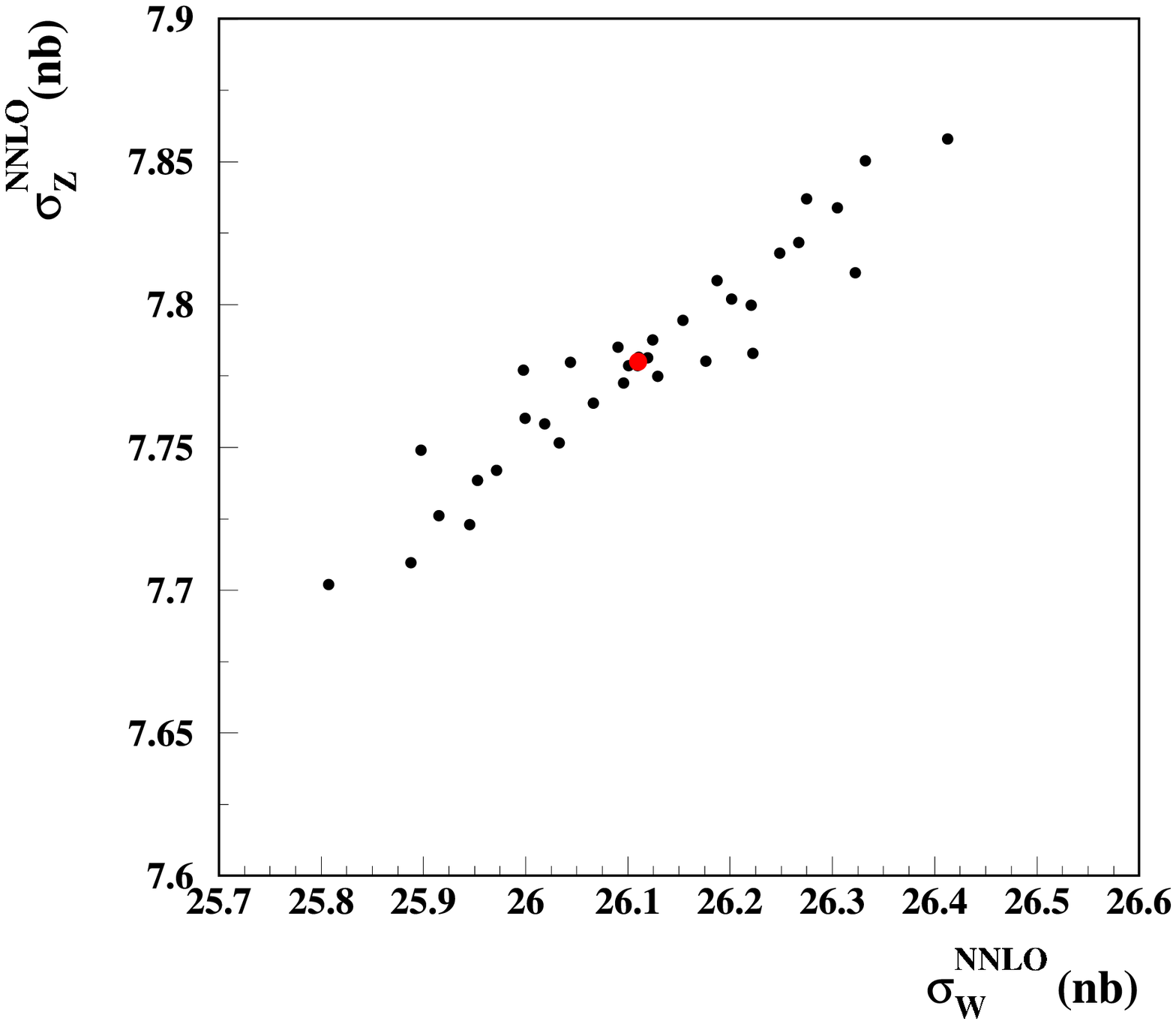} \\
\vspace*{-0.2in} 
\caption{Variation of the $k$-factors for inclusive $W$ and $Z$ production under different choices of parton distribution functions, from the Alekhin parton distribution error set~\cite{Alekhin:2005gq}.  The correlation of the uncertainty on these two $k$-factors due to uncertainty in the parton distribution functions is 0.955.}
\label{fig:alekhin_kWkZ_pdfs}
\end{figure}

\paragraph*{\tt{0001}.}
The integrated luminosity of the analysis sample has a close relationship with the theoretically determined values of inclusive $W$ and $Z$ production at the Tevatron.  Figure~\ref{fig:alekhin_kWkZ_pdfs} shows the variation in calculated inclusive $W$ and $Z$ $k$-factors under changes in the assumed parton distribution functions.  Each point represents a different $W$ and $Z$ inclusive cross section determined using modified parton distribution functions.  The use of 16 bases to reflect systematic uncertainties results in 32 black dots in Fig.~\ref{fig:alekhin_kWkZ_pdfs}.  The uncertainties in the $W$ and $Z$ cross sections due to variations in the renormalization and factorization scales are nearly 100\% correlated; varying these scales affects both the $W$ and $Z$ inclusive cross sections in the same way.  The uncertainties in the parton distribution functions and the choice of renormalization and factorization scales represent the dominant contributions to the theoretical uncertainty in the total inclusive $W$ and $Z$ cross section calculations at the Tevatron.  The \cdfSpecificHref{\codeBrowserURL{Vista/src/constraints.cc}}{term} in $\chi^2_{\text{constraints}}$ that reflects our knowledge of the theoretical prediction of the inclusive $W$ and $Z$ cross sections explicitly acknowledges this high degree of correlation.

Theoretical constraints on all other $k$-factors are assumed to be uncorrelated with each other, not because the uncertainties of these calculations are indeed uncorrelated, but rather because the correlations among these computations are poorly known.

\begin{figure}
\includegraphics[width=3.5in]{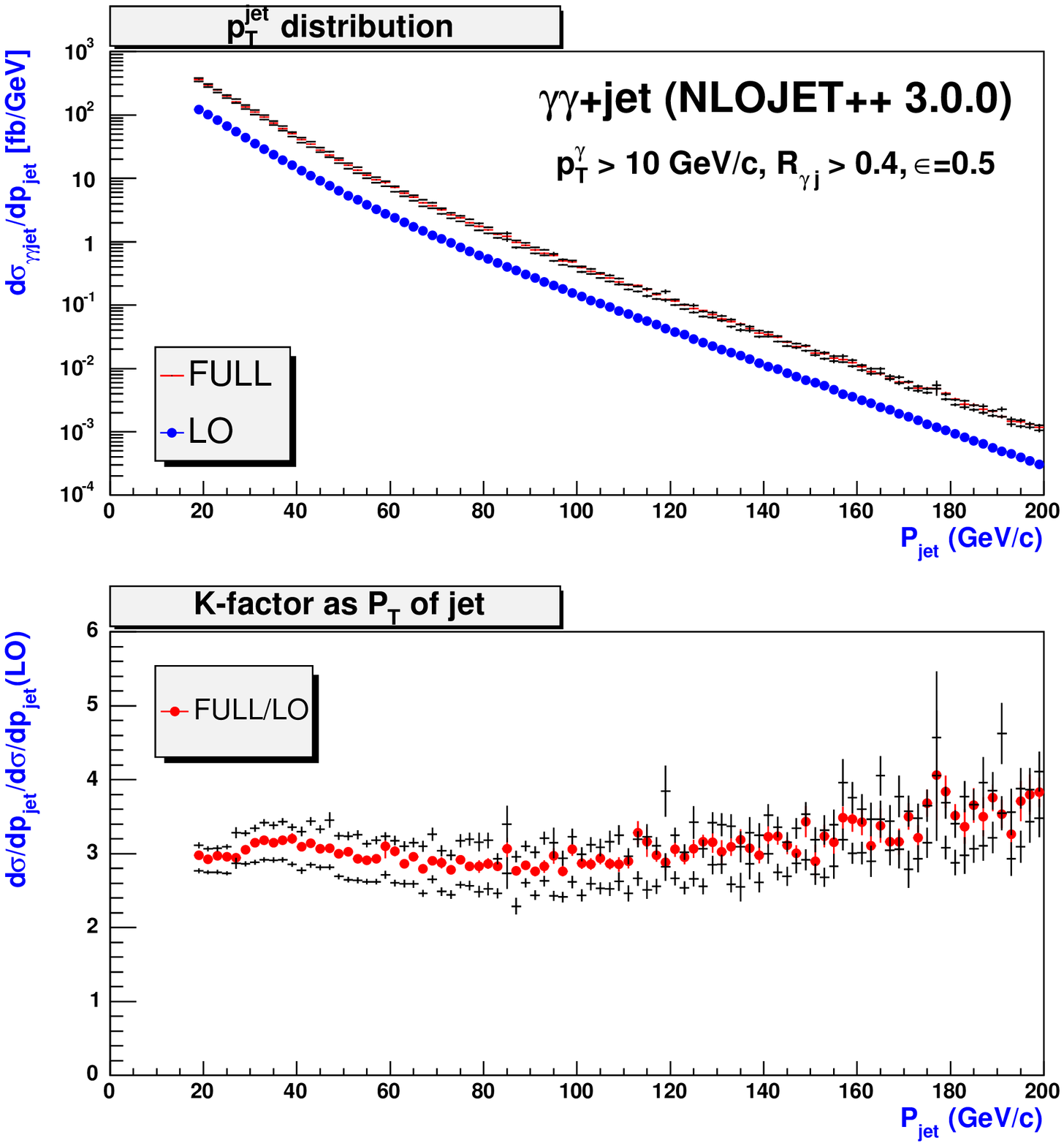}
\caption{Calculation of the $\gamma\gamma j$ $k$-factor, as a function of jet transverse momentum.  The effect of changing the factorization scale by a factor of two in either direction is also shown (small black points with error bars).}
\label{fig:soonJun_kfactor_2ph1j}
\end{figure}

\paragraph*{\tt{0002}, \tt{0003}.}
The {\tt{cosmic}}~$\gamma$ and {\tt{cosmic}}~$j$ backgrounds are estimated using events recorded in the CDF data with one or more reconstructed photons and with two or fewer reconstructed tracks.  The use of events with two or fewer reconstructed tracks is a new technique for estimating these backgrounds.  These correction factors are primarily constrained by the number of events in the \Vista\ $\gamma\pmiss$ and $j\pmiss$ final states.  The values are related to (and consistent with) the fraction of bunch crossings with one or more inelastic $p\bar{p}$ interactions, complicated slightly by the requirement that any jet falling in the final state $j\pmiss$ has at least 5~GeV of track $p_T$ within a cone of 0.4 relative to the jet axis.
\paragraph*{{\tt{0004}}, {\tt{0005}}, {\tt{0006}}, {\tt{0007}}.}
The NLOJET++ calculation of the $\gamma j$ inclusive $k$-factor constrains the cross section weighted sum of the $\gamma j$, $\gamma 2j$, $\gamma 3j$, and $\gamma 4j$ correction factors to $1.25\pm0.15$~\cite{NLOJET:Nagy:2003tz,NLOJET:Nagy:2001xb}.
\paragraph*{{\tt{0008}}, {\tt{0009}}, {\tt{0010}}.}
The DIPHOX calculation of the inclusive $\gamma\gamma$ cross section at NLO constrains the weighted sum of these correction factors to $2.0\pm0.15$~\cite{Diphox:Binoth:1999qq}.  From Table~\ref{tbl:CorrectionFactorDescriptionValuesSigmas}, the $\gamma\gamma j$ $k$-factor ({\tt{0009}}) appears anomalously large.  Figure~\ref{fig:soonJun_kfactor_2ph1j} shows a calculation of this $\gamma\gamma j$ $k$-factor using NLOJET++~\cite{NLOJET:Nagy:2003tz,NLOJET:Nagy:2001xb} as a function of summed transverse momentum.  The NLO correction to the LO prediction is found to be large, and not manifestly inconsistent with the value for this $k$-factor determined from the \Vista\ fit.  The cross section for $\gamma\gamma 2j$ production has not been calculated at NLO.
\paragraph*{{\tt{0011}}, {\tt{0012}}, {\tt{0013}}, {\tt{0014}}.}
These correction factors correspond to $k$-factors for $W$ production in association with zero, one, two, and three or more jets, respectively.  A linear combination of these correction factors is constrained by the requirement that the inclusive $W$ production cross section is consistent with the NNLO calculation of Ref.~\cite{Alekhin:2005gq}.  The values of these correction factors, and their trend of decreasing as the number of jets increases, depends heavily on the choice of renormalization and factorization scales.  The individual correction factors are not explicitly constrained by a NLO calculation.
\paragraph*{{\tt{0015}}, {\tt{0016}}, {\tt{0017}}.}
These correction factors correspond to $k$-factors for $Z$ production in association with zero, one, and two or more jets, respectively.  A linear combination of these correction factors is constrained by the requirement that the inclusive $Z$ production cross section is consistent with the NNLO calculation of Ref.~\cite{Alekhin:2005gq}.  
\paragraph*{{\tt{0018}}, {\tt{0019}}.}
The two $k$-factors for dijet production correspond to two bins in $\hat{p}_T$, the $p_T$ of the hard two to two scattering in the parton center of mass frame.  These correction factors are constrained by a NLO calculation~\cite{2j_kfactor:Stump:2003yu}, and show expected behavior as a function of $\hat{p}_T$.
\paragraph*{{\tt{0020}}, {\tt{0021}}.}
The two $k$-factors for 3-jet production, corresponding to two bins in $\hat{p}_T$, are unconstrained by any NLO calculation, but show reasonable behavior as a function of $\hat{p}_T$.
\paragraph*{{\tt{0022}}, {\tt{0023}}.}
The $k$-factors for 4-jet production, corresponding to two bins in $\hat{p}_T$, are unconstrained by any NLO calculation, but show reasonable behavior as a function of $\hat{p}_T$.
\paragraph*{\tt{0024}.}
The $k$-factor for the production of five or more jets, constrained primarily by the \Vista\ low-$p_T$ $5j$ final state, is found to be close to unity.

\subsubsection{Identification efficiencies}

The correction factors in this section, although billed as ``identification efficiencies,'' are truly ratios of the identification efficiency in the data relative to the identification efficiency in \CdfSim.  A correction factor value of unity indicates a proper modeling of the overall identification efficiency by \CdfSim; a correction factor value of 0.5 indicates that \CdfSim\ overestimates the overall identification efficiency by a factor of two.  

\paragraph*{{\tt{0025}}.}
The central electron identification efficiency scale factor is close to unity, indicating the central electron efficiency measured in data is similar (to within 1\%) to the central electron efficiency in the CDF detector simulation.  This reflects an emphasis within CDF on tuning the detector simulation for central electrons.  The determination of this correction factor is dominated by the \Vista\ final states $e\pmiss$ and $e^+e^-$, where one of the electrons has $\abs{\eta}<1$.
\paragraph*{{\tt{0026}}.}
The plug electron identification efficiency scale factor is several percent less than unity, indicating that the CDF detector simulation slightly overestimates the electron identification efficiency in the plug region of the CDF detector.  The determination of this correction factor is dominated by the \Vista\ final states $e\pmiss$ and $e^+e^-$, where one of the electrons has $\abs{\eta}>1$.
\paragraph*{{\tt{0027}}, {\tt{0028}}.}
To reduce backgrounds hypothesized to arise from pion and kaon decays in flight with a substantially mismeasured track, a very good track fit in the CDF tracker is required.  Partially due to this tight track fit requirement, CDF muon identification efficiencies in the regions $\abs{\eta}<0.6$ and $0.6<\abs{\eta}<1.5$ are overestimated in the CDF detector simulation by over 10\%.  The determination of the identification efficiencies $\poo{\mu}{\mu}$ is dominated by the \Vista\ final states $\mu\pmiss$ and $\mu^+\mu^-$.
\paragraph*{\tt{0029}.}
The central photon identification efficiency scale factor is determined primarily by the number of events in the \Vista\ final states $j\gamma$ and $\gamma\gamma$.  The uncertainty on this correction factor is highly correlated with the uncertainties on the $\gamma j$ $k$-factor, the $\poo{j}{\gamma}$ fake rate, and the $\gamma\gamma$ $k$-factor.
\paragraph*{\tt{0030}.}
The plug photon identification efficiency scale factor is determined primarily by the number of events in the \Vista\ final state $\gamma\gamma$.  The uncertainty on this correction factor is highly correlated with the uncertainty on the plug $\poo{j}{\gamma}$ fake rate.
\paragraph*{\tt{0031}.}
The $b$-jet identification efficiency is determined to be consistent with the prediction from \CdfSim.

\subsubsection{Fake rates}

\paragraph*{\tt{0032}.}
The fake rate $\poo{e}{\gamma}$ for electrons to be misreconstructed as photons in the plug region of the detector is added on top of the significant number of electrons misreconstructed as photons by \CdfSim.
\paragraph*{\tt{0033}.}
In \Vista, the contribution of jets faking electrons is modeled by applying a fake rate $\poo{j}{e}$ to Monte Carlo jets.  \Vista\ represents the first large scale Tevatron analysis in which a completely Monte Carlo based modeling of jets faking electrons is employed.  Significant understanding of the physical mechanisms contributing to this fake rate has been achieved, as summarized in Appendix~\ref{sec:MisidentificationMatrix}.  Consistency with this understanding is required; for example, $\poo{j}{e} \approx \poo{j}{\gamma}\poo{\gamma}{e}$.  The value of this correction factor is determined primarily by the number of events in the \Vista\ final state $ej$, where the electron is identified in the central region of the CDF detector.  It is notable that this fake rate is independent of global event properties, and that a consistent simultaneous understanding of the $ej$, $e2j$, $e3j$, and $e4j$ final states is obtained.
\paragraph*{\tt{0034}.}
The value of the fake rate $\poo{j}{e}$ in the plug region of the CDF detector is roughly one order of magnitude larger than the corresponding fake rate $\poo{j}{e}$ in the central region of the detector, consistent with an understanding of the relative performance of the detector in the central and plug regions for the identification of electrons.  This correction factor is determined primarily by the number of events in the \Vista\ final state $ej$, where the electron is identified in the plug region of the CDF detector.
\paragraph*{\tt{0035}.}
In \Vista, the contribution of jets faking muons is modeled by applying a fake rate $\poo{j}{\mu}$ to Monte Carlo jets.  \Vista\ represents the first large scale Tevatron analysis in which a completely Monte Carlo based modeling of jets faking muons is employed.  The value obtained from the \Vista\ fit is seen to be roughly one order of magnitude smaller than the fake rate $\poo{j}{e}$ in the central region of the detector, consistent with our understanding of the physical mechanisms underlying these fake rates, as described in Appendix~\ref{sec:MisidentificationMatrix}.  The value of this correction factor is determined primarily by the number of events in the \Vista\ final state $j\mu$.
\paragraph*{{\tt{0036}}.}
The fake rate $\poo{j}{b}$ has $p_T$ dependence explicitly imposed.  The number of tracks inside a typical jet, and hence the probability that a secondary vertex is (mis)reconstructed, increases with jet $p_T$.  The values of these correction factors are consistent with the mistag rate determined using secondary vertices reconstructed on the other side of the beam axis with respect to the direction of the tagged jet \cite{bTagging}.  The value of this correction factor is determined primarily by the number of events in the \Vista\ final states $bj$ and $bb$.
\paragraph*{{\tt{0037}}, {\tt{0038}}.}
The fake rate $\poo{j}{\tau}$ decreases with jet $p_T$, since the number of tracks inside a typical jet increases with jet $p_T$.  The values of these correction factors are determined primarily by the number of events in the \Vista\ final state $j\tau$.
\paragraph*{{\tt{0039}}, {\tt{0040}}.}
The fake rate $\poo{j}{\gamma}$ is determined separately in the central and plug regions of the CDF detector.  The values of these correction factors are determined primarily by the number of events in the \Vista\ final states $j\gamma$ and $\gamma\gamma$.  The value obtained for {\tt{0039}} is consistent with the value obtained from a study using detailed information from the central preshower detector.  The fake rate determined in the plug region is noticeably higher than the fake rate determined in the central region, as expected.

\subsubsection{Trigger efficiencies}

\paragraph*{{\tt{0041}}.}
The central electron trigger inefficiency is dominated by not correctly reconstructing the electron's track at the first online trigger level.
\paragraph*{{\tt{0042}}.}
The plug electron trigger inefficiency is due to inefficiencies in clustering at the second online trigger level.
\paragraph*{{\tt{0043}}, {\tt{0044}}.}
The muon trigger inefficiencies in the regions $\abs{\eta}<0.6$ and $0.6<\abs{\eta}<1.0$ derive partly from tracking inefficiency, and partly from an inefficiency in reconstructing muon stubs in the CDF muon chambers.\\
~\\
The value of these corrections factors are consistent with other trigger efficiency measurements made using additional information~\cite{Messina:2006zz}.

\subsubsection{Energy scales}

The \Vista\ infrastructure also allows the jet energy scale to be treated as a correction factor.  At present this correction factor is not used, since there is no discrepancy requiring it.

To understand the effect of introducing such a correction factor, a jet energy scale correction factor is added and constrained to $1\pm0.03$, reflecting the jet energy scale determination at CDF~\cite{jetEnergyScale:Bhatti:2005ai}.  The fit returns a value with a very small error, since this correction factor is highly constrained by the low-$p_T$ $2j$, $3j$, $e\,j$, and $e\,2j$ final states.  Assuming perfectly correct modeling of jets faking electrons, as described in Appendix~\ref{sec:MisidentificationMatrix}, this is a correct energy scale error.  The inclusion of additional correction factor degrees of freedom to reflect possible imperfections in this modeling of jets faking electrons increases the energy scale error.  The interesting conclusion is that the jet energy scale (considered as a lone free parameter) is very well constrained by the large number of dijet events; adjustment to the jet energy scale must be accompanied by simultaneous adjustment of other correction factors (such as the dijet $k$-factor) in order to retain agreement with data.


\section{\Sleuth\ details}
\label{sec:Sleuth:MinimumNumberOfEvents}

This appendix elaborates on the \Sleuth\ partitioning rule, and on the minimum number of events required for a final state to be considered by \Sleuth.

\subsection{Partitioning}
\label{sec:Sleuth:Partitioning}

\begin{table*}
\hspace*{-0.0in}\mbox{\tiny\begin{minipage}{7.0in} 
\begin{tabular}{p{1.5cm}p{5cm}} 
\Sleuth\ & \Vista\ Final States   \\ 
\hline 
\hline 
$b \bar{b}$ & $b$$j$, $b$2$j$, 2$b$$j$, 2$b$, 3$b$ \\ 
$b \bar{b} \ell^{+}\ell^{-}$ & $e^{+}$$e^{-}$$b$$j$, $e^{+}$$e^{-}$$b$2$j$, $\mu^{+}$$\mu^{-}$$b$$j$, $\mu^{+}$$\mu^{-}$$b$2$j$, $e^{+}$$e^{-}$2$b$ \\ 
$b \bar{b} \ell^{+}\ell^{-} 2j$ & $e^{+}$$e^{-}$$b$3$j$, $\mu^{+}$$\mu^{-}$$b$3$j$ \\ 
$b \bar{b} \ell^{+}\ell^{-} 2j \pslash$ & $\mu^{+}$$\mu^{-}$2$b$2$j$$\pslash$ \\ 
$b \bar{b} \ell^{+}\ell^{-} \pslash$ & $e^{+}$$e^{-}$$b$2$j$$\pslash$, $e^{+}$$e^{-}$$b$$j$$\pslash$, $\mu^{+}$$\mu^{-}$$b$$j$$\pslash$, $\mu^{+}$$\mu^{-}$$b$2$j$$\pslash$, $e^{+}$$e^{-}$2$b$$j$$\pslash$, $e^{+}$$e^{-}$2$b$$\pslash$, $\mu^{+}$$\mu^{-}$2$b$$\pslash$ \\ 
$b \bar{b} \ell^{+} 2j  \gamma \pslash$ & $e^{+}$$\gamma$$b$3$j$$\pslash$, $\mu^{+}$$\gamma$$b$3$j$$\pslash$ \\ 
$W b\bar{b} jj$ & $e^{+}$$b$3$j$$\pslash$, $\mu^{+}$$b$3$j$$\pslash$, $e^{+}$2$b$2$j$$\pslash$, $\mu^{+}$2$b$2$j$$\pslash$ \\ 
$b \bar{b} \ell^{+} \ell'^{+}$ & $e^{+}$$\mu^{+}$2$b$ \\ 
$b \bar{b} \ell^{+} \ell'^{-}$ & $e^{+}$$\mu^{-}$$b$$j$ \\ 
$b \bar{b} \ell^{+} \ell'^{-}  \pslash$ & $e^{+}$$\mu^{-}$$b$$j$$\pslash$, $e^{+}$$\mu^{-}$$b$2$j$$\pslash$ \\ 
$b \bar{b} \ell^{+} \gamma \pslash$ & $\mu^{+}$$\gamma$$b$2$j$$\pslash$ \\ 
$W b\bar{b}$ & $e^{+}$$b$$j$$\pslash$, $\mu^{+}$$b$$j$$\pslash$, $e^{+}$$b$2$j$$\pslash$, $\mu^{+}$$b$2$j$$\pslash$, $e^{+}$2$b$$\pslash$, $e^{+}$2$b$$j$$\pslash$, $\mu^{+}$2$b$$j$$\pslash$, $\mu^{+}$2$b$$\pslash$ \\ 
$b \bar{b} \ell^{+} \pslash \tau^{-}$ & $\mu^{+}$$\tau^{-}$$b$$j$$\pslash$ \\ 
$b \bar{b} \ell^{+} \tau^{+}$ & $e^{+}$$\tau^{+}$$b$$j$ \\ 
$b \bar{b} \ell^{+} \tau^{-}$ & $e^{+}$$\tau^{-}$$b$$j$, $e^{+}$$\tau^{-}$2$b$ \\ 
$b \bar{b} 2j$ & $b$3$j$, 2$b$2$j$ \\ 
$b \bar{b} 2j \gamma$ & $\gamma$$b$3$j$, $\gamma$2$b$2$j$ \\ 
$b \bar{b} 2j \gamma \pslash$ & $\gamma$$b$3$j$$\pslash$ \\ 
$b \bar{b} 2j \pslash$ & $b$3$j$$\pslash$, 2$b$2$j$$\pslash$ \\ 
$b \bar{b} \gamma \gamma 2j$ & 2$\gamma$$b$3$j$ \\ 
$\gamma b \bar{b}$ & $\gamma$$b$$j$, $\gamma$$b$2$j$, $\gamma$2$b$, $\gamma$2$b$$j$, $\gamma$3$b$ \\ 
$b \bar{b} \gamma \pslash$ & $\gamma$$b$$j$$\pslash$, $\gamma$$b$2$j$$\pslash$, $\gamma$2$b$$\pslash$ \\ 
$b \bar{b} \pslash$ & $b$2$j$$\pslash$, $b$$j$$\pslash$, 2$b$$j$$\pslash$, 2$b$$\pslash$ \\ 
$\gamma \gamma b \bar{b}$ & 2$\gamma$$b$$j$, 2$\gamma$$b$2$j$, 2$\gamma$2$b$ \\ 
$\ell^{+}\ell^{-}$ & $e^{+}$$e^{-}$, $e^{+}$$e^{-}$$j$, $\mu^{+}$$\mu^{-}$, $\mu^{+}$$\mu^{-}$$j$, $e^{+}$$e^{-}$$b$, $\mu^{+}$$\mu^{-}$$b$ \\ 
$\ell^{+}\ell^{-} 2j$ & $e^{+}$$e^{-}$2$j$, $e^{+}$$e^{-}$3$j$, $\mu^{+}$$\mu^{-}$2$j$, $\mu^{+}$$\mu^{-}$3$j$ \\ 
$\ell^{+}\ell^{-} 2j \gamma$ & $e^{+}$$e^{-}$$\gamma$2$j$, $e^{+}$$e^{-}$$\gamma$3$j$, $\mu^{+}$$\mu^{-}$$\gamma$2$j$, $\mu^{+}$$\mu^{-}$$\gamma$3$j$ \\ 
$\ell^{+}\ell^{-} 2j \pslash$ & $e^{+}$$e^{-}$2$j$$\pslash$, $e^{+}$$e^{-}$3$j$$\pslash$, $\mu^{+}$$\mu^{-}$2$j$$\pslash$, $\mu^{+}$$\mu^{-}$3$j$$\pslash$ \\ 
$\ell^{+}\ell^{-} \tau^{+} 2j \pslash$ & $e^{+}$$e^{-}$$\tau^{+}$2$j$ \\ 
$\ell^{+}\ell^{-} \ell' \gamma \pslash$ & $e^{+}$$\mu^{+}$$\mu^{-}$$\gamma$$j$ \\ 
$\ell^{+}\ell^{-} \ell' \pslash$ & $e^{+}$$\mu^{+}$$\mu^{-}$, $e^{+}$$e^{-}$$\mu^{+}$, $e^{+}$$e^{-}$$\mu^{+}$$\pslash$ \\ 
$\ell^{+}\ell^{-} \gamma$ & $e^{+}$$e^{-}$$\gamma$, $\mu^{+}$$\mu^{-}$$\gamma$, $e^{+}$$e^{-}$$\gamma$$j$, $\mu^{+}$$\mu^{-}$$\gamma$$j$ \\ 
$\ell^{+}\ell^{-} \gamma  \pslash$ & $e^{+}$$e^{-}$$\gamma$$j$$\pslash$, $e^{+}$$e^{-}$$\gamma$$\pslash$ \\ 
$\ell^{+}\ell^{-} \gamma \tau^{+}  \pslash$ & $e^{+}$$e^{-}$$\tau^{+}$$\gamma$ \\ 
$\ell^{+}\ell^{-} \pslash$ & $e^{+}$$e^{-}$$\pslash$, $e^{+}$$e^{-}$$j$$\pslash$, $\mu^{+}$$\mu^{-}$$\pslash$, $\mu^{+}$$\mu^{-}$$j$$\pslash$, $e^{+}$$e^{-}$$b$$\pslash$ \\ 
$\ell^{+}\ell^{-} \pslash  \tau^{+}$ & $e^{+}$$e^{-}$$\tau^{+}$, $e^{+}$$e^{-}$$\tau^{+}$$j$, $\mu^{+}$$\mu^{-}$$\tau^{+}$ \\ 
$\ell^{+}\ell^{-} 4j$ & $e^{+}$$e^{-}$4$j$, $\mu^{+}$$\mu^{-}$4$j$ \\ 
$\ell^{+}\ell^{-} 4j \pslash$ & $e^{+}$$e^{-}$4$j$$\pslash$ \\ 
$\ell^{+}\ell^{-} \tau^{+} 4j \pslash$ & $e^{+}$$e^{-}$$\tau^{+}$4$j$ \\ 
$\ell^{+} \ell'^{+} jj$ & $e^{+}$$\mu^{+}$3$j$ \\ 
$\ell^{+} \ell'^{+} \pslash jj$ & $e^{+}$$\mu^{+}$2$j$$\pslash$ \\ 
$\ell^{+} \ell'^{-} jj$ & $e^{+}$$\mu^{-}$2$j$ \\ 
$\ell^{+} \ell'^{-} \pslash jj$ & $e^{+}$$\mu^{-}$3$j$$\pslash$, $e^{+}$$\mu^{-}$2$j$$\pslash$ \\ 
$W \gamma jj$ & $\mu^{+}$$\gamma$2$j$$\pslash$, $e^{+}$$\gamma$2$j$$\pslash$, $\mu^{+}$$\gamma$3$j$$\pslash$, $e^{+}$$\gamma$3$j$$\pslash$ \\ 
$W jj$ & $e^{+}$2$j$$\pslash$, $\mu^{+}$2$j$$\pslash$, $e^{+}$3$j$$\pslash$, $\mu^{+}$3$j$$\pslash$ \\ 
$\ell^{+} \tau^{+} \pslash jj$ & $\mu^{+}$$\tau^{+}$3$j$$\pslash$ \\ 
$\ell^{+} \tau^{-} \pslash jj$ & $e^{+}$$\tau^{-}$2$j$$\pslash$, $e^{+}$$\tau^{-}$3$j$$\pslash$, $\mu^{+}$$\tau^{-}$3$j$$\pslash$, $\mu^{+}$$\tau^{-}$2$j$$\pslash$ \\ 
\end{tabular} 
\hspace{0.6cm}
\begin{tabular}{p{1.5cm}p{5cm}} 
\Sleuth\ & \Vista\ Final States   \\ 
\hline 
\hline 
$\ell^{+} \tau^{+} jj$ & $e^{+}$$\tau^{+}$2$j$, $\mu^{+}$$\tau^{+}$2$j$, $e^{+}$$\tau^{+}$3$j$ \\ 
$\ell^{+} \tau^{-} jj$ & $e^{+}$$\tau^{-}$2$j$, $e^{+}$$\tau^{-}$3$j$, $\mu^{+}$$\tau^{-}$2$j$, $\mu^{+}$$\tau^{-}$3$j$ \\ 
$\ell^{+} \ell'^{+}$ & $e^{+}$$\mu^{+}$, $e^{+}$$\mu^{+}$$j$ \\ 
$\ell^{+} \ell'^{+} \pslash$ & $e^{+}$$\mu^{+}$$j$$\pslash$, $e^{+}$$\mu^{+}$$\pslash$ \\ 
$\ell^{+} \ell'^{-}$ & $e^{+}$$\mu^{-}$, $e^{+}$$\mu^{-}$$j$ \\ 
$\ell^{+} \ell'^{-} \gamma \pslash$ & $e^{+}$$\mu^{-}$$\gamma$$j$$\pslash$ \\ 
$\ell^{+} \ell'^{-} \pslash$ & $e^{+}$$\mu^{-}$$\pslash$, $e^{+}$$\mu^{-}$$j$$\pslash$, $e^{+}$$\mu^{-}$$b$$\pslash$ \\ 
$W \gamma$ & $\mu^{+}$$\gamma$$\pslash$, $e^{+}$$\gamma$$\pslash$, $\mu^{+}$$\gamma$$j$$\pslash$, $e^{+}$$\gamma$$j$$\pslash$ \\ 
$\ell^{+} \tau^{-} \gamma$ & $e^{+}$$\tau^{-}$$\gamma$ \\ 
$W             $ & $e^{+}$$\pslash$, $\mu^{+}$$\pslash$, $e^{+}$$j$$\pslash$, $\mu^{+}$$j$$\pslash$, $e^{+}$$b$$\pslash$, $\mu^{+}$$b$$\pslash$ \\ 
$\ell^{+} \tau^{+} \pslash$ & $e^{+}$$\tau^{+}$$\pslash$, $\mu^{+}$$\tau^{+}$$\pslash$, $e^{+}$$\tau^{+}$$j$$\pslash$, $\mu^{+}$$\tau^{+}$$j$$\pslash$ \\ 
$\ell^{+} \tau^{-} \pslash$ & $e^{+}$$\tau^{-}$$\pslash$, $e^{+}$$\tau^{-}$$j$$\pslash$, $\mu^{+}$$\tau^{-}$$\pslash$, $\mu^{+}$$\tau^{-}$$j$$\pslash$ \\ 
$\ell^{+} \tau^{+}$ & $e^{+}$$\tau^{+}$, $e^{+}$$\tau^{+}$$j$, $\mu^{+}$$\tau^{+}$, $\mu^{+}$$\tau^{+}$$j$ \\ 
$\ell^{+} \tau^{-}$ & $e^{+}$$\tau^{-}$, $\mu^{+}$$\tau^{-}$, $e^{+}$$\tau^{-}$$j$, $\mu^{+}$$\tau^{-}$$j$, $e^{+}$$\tau^{-}$$b$ \\ 
$W \gamma 4j$ & $\mu^{+}$$\gamma$4$j$$\pslash$, $e^{+}$$\gamma$4$j$$\pslash$ \\ 
$W 4j$ & $e^{+}$4$j$$\pslash$, $\mu^{+}$4$j$$\pslash$ \\ 
$\ell^{+} 4j \tau^{-}$ & $e^{+}$$\tau^{-}$4$j$ \\ 
$W \gamma \gamma$ & $e^{+}$2$\gamma$$\pslash$, $\mu^{+}$2$\gamma$$\pslash$ \\ 
$jj$ & 2$j$, 3$j$ \\ 
$\gamma jj$ & $\gamma$2$j$, $\gamma$3$j$ \\ 
$\gamma \pslash jj$ & $\gamma$2$j$$\pslash$, $\gamma$3$j$$\pslash$ \\ 
$jj \pslash$ & 3$j$$\pslash$, 2$j$$\pslash$ \\ 
$\tau \pslash jj$ & $\tau^{+}$2$j$$\pslash$, $\tau^{+}$3$j$$\pslash$ \\ 
$\gamma \gamma jj$ & 2$\gamma$2$j$, 2$\gamma$3$j$ \\ 
$jj \gamma \gamma \pslash$ & 2$\gamma$2$j$$\pslash$ \\ 
$\gamma \gamma \gamma jj$ & 3$\gamma$2$j$ \\ 
$\gamma j$ & $\gamma$$j$, $\gamma$$b$ \\ 
$\gamma \pslash$ & $\gamma$$\pslash$, $\gamma$$j$$\pslash$, $\gamma$$b$$\pslash$ \\ 
$\tau \pslash \gamma$ & $\tau^{+}$$\gamma$$\pslash$, $\tau^{+}$$\gamma$$j$$\pslash$ \\ 
$j \pslash$ & $j$$\pslash$, $b$$\pslash$ \\ 
$\tau \pslash$ & $\tau^{+}$$j$$\pslash$, $\tau^{+}$$b$$\pslash$ \\ 
$b \bar{b} b \bar{b}$ & 3$b$$j$ \\ 
$W b \bar{b} b \bar{b}$ & $e^{+}$3$b$$j$$\pslash$ \\ 
$\ell^{+} \ell^{+}$ & 2$e^{+}$, 2$e^{+}$$j$, 2$\mu^{+}$ \\ 
$\ell^{+}\ell^{-}\ell^{+} jj \pslash$ & 2$e^{+}$$e^{-}$2$j$, 2$e^{+}$$e^{-}$3$j$ \\ 
$\ell^{+}\ell^{-}\ell^{+} \pslash$ & 2$e^{+}$$e^{-}$, 2$e^{+}$$e^{-}$$j$, 2$e^{+}$$e^{-}$$\pslash$ \\ 
$\ell^{+}\ell^{+} jj$ & 2$e^{+}$2$j$ \\ 
$\ell^{+}\ell^{+} \ell'^{-} \pslash$ & $e^{+}$2$\mu^{-}$$\pslash$ \\ 
$\ell^{+}\ell^{+} \gamma$ & 2$e^{+}$$\gamma$ \\ 
$\ell^{+}\ell^{+} \gamma  \pslash$ & 2$e^{+}$$\gamma$$\pslash$ \\ 
$\ell^{+}\ell^{+} \pslash$ & 2$e^{+}$$\pslash$, 2$e^{+}$$j$$\pslash$ \\ 
$\ell^{+}\ell^{+} 4j$ & 2$e^{+}$4$j$ \\ 
$4j$ & 4$j$ \\ 
$\gamma 4j$ & $\gamma$4$j$ \\ 
$\gamma 4j \pslash$ & $\gamma$4$j$$\pslash$ \\ 
$4j \pslash$ & 4$j$$\pslash$ \\ 
$\tau^{+} \pslash 4j$ & $\tau^{+}$4$j$$\pslash$ \\ 
$\gamma \gamma 4j$ & 2$\gamma$4$j$ \\ 
$\gamma \gamma$ & 2$\gamma$, 2$\gamma$$j$, 2$\gamma$$b$ \\ 
$\gamma \gamma \pslash$ & 2$\gamma$$\pslash$, 2$\gamma$$j$$\pslash$ \\ 
$3 \gamma$ & 3$\gamma$, 3$\gamma$$j$ \\ 
\end{tabular} 
\end{minipage}}
\caption{Correspondence between \Sleuth\ and \Vista\ final states.  The first column shows the \Sleuth\ final state formed by merging the populated \Vista\ final states in the second column.  Charge conjugates of each \Vista\ final state are implied.}
\label{tbl:sleuthFinalStateContentIndex}
\end{table*}

Table~\ref{tbl:sleuthFinalStateContentIndex} lists the \Vista\ final states associated with each \Sleuth\ final state.


\subsection{Minimum number of events}

This section expands on a subtle point in the definition of the \Sleuth\ algorithm:  for purely practical considerations, only final states in which three or more events are observed in the data are considered.  

Suppose $\scriptP_{e^+e^-b\bar{b}} = 10^{-6}$; then in computing $\twiddleScriptP$ all final states with $b>10^{-6}$ must be considered and accounted for.  (A final state with $b=10^{-7}$, on the other hand, counts as only $\approx 0.1$ final states, since the fraction of hypothetical similar experiments in which $\scriptP<10^{-6}$ in this final state is equal to the fraction of hypothetical similar experiments in which one or more events is seen in this final state, which is $10^{-7}$.)  This is a large practical problem, since it requires that all final states with $b>10^{-6}$ be enumerated and estimated, and it is difficult to do this believably.

To solve this problem, let \Sleuth\ consider only final states with at least ${d_{\text{min}}}$ events observed in the data.  The goal is to be able to find $\twiddleScriptP < 10^{-3}$.  There will be some number $N_{\text{fs}}(b_{\text{min}})$ of final states with expected number of events $b>b_{\text{min}}$, writing $N_\text{fs}$ explicitly as a function of $b_{\text{min}}$; thus $b_{\text{min}}$ must be chosen to be sufficiently large that all of these $N_{\text{fs}}(b_{\text{min}})$ final states can be enumerated and estimated.  The time cost of simulating events is such that the integrated luminosity of Monte Carlo events is at most 100 times the integrated luminosity of the data; this practical constraint restricts $b_{\text{min}}>0.01$.  The number of \Sleuth\ Tevatron Run II final states with $b>0.01$ is $N_{\text{fs}}(b_{\text{min}}=0.01) \approx 10^3$.  

For small $\scriptP_{\text{min}}$, keeping the first term in a binomial expansion yields $\twiddleScriptP = \scriptP_{\text{min}} N_{\text{fs}}(b_{\text{min}})$, where $\scriptP_{\text{min}}$ is the smallest $\scriptP$ found in any final state.  From the discussion above, the computation of $\twiddleScriptP$ from $\scriptP_{\text{min}}$ can only be justified if $\scriptP_{\text{min}} > ({b_{\text{min}}}^{d_{\text{min}}})$; if otherwise, final states with $b<b_{\text{min}}$ will need to be accounted for.  Thus $\twiddleScriptP$ can be confidently computed only if $\twiddleScriptP > ({b_{\text{min}}}^{d_{\text{min}}}) N_{\text{fs}}(b_{\text{min}})$.

Solving this inequality for $d_{\text{min}}$ and inserting values from above,
\begin{equation}
d_{\text{min}}  \geq  \frac{\log_{10}{\twiddleScriptP} - \log_{10}{N_{\text{fs}}(b_{\text{min}})}}{\log_{10}{b_{\text{min}}}} \approx  \frac{-3 - 3}{-2} = 3.
\end{equation}
A believable trials factor can be computed if $d_{\text{min}} \geq 3$.


At the other end of the scale, computational strength limits the maximum number of events \Sleuth\ is able to consider to $\lesssim 10^4$.  Excesses in which the number of events exceed $10^4$ are expected to be identified by \Vista's normalization statistic.

For each final state, pseudo experiments are run until $\scriptP$ is determined to within a fractional precision of 5\% or a time limit is exceeded.  If the time limit is exceeded before $\scriptP$ is determined to within the desired fractional precision of 5\%, \Sleuth\ returns an upper bound on $\scriptP$, and indicates explicitly that only an upper bound has been determined.  For the data described in this article, the desired precision is obtained.


\bibliography{prd1}

\end{document}